\documentclass[a4paper,10pt,onecolumn]{article}
\usepackage[english]{babel}
\usepackage{fancyhdr} 
\usepackage{url} 
\usepackage{caption} 
\usepackage{subcaption} 
\usepackage{natbib} 
\usepackage{multirow} 
\usepackage{float} 
\usepackage{amssymb}
\usepackage[intlimits]{amsmath} 
\usepackage{amsthm} 
\usepackage[german]{nomencl}
\usepackage{booktabs} 
\usepackage{geometry}
\usepackage{enumitem}
\usepackage{array}
\makenomenclature
\usepackage[pdfborder={000},colorlinks=true,linkcolor=blue,citecolor=blue]{hyperref}
\usepackage{listings}
\usepackage[usenames,dvipsnames]{color}
\usepackage{tikz}

\usetikzlibrary{mindmap,trees, backgrounds}

\definecolor{b}{rgb}{0,0,.8}	
\definecolor{g}{rgb}{0,.6,0}	
\definecolor{n}{rgb}{0,0,0}	
\definecolor{h}{rgb}{0.4,0.2,0.2}	
\definecolor{v}{rgb}{0.2,0.6,0}

\newcommand{\C}{{\mathbb C}}

\newcommand{\E}{{\mathbb E}}

\newcommand{\R}{{\mathbb R}}

\newcommand{\Z}{{\mathbb Z}}

\newcommand{\GG}{{\mathcal{G}}}

\newcommand{\II}{{\mathcal{I}}}

\newcommand{\WW}{{\mathcal{W}}}

\newcommand{\bsF}{\boldsymbol F}

\newcommand{\bsP}{\boldsymbol P}

\newcommand{\bsX}{\boldsymbol X}

\newcommand{\bsnull}{\boldsymbol 0}

\newcommand{\bsbeta}{\boldsymbol \beta}

\newcommand{\bseps}{\boldsymbol \varepsilon}

\newcommand{\bspsi}{\boldsymbol \psi}
\newcommand{\bsphi}{\boldsymbol \phi}
\newcommand{\bsGamma}{\boldsymbol \Gamma}
\newcommand{\bsPhi}{\boldsymbol \Phi}
\newcommand{\bsLambda}{\boldsymbol \Lambda}

\newcommand{\eps}{{\varepsilon}}

\DeclareMathOperator*{\argmin}{arg\,min}

\newcommand{\ov}\overline
\newcommand{\what}{\widehat}
\newcommand{\wtilde}{\widetilde}

\newcommand{\rig}\right
\newcommand{\lef}\left
\newcommand{\nf}\normalfont

\definecolor{rickgreen}{rgb}{0,0.6,0}

\begin{document}
\title{Forecasting Electricity Spot Prices using Lasso: \\ On Capturing the Autoregressive Intraday Structure}

\author{Florian~Ziel}

\maketitle

\begin{abstract}
In this paper we present a regression based model for day-ahead electricity spot prices. We estimate the considered linear regression model by the lasso estimation method. The lasso approach allows for many possible parameters in the model, but also shrinks and sparsifies the parameters automatically to avoid overfitting. Thus, it is able to capture the autoregressive intraday dependency structure of the electricity price well. 
We discuss in detail the estimation results which provide insights to the intraday behavior of electricity prices. We perform an out-of-sample forecasting study for several European electricity markets. 
The results illustrate well that the efficient lasso based estimation technique can exhibit advantages 
from two popular model approaches.
\end{abstract}


\section{Introduction and motivation} \label{Introduction}

In day-ahead electricity spot price forecasting there exists a vast range of differently structured models.
A recent overview with a classification is given in \cite{weron2014electricity}.

Most of the day-ahead electricity spot prices are traded on an hourly (or half-hourly) basis. 
This leads to the obvious fact, that we have 24 (or 48) observations a day if we ignore the 
clock change issue. Subsequently we only consider hourly traded prices, but for half-hourly data the adjusted facts hold as well.
All electricity price models rely heavily on this fact of having 24 observations a day. 
These models can be divided into two classes independent of whether the model is designed e.g. for point or for probabilistic forecasts.

Modeling approaches which belong to the first class construct 24 separate models, one model for each hour of the day. 
These models can be e.g. a standard time series method (e.g. \cite{paraschiv2014impact}, \cite{nowotarski2014computing},\cite{maciejowska2015short})
or a regime switching model (e.g. \cite{karakatsani2008intra}). 
In general, the model approach is usually motivated by the fact that there is 
one auction each day (mostly in the morning or at lunch time) for the next 24 hours\footnote{Once a year for the last Sunday March only 23 and for last Sunday in October 25 hours due to daylight saving time}.
Hence, we observe the 24 traded prices at once. Thus, modeling a 24-dimensional process 
by 24 small models is an obvious way to proceed.

The second class contains electricity spot price models that 
propose a single model for the full observed data.
Here a 24 hours ahead forecast is computed. This approach is motivated 
by seeing the price process as one time series that evolves over time as related processes like electricity load or renewable energy 
generation.
As in the first model class it can be of various model types, 
like time series based approaches (e.g. \cite{haldrup2010vector}, \cite{cruz2011effect}, \cite{keles2012comparison}, \cite{paraschiv2015spot} or \cite{ziel2015efficient}) 
or machine learning models (e.g. \cite{conejo2005day}, \cite{singhal2011electricity} or \cite{chaabane2014hybrid}).


From the applied point of view both model approaches are valid as they can have a good forecasting performance in particular situations.
Nevertheless, the first one seems to be more plausible, 
as it matches more directly the data generating process. However, in the first class with 24 models for each hour
we have usually less observations and consequently 24 simple and low parametrized models (see e.g. \cite{weron2008forecasting} or \cite{maciejowska2015probabilistic}). 
More crucially, most methods assume the same model structure for all 24 hours.
The model structure is often assumed in such a way, that 
the price at hour $i$ depends on the price of the previous day at hour $i$, but not on a different hour $j$.
Many models do not allow for such a dependency to simplify the model structure, even though 
there might be strong dependency present. 
Here models of the second class can have an advantage.
As they use more observations for the estimation the resulting model 
tends to have more involved parameters that can theoretically capture the dependence, if the model design allows for it. 
Interestingly the vast majority of machine learning models (esp. artificial neural networks) 
fall into this second model class that estimates one big model, see e.g. \cite{conejo2005day}, \cite{singhal2011electricity}, \cite{catalao2011hybrid}, \cite{shafie2011price}, \cite{hu2014novel} or \cite{abedinia2015electricity}). 
It seems that these estimation techniques rely much more on the fact of having a lot of observations than 
time series based approaches.

To better motivate and understand this dependency problem we introduce $P_{d,h}$ as electricity price 
at day $d$ and hour $h$. Throughout the paper we ignore we clock change issue, and interpolate the missing 2am
hour in March and average the double occurring 2am hour in October so that there are 24 observable prices each day.

\begin{figure*}[hbt!]
\centering
\begin{subfigure}[b]{.49\textwidth}
 \includegraphics[width=1\textwidth]{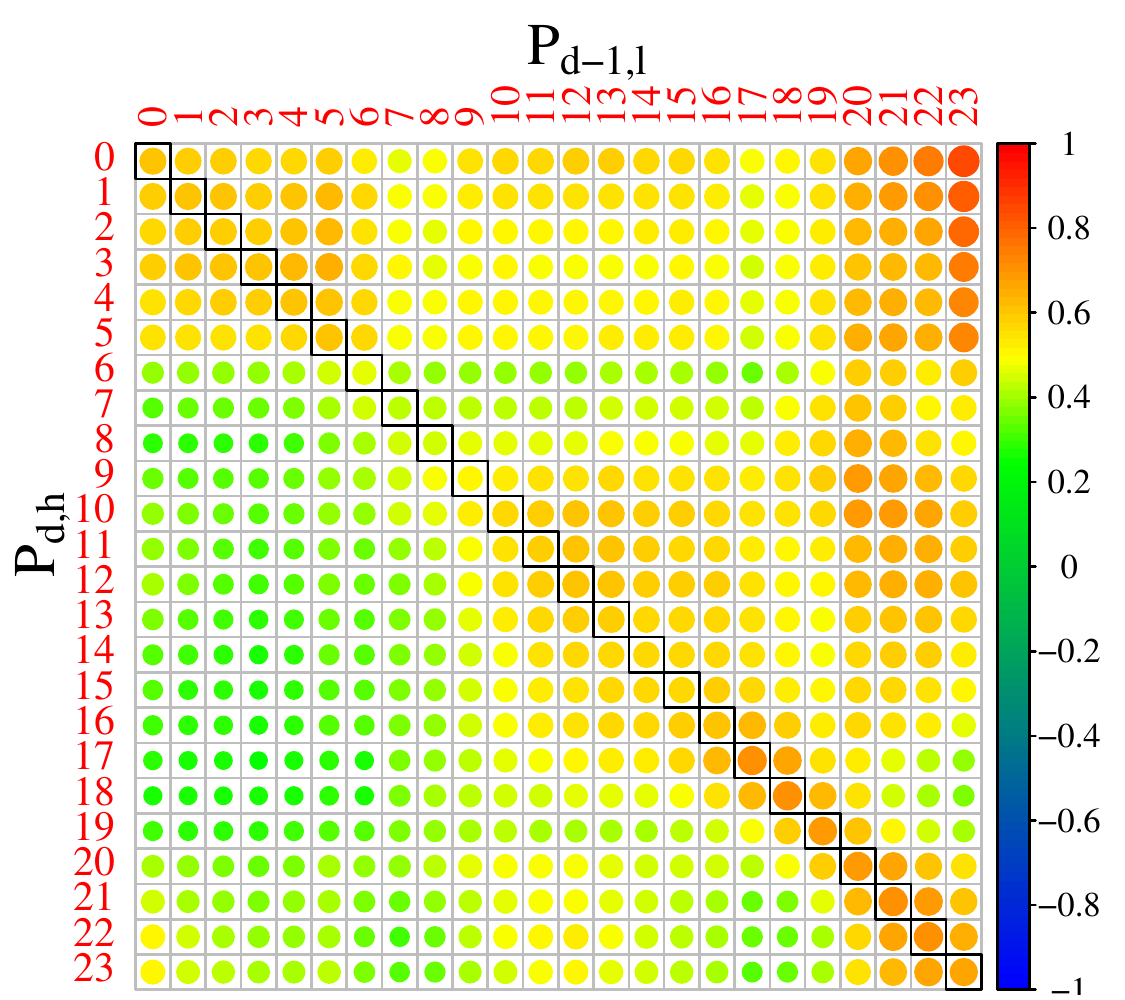} 
  \caption{EPEX spot price for Germany and Austria}
  \label{fig_epex}
\end{subfigure}
\begin{subfigure}[b]{.49\textwidth}
 \includegraphics[width=1\textwidth]{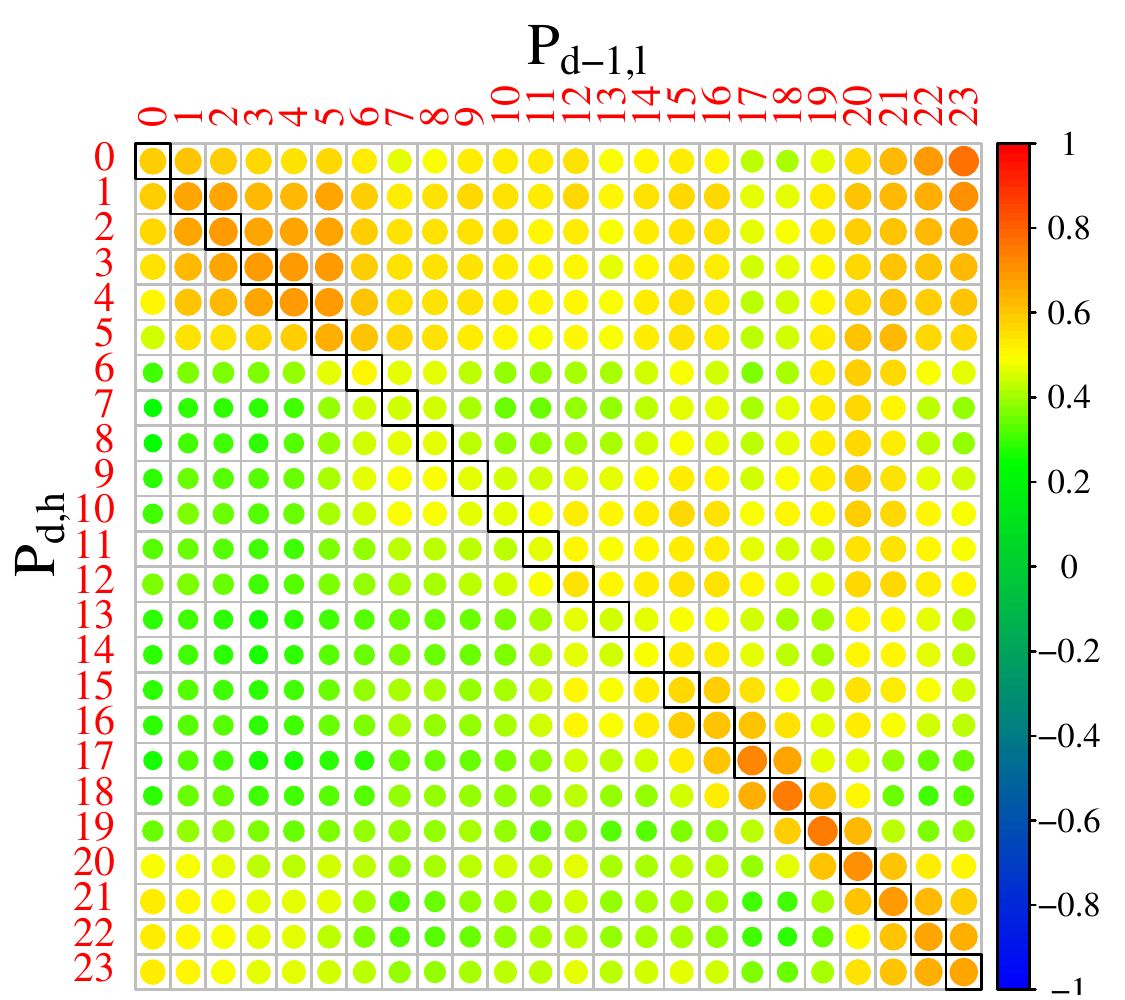} 
  \caption{APX spot price for Netherlands}
  \label{fig_apx}
\end{subfigure}
\caption{Sample correlations of $\C\text{or}( P_{d,h}, P_{d-1,l})$ for $h,l = 0,\ldots, 23$ and two markets 
from 17.12.2009 to 12.08.2014.}
 \label{fig_intro_corr}
\end{figure*}

Now we have a closer look at the linear dependency structure between the price $P_{d,h}$ at day $d$ and hour $h$
and the price $P_{d-1,l}$ at the previous day $d-1$ and hour $l$. 
The corresponding sample correlations of the correlations $\C\text{or}( P_{d,h}, P_{d-1,l})$ are exemplary visualized in Figure \eqref{fig_intro_corr}
for two selected markets, namely the EPEX spot price for Germany and Austria, and the 
APX spot price for Netherlands.

The highlighted diagonal elements measure the linear dependency between the electricity price at hour $h$ 
to its previous day value at the same hour $h$. 
We observe relatively high correlations on the diagonal. 
But, only for some hours in the night, the late afternoon and the evening they are the largest of each row.
For most of the hours from midnight to the afternoon this is not the case. 
They exhibit their largest correlation with hours in the late evening.
The most distinct lagged cross-hour dependency concerns
$Y_{d, 0}$. It depends much more on the last hour price $Y_{d-1,23}$ of the previous day than on the same hour price $Y_{d-1,0}$.

It is clear that this complex dependency structure should be covered by an appropriate spot price model. 
As mentioned before, most of the models of the first class do not capture this dependency structure, whereas
the second one in principle does. But, the second class that uses one big model has the problem, that e.g. the forecast $\what{Y}_{d+1,23}$ will depend on the previous hour predictions $\what{Y}_{d+1, 0}, \ldots, \what{Y}_{d+1, 22}$. Thus 
the prediction error $\what{\eps}_{d+1,h} = Y_{d+1, h} - \what{Y}_{d+1, h}$ tend to increase in $h$ as we have to compute 
24 hours ahead forecast. 
And even though models of the second class can capture the cross-hour dependency, we observe that the full dependency pattern of Figure \ref{fig_intro_corr} is time-varying.
Thus, it is required for models of the second class to utilize time-varying parameters for the autoregressive effect.
This is usually not included in many models of this model class, see e.g. \cite{conejo2005day},
\cite{singhal2011electricity}, \cite{catalao2011hybrid}, \cite{shafie2011price}, \cite{hu2014novel} or \cite{abedinia2015electricity}.

Interestingly, when we compare forecasting performance of relatively simple models of both model classes,
 models of the second class perform better for the first half of the day, whereas 
models of the first class are better in the second half of the day. This observation is also part of the results of the empirical section.
Nevertheless, it is not clear which modeling method is superior in terms of forecasting performance. 

In this paper we introduce a simple linear model approach that falls into the first model class. 
It contains 24 simple structured models and captures the mentioned dependency pattern directly. Thus, it joins 
the advantages of both model classes which leads to an improved overall forecasting performance.
The model captures the full linear autoregressive dependency structure as seen in 
Figure \ref{fig_intro_corr}, especially time-varying cross-hour dependencies.
Furthermore, it also allows for a time-varying day-of-the-week effect that will be discussed more 
detailed in the next section. In electricity price forecasting literature such flexibility is a clear novelty even though there are models 
 available that cover the same feature. For example, an autoregressive principal component analysis based approach as used in
\cite{maciejowska2013forecasting} and \cite{maciejowska2015short} can cover this behavior. However, there the impact is not modeled directly but indirectly
through the underlying dimension reduction approach. 

We restrict the model to these two modeling aspects (time-varying cross-hour dependencies and time-varying day-of-the-week effects) to highlight the advantage of the modeling and estimation approach.
Especially, we do not focus on well known features like volatility clustering, non-linearity effects (esp. price spikes), long-term trends, structural breaks, public holidays effects, effects of other external regressors
like renewable energy feed-in, or the day-light saving time effect.
However extensions to all these directions are possible.
We discuss this issue briefly in the model section and final discussion.

The crucial part of the estimation relies on high-dimensional statistics. 
We utilize the lasso (least absolute shrinkage and selection operator) method introduced by \cite{tibshirani1996regression},
as a regularized estimation technique.
Due to the shrinkage properties of the estimators we can handle models with many potential parameters.
As the estimation procedure has the sparsity property,
many of the potential parameters are not included in the finally estimated model. This property is great to avoid the 
overfitting problem. 

The considered estimation technique is very fast. The estimation and forecasting of the 24 models is feasible
in a few seconds without big effort.
Thus, the model is also suitable as being an element of a forecast combination approach, 
see e.g. \cite{bordignon2013combining} or \cite{maciejowska2015probabilistic}. 
For such applications fast estimation methods are clearly preferable.

The lasso based approach is not completely new in electricity price forecasting literature. 
It is used by \cite{ziel2015efficient} and \cite{ludwig2015putting} for electricity price forecasting.
Both models use the variable selection property, but consider a model of the second class and share the mentioned disadvantages.
The approach of \cite{ziel2015efficient} uses time-varying coefficients to capture the mentioned time-varying changes in the dependency structure. 
However, the time-varying coefficients are modeled using B-splines, that only allow for smooth changes over the day.
In contrast, our model belongs to the first model class and allows for more flexibility and abrupt changes in the cross-hour dependency. 
Additionally, \cite{ziel2015efficient} incorporate the impact of load and wind and solar net feed-in such as volatility clustering, public holiday and daylight saving time effects.
In \cite{ludwig2015putting} the parameters are assumed to be constant over time, so the lasso estimation works just as features selection
 and ignores time-varying dependencies. They use wind speed and temperature data to model exogenous impacts.
In recent research of \cite{dudek2016pattern} the lasso approach is applied to short-term electricity load forecasting. 

The paper is structured as follows:
In the second section we introduce the time series model and estimation method.
In the third one we discuss several benchmarks for the empirical study which is performed in the fourth section.
For the out-of-sample sample forecasting study we consider 10 different hourly European day-ahead electricity spot prices.
We close with a summary and some conclusions.

\section{Time series model and estimation technique}

\begin{figure*}[hbt!]
\centering
\begin{subfigure}[b]{.49\textwidth}
 \includegraphics[width=1\textwidth]{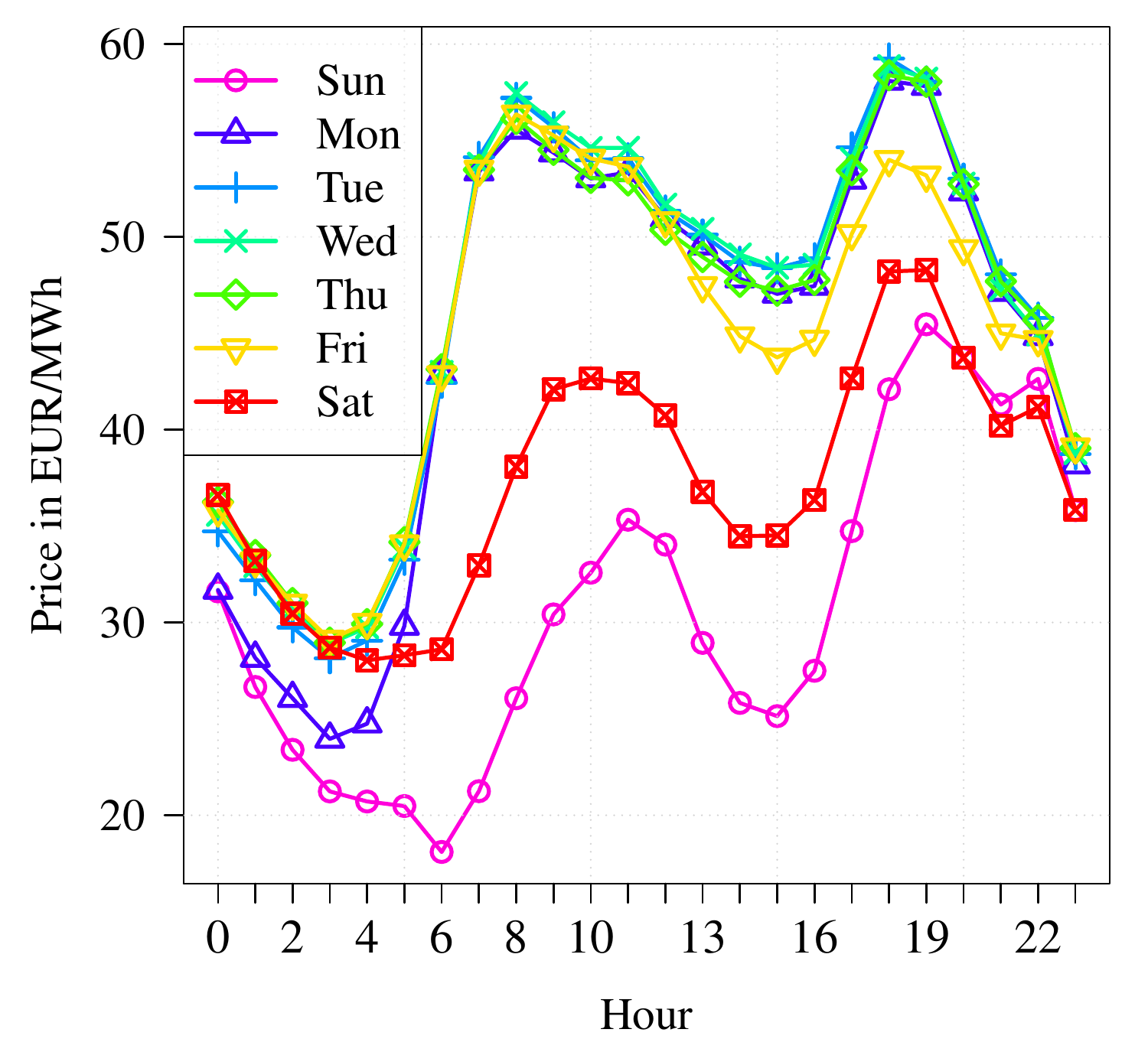} 
  \caption{EPEX Germany and Austria}
  \label{fig_wm_epex}
\end{subfigure}
\begin{subfigure}[b]{.49\textwidth}
 \includegraphics[width=1\textwidth]{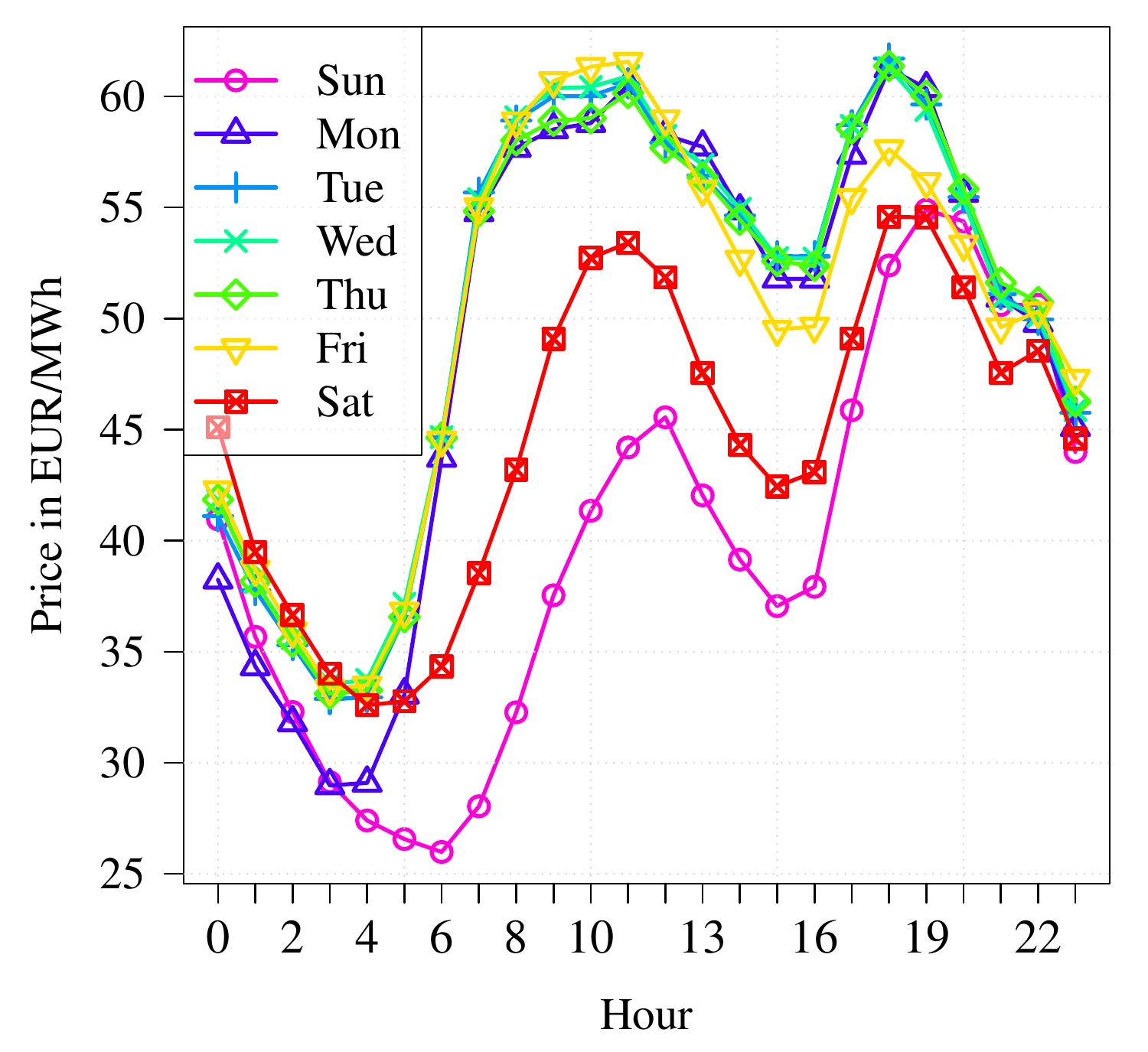} 
  \caption{APX Netherlands}
  \label{fig_wm_apx}
\end{subfigure}
\caption{Weekly sample mean of $P_{d,h}$ for two markets. }
 \label{fig_wm}
\end{figure*}

The time series model that we consider is an autoregressive regression approach, that is similar to the one used in \cite{weron2008forecasting} or 
\cite{maciejowska2015probabilistic}.
In \cite{weron2008forecasting} the electricity price at hour $h$ depends 
linearly on the price at the same hour lagged by $1$, $2$ and $7$ days, such as 
dummies on the weekdays Sunday, Monday and Saturday.
However, the choice of the lags $1$, $2$ and $7$ such as the selection of the weekday dummies is the same for all 24 hours
and it is very restrictive. 
Concerning the weekday structure it is advisable to take a closer look at the 
weekly mean price of the electricity price.
The weekly sample mean of the German/Austrian EPEX spot price and the Dutch APX spot price
is given in Figure \ref{fig_wm}.

There we see the $7\times24=168$ sample mean values grouped by the 7 weekdays and 24 hours of the day.
We can observe that the sample mean of the Sunday and Saturday values is clearly different from the
mean on working days. On the working days from Monday to Friday the mean behavior is relatively similar. 
But for the morning hours on Monday and the afternoon and evening hours on Friday we observe clear deviations. 
This transition towards and from the weekend is known for the German/Austrian EPEX and modeled by \cite{ziel2015efficient}
by an approach of the second model class using time-varying parameters.
Most importantly it suggests that the impact of the weekday dummies in an autoregressive regression model
might be different for different hours of the day.
Hence, we require flexibility in the model that is able tackle these possible changes in the day-of-the-week dependency structure.

For the time series model that we introduce for the electricity price $P_{d,h}$
we assume that the processes $(P_{d,h})_{d\in \Z}$ are stationary for each hour $h$. 
Furthermore, we do not directly model $P_{d,h}$ itself, but its
zero mean version $Y_{d,h} = P_{d,h} - \mu_h$ with $\mu_h = \E(P_{d,h})$. We can estimate $\mu_h$ 
easily by the sample mean.

To model the mentioned weekday effects we define the weekday indicators
$$W_{k}(d) = \begin{cases}
              1 &, \WW(d) = k \\
              0 &, \WW(d) \neq k \\
             \end{cases}.
$$
Here $\WW(d)$ is a function that gives a number which represent each day of the week. We use without loss
of generality the enumeration $k=0$ for a Sunday, $k=1$ for a Monday, $\ldots$, $k=6$ for a Saturday. 
As we want to point out the effect of the weekdays, we propose two models, one without
the weekday effects, the other one with possible weekday effects.

For each hour $h$ the two considered time series models for $Y_{d,h}$ are given by
\begin{align}
 Y_{d,h} &= \sum_{l=1}^{24} \sum_{k \in \II(h,l)} \phi_{h,k,l} Y_{d-k,l} + \eps_{d,h} \ \ \text{ and} 
 \label{eq_model1} \\
 Y_{d,h} &= \sum_{l=1}^{24} \sum_{k \in \II(h,l)} \phi_{h,k,l} Y_{d-k,l} 
 + \sum_{k=0}^6 \psi_{h,k} W_k(d) + \eps_{d,h}
 \label{eq_model2}
\end{align}
with parameters $\phi_{h,k,l}$ and $\psi_{h,k}$, 
 $\II(h,l)$ as an index set of possible lags and $\eps_{d,h}$ as error term.
 We assume that the error process $(\eps_{d,h})_{d\in \Z}$ is uncorrelated 
 with constant variance $\sigma_h^2$. The parameters $\phi_{h,k,l}$ 
 model the linear autoregressive impact of the past prices
 and $\psi_{h,k}$ a weekday effect.

 
 The index sets $\II(h,l)$ are very important for the model. They 
specify the possible model structure. 
Of course larger sets $\II(h,l)$ increase the probability of overfitting,
even though the used regularization technique reduces this problem a lot. 
If we choose $\II(h,l)$ too small we might miss important information in the dependency structure.
Therefore we should choose $\II(h,l)$ reasonably, in the sense that 
we should only include those lags where we, as a modeler, think that a dependence is plausible. 
In our situation we choose 
\begin{align}
\II(h,l) = \begin{cases}
\{1, \ldots, 36\} &, h= l\\
\{1, \ldots, 8\} &, h \neq l             
             \end{cases}. 
\end{align}
Thus, we include a possible relationship between $Y_{d,h}$ and the previous 36 days (five weeks plus a day) at the same hour $h$,
but also a dependence to the past 8 days of the electricity prices of other hours. 
             
Furthermore, model equation \eqref{eq_model2} (also \eqref{eq_model1}) is a regression model, so it is very modular. It is extremely easy
to add new regressors, like lagged electricity load or renewable energy feed-in, seasonal cycles, trends, or event dummies. 
These additional regressors can be used to cover some of the ignored stylized facts, 
like long-term trends, structural breaks, public holidays and day-light saving time effects.
Even non-linear impacts in $Y_{d-k,h}$ like in a threshold-AR process are possible,
as long as the relationship is linear in the parameters. Another direction of extensions 
concerns the volatility of the residuals. 
\cite{ziel2015efficient} and \cite{ziel2015iteratively} 
show that an extension to AR-ARCH type processes is possible as well, using an iterative lasso estimation technique.


For the estimation of the parameters in \eqref{eq_model1} or \eqref{eq_model2} 
we will use the corresponding ordinary least square (OLS) representation.
We denote it by
\begin{align}
  Y_{d,h} =  \bsX_{d,h}' \bsbeta_h + \eps_{d,h}.
  \label{eq_model_ols}
\end{align}
with the vector of regressors $\bsX_{d,h} = (\bsX_{d,h,1}, \ldots, \bsX_{d,h, m_h})'$ of length $m_h$ 
and parameter vector $\bsbeta_h = (\beta_{h,1} , \ldots, \beta_{h,m_h})' $ of length $m_h$.
Furthermore, we require for the lasso estimation technique 
the standardized version of \eqref{eq_model_ols}.
Thus, we introduce it with
\begin{align}
  \wtilde{Y}_{d,h} =  \wtilde{\bsX}_{d,h}' \wtilde{\bsbeta}_h + \wtilde{\eps}_{d,h} .
  \label{eq_model_ols_stand}
\end{align}
 Here $\wtilde{Y}_{d,h}$ and the elements of $\wtilde{\bsX}_{d,h}$ 
are scaled in such a way that they have a variance of 1. In the empirical part we scale using the corresponding sample 
standard deviations. If we have $\wtilde{\bsbeta}_h$ we can calculate the parameter vector $\bsbeta_h$
of \eqref{eq_model_ols} easily by rescaling.

The mentioned lasso estimation technique is a penalized regression approach that penalizes the parameters on the sum of the absolute values.
Let $D$ be the number of observable days. Then the lasso estimator $\what{\wtilde{\bsbeta}}_h$ for $\wtilde{\bsbeta}_h$ is given by
\begin{align}
 \what{\wtilde{\bsbeta}}_h  = \argmin_{\bsbeta \in \R^{m_h}} \sum_{d=1}^D \left(\wtilde{Y}_{d,h} -  \wtilde{\bsX}_{d,h}' \bsbeta \right)^2
 + \lambda_h \sum_{i=1}^{m_h} |\beta_{i}|
\end{align}
where $\lambda_h\geq 0$ is a tuning parameter. Note that for $\lambda_h=0$ we receive 
the standard OLS estimator. 
For extremely huge tuning parameters $\lambda_h$ the 
estimator will return the estimate $\what{\wtilde{\bsbeta}}_h = \bsnull = (0, \ldots, 0)'$ where all parameters are shrinked to zero. 
In general the lasso estimator is a biased estimator. However, it can be tuned in such a way that the out-of-sample performance 
is better than the OLS estimator performance.
The lasso estimator has an automatic sparsity property,
so many of the potential parameters are not included in the finally estimated model. The larger $\lambda_h$ the less parameters are included 
in the estimated model. But lasso does not only sparsify, it also regularizes the model.
So the estimated parameter that are not estimated to be exactly zero are usually smaller (in absolute values) than the 
corresponding OLS solution with the same parameters. To be precise: the 
in-sample residual sum of square (RSS) is larger for the lasso solution than for the corresponding OLS solution
with the same non-zero parameters.
From a Baysian point of view lasso acts as a prior on the parameters that is centered at zero.
 For more information on the lasso algorithm from the theoretical and applied point of view we recommend 
the recently published lasso overview book \cite{hastie2015statistical}.

As estimation algorithm the very fast coordinate descent algorithm of \cite{friedman2007pathwise} is available.
It solves the lasso problem on a given grid $\bsLambda$ of $\lambda$-values that 
is commonly chosen as a exponential grid. For this exponential grid $\bsLambda$
it is important that the corresponding solutions cover the full range from the $\what{\bsbeta} = \bsnull$ solution to
a solution that contains all possible parameters as the OLS solution.  In detail we
define the grid as $\bsLambda = \{ 2^k | k \in \GG \} \cup \{0\} $ where $\GG$ is an equidistant grid from 4 to -15 of length 50.
We use the \texttt{R} package \texttt{glmnet} for implementation, see e.g. \cite{friedman2010regularization}
for more details.

We choose the tuning parameters $\lambda_h\in \bsLambda$ for each hour $h$ by minimizing the Bayesian information criterion (BIC)
for each of the 24 single models. 
The BIC is a very conservative information criteria that avoids overestimation. 
However other information criteria, like the Akaike information criterion (AIC) or cross-validation based criteria, are possible as well.

\section{Benchmarks models}

As benchmarks we allow only those processes that model linear dependency structure in the data.
These models are usually those ones that can be estimated and forecasted in a very fast way, like the proposed 
model with the corresponding estimation method.
Furthermore, we take suitable representatives of the introduced two model classes.
As we consider the lasso based model without \eqref{eq_model1} and with \eqref{eq_model2} day-of-the-week effects,
we consider for the benchmark models a similar weekday modeling extension to keep comparability.

%

\subsection{24 univariate AR(p) models}

The model of 24 univariate AR(p) models is given by
\begin{align}
P_{d,h} = \phi_{h,0} + \sum_{k=1}^p \phi_{h, k} (P_{d-k,h} - \phi_{h,0} )+ \eps_{d,h} 
  \label{eq_24ar}
\end{align}
where $(\eps_{d,h})_{d\in \Z}$ is i.i.d. with $\eps_{d,h} \sim N(0, \sigma_h^2)$. 
It models the linear autoregressive dependence structure of each price to the corresponding hour well.
We estimate the model by solving the Yule-Walker equations, which solution is guaranteed to be stationary. 
We select the model orders $p_h$ by minimizing the Akaike information criteria (AIC) given 
a maximal upper bound of $p_h$ by $p_{\max} = 50$.
For the model version with weekday effects we consider the equation
\begin{align}
  P_{d,h} = & \phi_{h,0} + \sum_{j=0}^6 \psi_{h,j} W_j(d) +
\sum_{k=1}^p \phi_{h, k} \left( P_{d-k,h} - \sum_{j=0}^6 \psi_{h,j} W_j(d-k) - \phi_{h,0} \right) + \eps_{d,h}  .
  \label{eq_24ar_wd}
\end{align}
We estimate model \eqref{eq_24ar_wd} in a two-step approach. First we estimate the day-of-the-week coefficients $\psi_{h,j}$
by OLS and solve the Yule-Walker equations for the resulting residuals. 
As for model \eqref{eq_24ar} we
minimize the AIC with an upper bound for the order $p_h$ of $p_{\max} = 50$. Note that model 
\eqref{eq_24ar_wd} is just one possibility to include the day-of-the-week effects. Other ones are possible as well, 
but give similar results.
We consider the chosen option as it allows for the application of the Yule-Walker equations 
as equivalently done for \eqref{eq_24ar}.

\subsection{Autoregressive expert models}

As mentioned in many applications, like in \cite{weron2008forecasting} or \cite{maciejowska2015probabilistic},
simple autoregressive models with a fixed regression structure are used. As the assumed regression structure is usually
build on some prior knowledge of experts, we refer these models as expert models.
We consider two expert models as benchmarks, a simple expert model and a weekday based expert model.
They are given by
\begin{align}
  P_{d,h} = \phi_{h,0} &+ \phi_{h, 1} P_{d-1,h} + \phi_{h, 2} P_{d-2,h}+ \phi_{h, 7} P_{d-7,h}+ \eps_{d,h} ,
  \label{eq_exp_arp}
\\
 P_{d,h} = \phi_{h,0} &+ \phi_{h, 1} P_{d-1,h} + \phi_{h, 2} P_{d-2,h}+ \phi_{h, 7} P_{d-7,h}+ \psi_0 W_0(d)+ \psi_1 W_1(d)+ \psi_6 W_6(d) + \eps_{d,h} . 
  \label{eq_exp_arp_wd}
\end{align}
The latter one matches the first one but adds linear weekday effects on Sunday, Monday and Saturday.
Here we assume for the error process again that 
$(\eps_{d,h})_{d\in \Z}$ is i.i.d. with $\eps_{d,h} \sim N(0, \sigma_h^2)$. 
We estimate the parameters by solving the corresponding least square problem.
The chosen model of \cite{weron2008forecasting} is just a representative of an expert model.
There are several other versions available as well, they tend to slightly differ in some model components.
For example \cite{karakatsani2008forecasting} propose a model with a linear price dependency 
on a daily lag of order 1 and 7, but not at lag 2. Note that model \eqref{eq_exp_arp}
is nested in model \eqref{eq_24ar}, which is nested in the proposed model \eqref{eq_model1}.
Thus, we expect similar behavior, if the lags of the nested models capture the relevant features.
These models with a simpler structure have the advantage of being easier to interpret and implement. Furthermore,
the overestimation issue is negligible.

\subsection{Univariate AR(p) models}

This is a model of the second model class. \cite{ziel2015efficient} and \cite{ziel2015forecasting} show that the
univariate AR(p) model can be a powerful forecast method. It performs often even better than several modern models.
Denote $P_t = P_{d,h}$ with $t=24d + j$ for $j\in \{0, \ldots, 23\}$ a univariate time series
representation of electricity price at day $d$ and hour $h$.
The considered model and its weekday effect version are given by 
\begin{align}
P_t =& \phi_0 + \sum_{k=1}^p \phi_k (P_{t-k}- \phi_0) + \eps_t \label{eq_arp}, \\
P_t =& \phi_0 + \sum_{j=0}^6 \psi_{j} W_j(\tilde{d}(t)) + \sum_{k=1}^p \left(\phi_k P_{t-k} - \sum_{j=0}^6 \psi_{j} W_j(\tilde{d}(t)-k) - \phi_0 \right) + \eps_t \label{eq_arp_wd}
\end{align}
where $(\eps_{d,h})_{d\in \Z}$ is i.i.d. with $\eps_{d,h} \sim N(0, \sigma_h^2)$ and $\tilde{d}(t) = \lfloor t/24 \rfloor$
gives the day $d$ that corresponds to time point $t$.
Again we estimate the model \eqref{eq_arp} by solving the Yule-Walker equations and 
minimize the AIC with respect to a maximal order of $p_{\max} = 700$.
Note that a maximal oder larger than $168$ is required to capture any weekly dependency structure.
As for model \eqref{eq_24ar_wd} we estimate the weekday version \eqref{eq_arp_wd} by the two-step approach.
We solve the OLS regression for the day-of-the-week parameters and solve the Yule-Walker equations of the residuals by
minimizing the AIC with maximal order of $p_{\max} = 700$. 

\subsection{PCA based autoregressive models}

The model with 24 univariate AR(p) models is a special case of the
24-dimensional AR(p) process given by 
\begin{equation}
\bsP_d = \bsphi_0 + \sum_{k=1}^p \bsPhi_k ( \bsP_{d-k} - \bsphi_0 ) + \bseps_d
\label{eq_mAR} 
\end{equation}
with $24$-dimensional intercept vector $\bsphi_0$, 24$\times$24-dimensional autoregressive parameter matrices $\bsPhi_k$,
$\bsP_d = (P_{d,0}, \ldots, P_{d, 23})'$ and $\bseps_d \sim N_{24}(\bsnull, \boldsymbol \Sigma)$.
Such a general 24-dimensional AR(p)-process has $24+24\times24 \times p = 24+ 576p$ parameters.
This is usually quite a lot and leads quickly to models with too many parameters. 
Note that model \eqref{eq_mAR} is able to capture the lagged cross-hour dependency structure as seen in Figure \ref{fig_intro_corr}.
The proposed model \eqref{eq_model1} is nested in \eqref{eq_mAR} (and vice versa) with an appropriate choice of $\II(h,l)$. 
As the lasso estimation technique overcomes the overestimation problem the same features can be covered. 
Another approach that deals with this over-parametrization problem is principal component analysis (PCA) based.

The approach is based on a dimension reduction of the 24-dimensional process by PCA.
\cite{maciejowska2013forecasting} and \cite{maciejowska2015short} use it and a generalization 
to forecast the UK APX electricity spot prices. They show that the PCA approach can be suitable to capture 
the intraday relationship between the electricity prices.

The principal component decomposition of $\bsP_{d}$ is given by
\begin{align}
 \bsP_{d} = \bsGamma \bsF_{d}  
\end{align}
with $\bsGamma$ as 24$\times$24-dimensional loading matrix
and $\bsF_{d}$ as factor scores at day $d$.  As usual in PCA analysis we assume that 
without loss of generality the columns of $\bsGamma$ are ordered by the absolute value of their corresponding eigenvalue.
For the dimension reduction we consider only the first $K$ principal components of the scores
which we denote by $\bsF_{K,d}$. We denote the corresponding $K\times$24-dimensional reduced loading matrix 
by $\bsGamma_K$.

For forecasting the scores we employ a multivariate AR (VAR) model to the $K$-dimensional vector $\bsF_{K,d}$.
Thus, we assume that the factor scores follow a process that is given by 
\begin{align}
\bsF_{K,d} = \bsphi_{K,0} + \sum_{k=1}^p \bsPhi_{K,k} ( \bsF_{K,d-k} - \bsphi_{K,0}) + \bseps_{K,d} 
\label{eq_pca_mod1}
\end{align}
with $K$-dimensional intercept vector $\bsphi_{K,0}$, $K$ $\times$ $K$ dimensional autoregressive parameter matrices $\bsPhi_{K,k}$ 
and $\bseps_{K,d}$ as $K$-dimensional zero-mean error term vector. Now the dimension reduced model has only $K + K\times K \times p$ parameters.
As it holds  
$\bsP_{d} =  \bsGamma_K \bsphi_{K,0} + \sum_{k=1}^p \bsGamma_K \bsPhi_{K,k} (\bsF_{K,d-k} - \bsphi_{K,0} ) + \bsGamma_K \bseps_{K,d}$
model \eqref{eq_pca_mod1} is a special case of model \eqref{eq_mAR}, but also of the proposed model \eqref{eq_model1} if $\II(h,l)$ is chosen correctly.

Again we estimate model \eqref{eq_pca_mod1} by solving the multivariate Yule-Walker equations and 
minimizing the AIC with respect to a maximal order of $p_{\max} = 50$.
We perform the estimation for $K\in \{2, \ldots , 12\}$. 
Given a forecast $\what{\bsF}_{K,d}$ of the factor scores for day $d$ and an estimate $\what{\bsGamma}$ of the loading matrix 
we can easily get the price forecast by $\what{\bsP}_{d} =  \what{\bsGamma}_K \what{\bsF}_{K,d}$
using the plug-in principle.

The day-of-the-week version of model \eqref{eq_pca_mod1} is given by
\begin{align}
\bsF_{K,d} =& \bsphi_{K,0} + \sum_{j=0}^6 \bspsi_{j} W_j(d) 
+ \sum_{k=1}^p \left( \bsPhi_{K,k} \bsF_{K,d-k} - \sum_{j=0}^6 \bspsi_{j} W_j(d-k) \right) + \bseps_{K,d} 
\label{eq_pca_modwd}
\end{align}
where $\bspsi_{j}$ are the $K$-dimensional vectors of weekday parameter coefficient, 
and $\bsphi_{K,0}$ is again the $K$-dimensional intercept vector, $\bsPhi_{K,k}$ the $K\times K$ autoregressive parameter matrices
and $\bseps_{K,d}$ as $K$-dimensional error process.
As for models \eqref{eq_24ar_wd} and \eqref{eq_arp_wd} we estimate the day-of-the-week model \eqref{eq_arp_wd} by the two-step approach.
First we solve the OLS problem for the weekday parameters and take the Yule-Walker equations solution of the residuals by
minimizing the AIC with maximal possible order of $p_{\max} = 50$ 
and for $K\in \{2, \ldots , 12\}$.


\section{Empirical study and results}

We perform an out-of-sample forecast study by using a rolling window.
As in-sample period we select always $D=730 = 2\times 365$ days (usually two years) of data for the estimation. 
Given the estimated model we perform an out-of-sample forecast for the next 24 values.
The considered data range is from 17.12.2009 to 12.08.2014. Thus, the full out-of-sample period is from 
17.12.2011 to 12.08.2014 which is exactly $N=969$ days.

As forecasting measure we consider the overall MAE and RMSE, as they are also suitable on markets with negative prices.
The MAE is a more robust measure, whereas the RMSE is the optimal measure for least square problems and more sensitive to outliers.
They are given by
\begin{align}
 \text{MAE} &= \frac{1}{24N} \sum_{n=1}^N \sum_{h=0}^{23} |P_{D+n,h} - \what{P}_{D+n,h}|  , \\
\text{RMSE} &= \sqrt{\frac{1}{24N}\sum_{n=1}^N \sum_{h=0}^{23} |P_{D+n,h} - \what{P}_{D+n,h}|^2 }  
\end{align}
where $\what{P}_{D+n,h}$ denotes the forecasted price of day $D+n$.
However, to better understand how the models perform over the day, we use 
the hourly MAE and RMSE at hour $h$, denoted by $\text{MAE}_h$ and $\text{RMSE}_h$.
They are defined through
\begin{align}
 \text{MAE}_h &= \frac{1}{N} \sum_{n=1}^N |P_{D+n,h} - \what{P}_{D+n,h}| , \\
\text{RMSE}_h &= \sqrt{\frac{1}{N}\sum_{n=1}^N |P_{D+n,h} - \what{P}_{D+n,h}|^2 } .  
\end{align}
We conduct the out-of-sample study for a selected number of electricity markets and all considered models.
For the PCA based models in \eqref{eq_pca_mod1} we only report the model with minimal MAE for each market.
The selected electricity prices are 
the German/Austrian EXAA price, the German/Austrian and Swiss EPEX prices, the 
Belgian BELPEX price, the Dutch APX price, the Danish West, Danish East and Sweden(4) Nordpool prices,
the Polish POLPX price and the Czech OTE price. A summary of the considered prices with the subsequently used abbreviations is 
given in Table \ref{tab_bench}.
Note that all prices are hourly day-ahead electricity prices.
For comparison issues we converted all prices in EUR/MWh. If the reported currency is not EUR we consider the corresponding end of day exchange rate of the local currency 
to convert the prices to EUR/MWh.
A selection of different electricity prices is suitable to check if the intraday
structure on different electricity markets is similarly or not.
It also acts as a robustness check for the proposed lasso based estimation method.

\begin{table*}[tbh]
\normalsize
\centering
\begin{tabular}{rllcc}
 \textbf{Exchange} & \textbf{Region}& \textbf{Abbreviation} 
 & $\text{PCA}^*$ &  $\text{PCA}^*\text{-wd}$\\ \hline
 EXAA &Germany\&Austria & EXAA.DE\&AT  
 & 		7&8\\
 EPEX &Germany\&Austria& EPEX.DE\&AT  
 &		6&10\\
 EPEX &Switzerland  & EPEX.CH  
 &			8&6\\
  BELPEX & Belgium &BELPEX.BE 
  &			6&6\\ 
 APX & Netherlands & APX.NL 
 &		6&8\\ 

 Nordpool & Denmark (West) & NP.DK.West  
 &	7&3\\
 Nordpool & Denmark (East) & NP.DK.East 
 &	4&4\\
 Nordpool & Sweden (4) & NP.SW  
 &		6&4\\
  POLPX & Poland (Auction I) & POLPX.PL   
  & 8&8\\
 OTE &  Czech Republic& OTE.CZ  
 & 7&9\\ 
 \end{tabular}
\caption{Summary of the considered electricity markets with abbreviations and 
optimal number of factors for the PCA based models.}
\label{tab_bench}
\end{table*}
For the PCA-based models the optimal number of factors $K$ are between 3 and 10. They are also
listed in Table \ref{tab_bench}. Note that \cite{maciejowska2015short} get an optimal PCA-based model
with 3 principal components for the half-hourly volume-weighted APX UK spot log-price. They also provide 
a nice interpretation for the factors. 

Subsequently, we denote model \eqref{eq_model1} by lasso, 
\eqref{eq_24ar} by 24d.AR, 
\eqref{eq_exp_arp} by exp.AR, 
\eqref{eq_arp} by AR(p) and
\eqref{eq_pca_mod1} by PCA*.  For the day-of-the-week model version we add the suffix '-wd' to the model abbreviation.
The computed MAE and RMSE values are given in Table \ref{tab_mae_rmse}. There we also report the 
estimated standard deviations of MAE and RMSE which are computed using residual based bootstrap with a bootstrap sample size of 10000.
These standard deviations can be used as significance criterion as in \cite{ziel2015forecasting}.
For example if the MAE and RMSE would be normally distributed, then the 1.96-sigma range around the mean matches 
the symmetric 95\% confidence interval. Even though the distribution is likely not normal we can use the Chebyshev inequality or extensions
to make non-parametric statements about the confidence interval.
In Table \ref{tab_mae_rmse} we underlined all models that are not significantly worse than  the best model, where we take the 2-sigma range of the best model as significance indicator.

\setlength{\tabcolsep}{1.3mm}
\begin{table*}[ht]
\small
\centering
\begin{tabular}{rcccccccccc}
  \hline
MAE & lasso & lasso-wd & 24d.AR & 24d.AR-wd & exp.AR & exp.AR-wd & AR(p) & AR(p)-wd & PCA* & PCA*-wd \\ 
  \hline
  \multirow{2}{*}{EXAA.DE\&AT} & 4.89 & \textbf{4.61} & 5.30 & 4.96 & 6.09 & 5.27 & 4.87 & 4.68 & 5.11 & 4.72 \\ 
   & \small (.033) & \small (.031) & \small (.035) & \small (.034) & \small (.041) & \small (.035) & \small (.033) & \small (.032) & \small (.034) & \small (.032) \\ \hline
  \multirow{2}{*}{EPEX.DE\&AT} & 6.46 & \textbf{6.01} & 6.73 & 6.32 & 7.43 & 6.69 & 6.39 & 6.17 & 6.62 & \underline{6.09} \\ 
   & \small (.050) & \small (.048) & \small (.054) & \small (.052) & \small (.058) & \small (.054) & \small (.050) & \small (.049) & \small (.052) & \small (.050) \\ \hline
  \multirow{2}{*}{EPEX.CH} & 5.44 & \textbf{5.15} & 5.82 & 5.36 & 6.68 & 5.65 & 5.37 & \underline{5.23} & 5.93 & 5.38 \\ 
   & \small (.047) & \small (.045) & \small (.048) & \small (.045) & \small (.054) & \small (.047) & \small (.045) & \small (.043) & \small (.052) & \small (.048) \\ \hline
  \multirow{2}{*}{BELPEX.BE} & \textbf{6.75} & \underline{6.78} & 6.91 & 7.19 & 7.26 & 7.38 & 7.21 & 7.48 & 7.05 & 7.27 \\ 
   & \small (.052) & \small (.052) & \small (.053) & \small (.054) & \small (.056) & \small (.056) & \small (.053) & \small (.053) & \small (.052) & \small (.054) \\ \hline
  \multirow{2}{*}{APX.NL} & 5.11 & \textbf{4.95} & 5.26 & 5.04 & 5.62 & 5.30 & 5.13 & \underline{5.01} & 5.25 & 5.03 \\ 
   & \small (.034) & \small (.033) & \small (.035) & \small (.033) & \small (.038) & \small (.035) & \small (.034) & \small (.033) & \small (.034) & \small (.033) \\ \hline
  \multirow{2}{*}{NP.DK.West} & \underline{6.89} & \textbf{6.62} & 7.21 & \underline{6.95} & 8.03 & 7.86 & 7.61 & 7.42 & \underline{6.83} & \underline{6.78} \\ 
   & \small (.193) & \small (.191) & \small (.187) & \small (.186) & \small (.298) & \small (.328) & \small (.194) & \small (.194) & \small (.188) & \small (.187) \\ \hline
  \multirow{2}{*}{NP.DK.East} & 6.31 & \textbf{6.01} & 6.42 & 6.18 & 6.66 & 6.40 & \underline{6.13} & 6.07 & 6.50 & 6.35 \\ 
   & \small (.069) & \small (.067) & \small (.055) & \small (.054) & \small (.057) & \small (.057) & \small (.053) & \small (.052) & \small (.056) & \small (.055) \\ \hline
  \multirow{2}{*}{NP.SW4} & 3.28 & \textbf{3.20} & 3.67 & 3.60 & 3.73 & 3.54 & 3.48 & 3.52 & 3.78 & 3.74 \\ 
   & \small (.038) & \small (.038) & \small (.039) & \small (.037) & \small (.039) & \small (.038) & \small (.039) & \small (.038) & \small (.039) & \small (.039) \\ \hline
  \multirow{2}{*}{POLPX.PL} & 2.78 & \underline{2.68} & 2.82 & \textbf{2.65} & 3.19 & 2.73 & 2.83 & 2.77 & 3.04 & 2.81 \\ 
   & \small (.030) & \small (.029) & \small (.027) & \small (.027) & \small (.030) & \small (.028) & \small (.027) & \small (.027) & \small (.030) & \small (.028) \\ \hline
  \multirow{2}{*}{OTE.CZ} & 5.49 & \textbf{5.10} & 5.67 & 5.26 & 6.52 & 5.63 & 5.34 & \underline{5.11} & 5.60 & 5.20 \\ 
   & \small (.039) & \small (.036) & \small (.041) & \small (.039) & \small (.047) & \small (.042) & \small (.038) & \small (.037) & \small (.040) & \small (.038) \\ 
   \hline
     \hline
 RMSE & lasso & lasso-wd & 24d.AR & 24d.AR-wd & exp.AR & exp.AR-wd & AR(p) & AR(p)-wd & PCA* & PCA*-wd \\ 
  \hline
  \multirow{2}{*}{EXAA.DE\&AT} &  7.02 &  \textbf{6.63} &  7.52 &  7.12 &  8.70 &  7.58 &  7.01 &  \underline{6.73} &  7.33 &  6.79 \\ 
   & \small (.070) & \small (.067) & \small (.068) & \small (.066) & \small (.081) & \small (.068) & \small (.067) & \small (.066) & \small (.072) & \small (.070) \\ \hline
  \multirow{2}{*}{EPEX.DE\&AT} &  10.06 & \textbf{ 9.47} &  10.70 &  10.24 & 11.63 &  10.71 &  \underline{10.02} &  \underline{9.70} &  10.45 &  \underline{9.82} \\ 
   & \small (.275) & \small (.282) & \small (.308) & \small (.323) & \small (.284) & \small (.301) & \small (.265) & \small (.272) & \small (.296) & \small (.324) \\ \hline
  \multirow{2}{*}{EPEX.CH} &  9.03 &  \underline{8.60} &  9.35 &  \underline{8.72} &  10.58 &  9.16 &  \underline{8.77} &  \textbf{8.48} &  9.93 &  9.18 \\ 
   & \small (.220) & \small (.227) & \small (.211) & \small (.211) & \small (.212) & \small (.201) & \small (.208) & \small (.210) & \small (.260) & \small (.267) \\ \hline
  \multirow{2}{*}{BELPEX.BE} &  \textbf{10.33} &  \underline{10.37} &  \underline{10.58} &  10.98 & 11.21 & 11.30 &  \underline{10.79} & 11.02 &  \underline{10.58} &  10.90 \\ 
   & \small (.239) & \small (.239) & \small (.234) & \small (.243) & \small (.232) & \small (.231) & \small (.229) & \small (.222) & \small (.210) & \small (.215) \\ \hline
  \multirow{2}{*}{APX.NL} &  7.26 &  \textbf{7.02} &  7.44 &  \underline{7.14} &  8.00 &  7.50 &  7.26 &  \underline{7.09} &  7.40 &  \underline{7.09} \\ 
   & \small (.079) & \small (.074) & \small (.077) & \small (.073) & \small (.087) & \small (.076) & \small (.078) & \small (.076) & \small (.081) & \small (.079) \\ \hline
  \multirow{2}{*}{NP.DK.West} &  \underline{30.21} & \underline{29.94} & \underline{29.47} & \underline{29.37} & 46.27 &  50.67 &  \underline{30.69} &  \underline{30.50} & \underline{29.61} & \textbf{29.35} \\ 
   & \small (5.845) & \small (5.884) & \small (5.934) & \small (5.956) & \small (7.033) & \small (8.386) & \small (5.662) & \small (5.697) & \small (6.004) & \small (6.016) \\ \hline
  \multirow{2}{*}{NP.DK.East} & 12.38 & 11.92 &  \underline{10.59} &  \underline{10.26} & 11.07 &  10.83 &  \underline{10.34} &  \textbf{10.06} &  10.77 &  \underline{10.52} \\ 
   & \small (.917) & \small (.826) & \small (.285) & \small (.297) & \small (.281) & \small (.289) & \small (.311) & \small (.311) & \small (.288) & \small (.297) \\ \hline
  \multirow{2}{*}{NP.SW4} &  \underline{6.76} &  \textbf{6.64} &  \underline{6.97} &  \underline{6.72} &  \underline{7.06} &  \underline{6.82} &  \underline{6.90} &  \underline{6.78} &  \underline{7.08} &  7.10 \\ 
   & \small (.223) & \small (.226) & \small (.225) & \small (.229) & \small (.217) & \small (.224) & \small (.241) & \small (.243) & \small (.220) & \small (.221) \\ \hline
  \multirow{2}{*}{POLPX.PL} &  5.37 &  5.17 &  \underline{5.06} &  \underline{4.94} &  5.61 &  \underline{5.08} &  \underline{4.98} &  \textbf{4.92} &  5.52 &  5.19 \\ 
   & \small (.156) & \small (.145) & \small (.127) & \small (.130) & \small (.133) & \small (.133) & \small (.114) & \small (.115) & \small (.139) & \small (.138) \\ \hline
  \multirow{2}{*}{OTE.CZ} &  8.08 &  \textbf{7.57} &  8.46 &  7.98 &  9.66 &  8.54 &  7.90 &  \underline{7.58} &  8.25 &  \underline{7.81} \\ 
   & \small (.140) & \small (.137) & \small (.155) & \small (.154) & \small (.157) & \small (.145) & \small (.121) & \small (.123) & \small (.136) & \small (.150) \\ 
   \hline
\end{tabular}
\caption{MAE and RMSE values for the considered markets and models. The corresponding 
estimated standard deviations are given in parenthesis. The best model is highlighted in bold font. 
All models that are not significantly worse than best (indicated by the 2-sigma range of the best model) 
are underlined. }
\label{tab_mae_rmse}
\end{table*}

In Table \ref{tab_mae_rmse} we observe that concerning the MAE the lasso based approaches perform very well. 
The proposed model (lasso-wd) in equation \eqref{eq_model2} with weekdays effects is always the best model,  except for the BELPEX and POLPX case.
But even there it not significantly worse than the best models which are the lasso  without weekday effects for BELPEX and 
the 24-dimensional AR model with weekday effects for the POLPX.
It is remarkable that model \eqref{eq_model2} with weekdays effects is sometimes significantly better than all other considered models.
In several cases (EPEX.AT\&AT, EPEX.CH, APX.NL, NP.DK.East, OTE.CZ) there is only one model that is not significantly worse. 
In contrast, concerning the RMSE values the picture is not that clear. Still, the lasso-wd model \eqref{eq_model2} 
is most of the time the best model. However, as mentioned the 
RMSE is sensitive to outliers. So for example the 
Nordpool Denmark West price has 5 extreme prices between 1900EUR/MWh and 2000EUR/MWh 
on June 7, 2013. This leads to a drastic increase in the RMSE, as no (considered) model 
is able to capture these extreme price spikes. 
 Concerning the RMSE the AR(p)-wd seems to perform relatively well and scores best for 3 of the 10 considered markets. 
The reason might be that the AR(p)-wd has a relatively low outlier sensitivity, as it is a member of the second model class which
 contains big single equation models and use all observations as input. 
 Having a lot of observations helps these models to learn that the outliers are a rare event. In contrast, all models of the first class that use 24 small models (as the proposed lasso) have the problem that an outlier gets much more attention as less observations are available.

\begin{figure*}[hbt!]
\centering
\begin{subfigure}[b]{.49\textwidth}
 \includegraphics[width=1\textwidth]{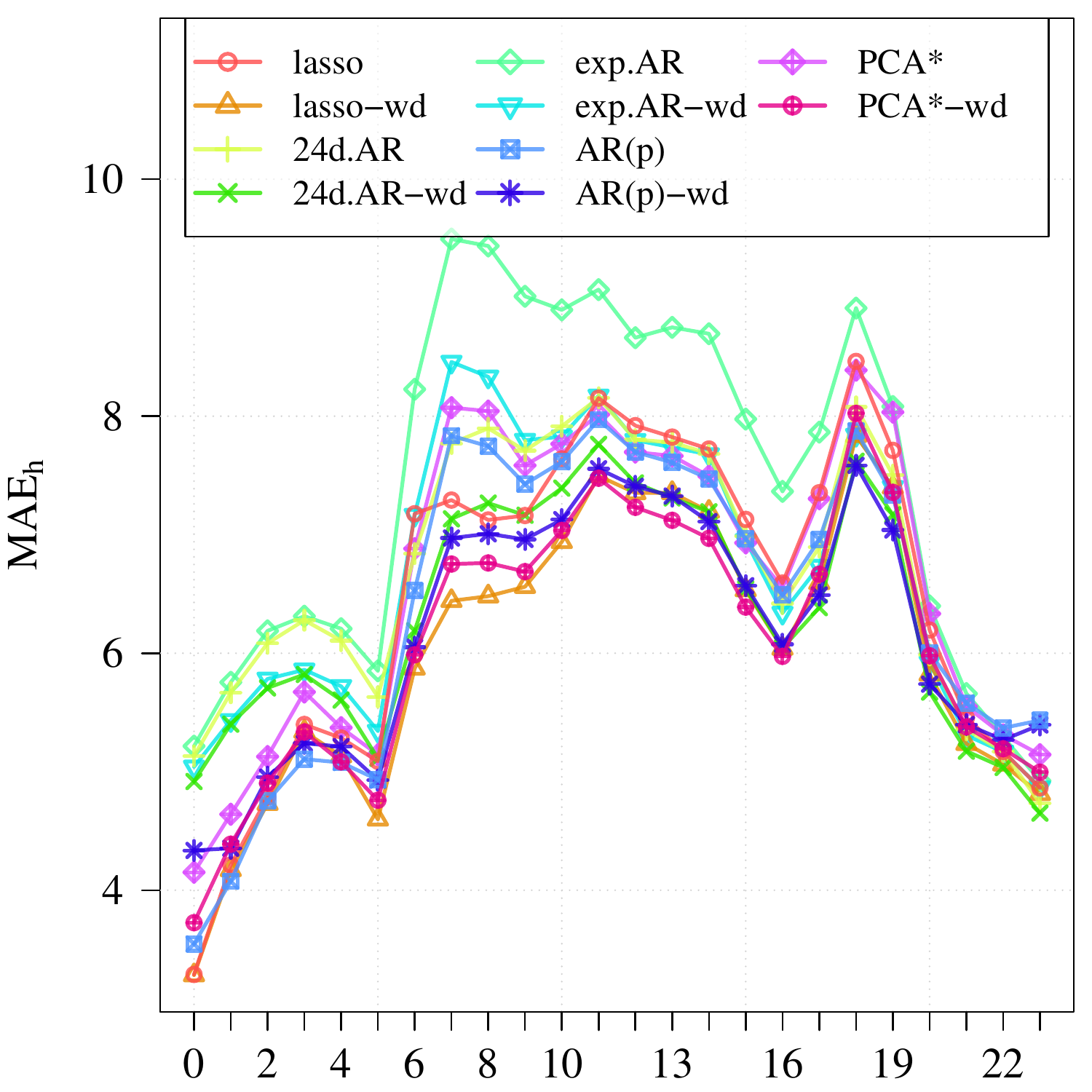} 
  \caption{$\text{MAE}_h$}
  \label{fig_maeh_epex_de_at}
\end{subfigure}
\begin{subfigure}[b]{.49\textwidth}
 \includegraphics[width=1\textwidth]{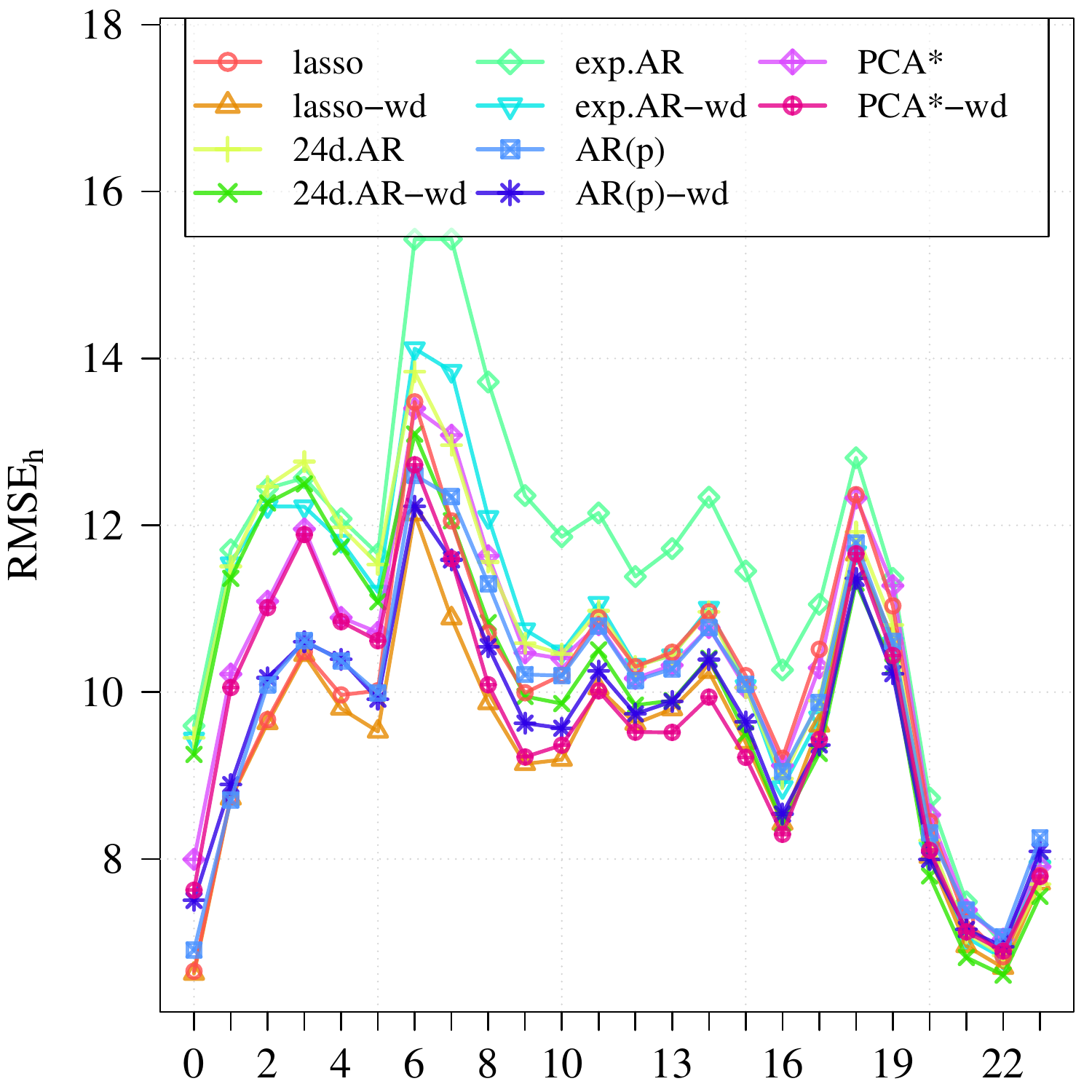} 
  \caption{$\text{RMSE}_h$}
  \label{fig_rmseh_epex_de_at}
\end{subfigure}
\caption{Hourly performance measures for EPEX spot price of Germany and Austria.}
 \label{fig_maeh_rmseh_epex_de_at}
\end{figure*}

For the $\text{MAE}_h$ and $\text{RMSE}_h$ we discuss detailed results only for the German/Austrian EPEX spot price, 
which is the largest day-ahead electricity market in Europe (regarding the traded volume).
The $\text{MAE}_h$ and $\text{RMSE}_h$ results are given in Figure  \ref{fig_maeh_rmseh_epex_de_at}.
Results for the other markets are similar. The corresponding graphs are given in the supplementary material. There we also present the 
correlation plots and the weekly mean graphs as in Figures  \ref{fig_intro_corr} and \ref{fig_wm} for all markets.

In Figure  \ref{fig_maeh_rmseh_epex_de_at} we observe that the daily pattern for the $\text{MAE}_h$ and $\text{RMSE}_h$ 
are similar. In general we see that the model with weekday effects estimated by lasso performs  
as one of the best models for all hours over the day. It seems that during the morning hours it outperforms
the other models most clearly. 
It is interesting that in the night hours from 0:00 to 5:00 the lasso based models perform similarly to 
the univariate AR(p). But this matches well the observation from the beginning that 
these hours are highly correlated with the last hours of the previous day. The univariate AR(p) and the lasso models
can capture this behavior. Moreover, in the evening hours (esp. 21:00 to 23:00) the 
24-dimensional AR model and the 24-dimensional autoregressive expert model perform as good as the lasso based models.
Again, this coincides with the observation from Figure \ref{fig_intro_corr}, 
that these hours are strongly correlated with the same hours of the previous days. The mentioned models are 
all able to cover this dependency well. Furthermore, we see a relatively well performance of the PCA*-wd model. We know from Table \ref{tab_mae_rmse} that for the German/Austrian EPEX price only 
the PCA*-wd model performs not significantly worse. The behavior is similar as both core models (ignoring the weekday effects) are nested in \eqref{eq_mAR}. 
However, regarding the cross-country Table \ref{tab_mae_rmse}, we know that in most cases the lasso-wd model outperforms the PCA*-wd model. The reason is likely that cross-hour dependency is modeled
directly instead of using the reduced PCA representation. It might happen that relevant information gets lost due to the dimension reduction technique.

Another interesting fact in Figure \ref{fig_maeh_rmseh_epex_de_at} is
that it seems that in the morning hours around 6:00 to 7:00 the lasso-wd performs much better in the
$\text{MAE}_h$ score than in the $\text{RMSE}_h$. However, having a closer look at the graph it turns out 
the AR(p) process catches up in terms of $\text{RMSE}_h$. The reason might be again that the AR(p) based models are less 
outlier sensitive than the other considered models. Interestingly this pattern does not hold for all considered markets, but we observe the same fact for EXAA.DE+AT, EPEX.CH, APX.NL, NP.DK.East and OTE.CZ.

Furthermore, we can see in Figure \ref{fig_maeh_rmseh_epex_de_at} that it seems that the weekday dummies in the model \eqref{eq_model2} improve the 
performance significantly compared to model \eqref{eq_model1}. We will understand this better when we analyze the intraday structure 
and the variable importance.

Hence, additionally to the out-of-sample study we have a closer look at the intraday structure of the electricity prices
by using the proposed model \eqref{eq_model2}.
As the lasso estimator sparsifies and regularizes, we can study the finally selected parameters and their magnitude.
For a measure of importance of a parameter $\beta_{h, i}$ resp. $\wtilde{\beta}_{h, i}$ we consider $\iota_{h,i}$, the  
fraction of the absolute standardized parameter of a model to the sum of all absolute standardized parameters:
\begin{equation}
 \iota_{h,i} = \frac{|\wtilde{\beta}_{h, i}|}{\sum_{j=1}^{m_h} |\wtilde{\beta}_{h, j}|} \\
 \label{eq_importance}
\end{equation}
Obviously, we have that $0\leq \iota_{h,i} \leq 1$ and $\sum_{i=1}^{m_h} \iota_{h,i} =1$. The larger 
$\iota_{h,i}$ the larger the relative impact to the electricity price of hour $h$.
We estimate $\iota_{h,i}$ for model \eqref{eq_model2} which includes the weekday dummies for all considered markets
by the corresponding plug-in estimator $\what{\iota}_{h,i}$ using the estimated standardized parameter vector
$\what{\wtilde{\bsbeta}}_h$. 
Given the estimated $\what{\iota}_{h,i}$ values, we can sort them decreasingly. 
Then the largest $\what{\iota}_{h,i}$ value matches the most relevant parameter in the model for hour $h$.

Again, we only discuss the results for the German/Austrian EPEX spot price in detail.
In Table \ref{tab_importance_EPEX_DE+AT}
we present the 7 most important variables for each model of hour $h$ and their $\what{\iota}_{h,i}$ value. 
The corresponding tables for the other electricity prices are given in the supplementary material.
\begin{table*}[ht]
\normalsize
\centering
\begin{tabular}{rrrrrrrr}
  \hline
 & Importance: 1 & Importance: 2 & Importance: 3 & Importance: 4 & Importance: 5 & Importance: 6 & Importance: 7 \\ 
 \hline
0 & 23@1 (70.6) & 19@1 ( 5.9) &  1@1 ( 5.7) & 18@1 ( 4.1) & 21@1 ( 3.6) &  4@1 ( 3.1) & Sat ( 2.2) \\ 
  1 & 23@1 (67.3) & 19@1 ( 8.6) &  3@1 ( 7.1) &  4@1 ( 5.9) & Sun ( 4.7) &  5@1 ( 2.6) &  6@1 ( 2.0) \\ 
  2 & 23@1 (56.3) &  4@1 ( 9.4) & 19@1 ( 8.9) &  3@1 ( 7.9) & Sun ( 6.6) & 22@7 ( 5.1) &  2@25 ( 1.6) \\ 
  3 & 23@1 (45.0) &  4@1 (14.3) &  3@1 (12.2) & 19@1 ( 8.5) & Sun ( 7.3) & 22@7 ( 4.3) &  3@6 ( 2.5) \\ 
  4 & 23@1 (44.8) &  4@1 (13.8) & 19@1 (10.3) & Sun (10.2) &  3@1 ( 7.4) & 22@7 ( 2.5) &  4@25 ( 2.4) \\ 
  5 & 23@1 (34.1) & Sun (15.5) &  5@1 ( 9.7) &  4@1 ( 8.2) & Sat ( 7.4) & 19@1 ( 7.2) & 22@7 ( 3.8) \\ 
  6 & Sun (20.4) & 23@1 (20.3) & Sat (12.8) &  6@1 ( 9.4) & Mon ( 8.7) & 20@1 ( 6.0) & 23@2 ( 4.1) \\ 
  7 & Sun (23.9) & Sat (15.9) & 23@1 (12.1) & Mon ( 7.8) & 20@1 ( 6.3) &  6@1 ( 5.3) & 19@1 ( 4.5) \\ 
  8 & Sun (28.3) & Sat (15.9) & 20@1 ( 8.5) & 23@1 ( 7.9) & 18@1 ( 7.0) & Mon ( 5.4) & 21@1 ( 5.2) \\ 
  9 & Sun (25.9) & 20@1 (13.3) & Sat (13.3) & 23@1 ( 7.0) & 22@1 ( 6.3) & Mon ( 6.1) & 18@1 ( 5.5) \\ 
  10 & Sun (21.8) & Sat (11.9) & 22@1 ( 9.6) & 20@1 ( 9.0) & Mon ( 6.6) & 13@1 ( 6.1) & 23@1 ( 5.6) \\ 
  11 & Sun (17.2) & 22@1 (12.3) & Sat (11.7) & 13@1 ( 9.2) & Mon ( 8.0) & 23@1 ( 5.8) & 11@7 ( 4.1) \\ 
  12 & Sun (15.6) & 13@1 (12.0) & Sat (10.7) & 22@1 (10.2) & Mon ( 8.1) & 23@1 ( 8.0) & 17@1 ( 4.9) \\ 
  13 & Sun (18.6) & Sat (11.2) & 13@1 (11.1) & 22@1 (10.0) & Mon ( 7.9) & 23@1 ( 7.3) & 17@1 ( 6.8) \\ 
  14 & Sun (18.7) & Sat (10.1) & Mon ( 8.2) & 23@1 ( 8.2) & 16@1 ( 8.1) & 22@1 ( 7.6) & 13@1 ( 4.2) \\ 
  15 & Sun (19.8) & 16@1 (12.5) & Sat ( 8.9) & Mon ( 8.0) & 23@1 ( 7.1) & 17@6 ( 6.0) & 17@1 ( 4.9) \\ 
  16 & Sun (19.3) & 17@1 (12.5) & Sat ( 8.8) & 16@1 ( 8.5) & Mon ( 8.4) & 23@1 ( 7.1) & 17@6 ( 6.7) \\ 
  17 & 17@1 (26.4) & Sun (16.1) & Mon (10.1) & 17@6 (10.0) & Sat ( 6.7) & 17@7 ( 6.3) & 18@1 ( 6.0) \\ 
  18 & 18@1 (33.8) & Sun (14.1) & Mon ( 9.9) & 18@6 ( 9.4) & Sat ( 6.0) & 18@2 ( 5.9) & 18@5 ( 4.6) \\ 
  19 & 19@1 (34.0) & Mon ( 9.4) & 19@6 ( 8.5) & Sun ( 8.2) & 19@2 ( 7.6) &  8@6 ( 5.0) & 19@7 ( 4.7) \\ 
  20 & 20@1 (23.9) & 20@2 (10.4) & Sat ( 9.5) & Mon ( 7.8) & 21@1 ( 7.5) & 20@6 ( 7.5) & Sun ( 7.0) \\ 
  21 & 21@1 (22.8) & Sat (10.9) & 21@2 ( 9.9) & Mon ( 6.2) & 21@28 ( 4.7) & Sun ( 4.6) & 23@1 ( 4.6) \\ 
  22 & 22@1 (30.6) & 22@2 ( 9.5) & Sat ( 7.7) & 23@3 ( 6.5) & 22@28 ( 6.1) & 22@5 ( 5.1) & 22@35 ( 5.0) \\ 
  23 & 23@1 (24.8) & 22@1 (17.5) & 23@3 (10.7) & 22@2 ( 9.6) & Sat ( 5.7) &  0@5 ( 5.0) & 12@7 ( 4.4) \\ 
\end{tabular}
\caption{Most relevant coefficients in model \eqref{eq_model2} for each hour of the German/Austrian EPEX price.
H@L represents the estimated parameter for $Y_{d-L,H}$ for a model on $Y_{d,h}$. Sun, Mon, $\ldots$, Sat represent the weekday dummies.
The corresponding $\what{\iota}_{h,i}$ value is given in parenthesis.}
\label{tab_importance_EPEX_DE+AT}
\end{table*}
In Table \ref{tab_importance_EPEX_DE+AT} we can observe 
an interesting pattern that we have similarly seen already in the introduction.
The most relevant parameters for the first 6 hours of the day is the last observed value of the previous day (which corresponds to $P_{d-1,23}$).
For the first hour this importance is very distinct. The last observation $P_{d-1,23}$ of the previous day contributes 70.6\% to the full model.
From the physical point of view this makes sense.
Obviously, the physical delivery of the auction result for a specific hour is based on related processes like the electricity load or wind energy production. These situations should not differ that much between 23:00 and the next day 0:00. So a large impact 
of the 23:00 price to the price of the next day at hour 0:00 is plausible. 
This observation seems to be very stable across all countries. So we observe for all considered markets
that the most relevant variable for the first 0:00 hour is always the last hour of the previous day. For all markets, except
for the BELPEX.BE, this is even the case for the first 3 hours (0:00, 1:00 and 2:00) of the day.
A similar result is recently established in load forecasting by \cite{dudek2016pattern}.
However, for electricity price forecasting this is not that obvious as we observe all 24 values at once. 
The last observation of the previous day is not more recent than the other ones.
 The fact that for the first hours the last values of the previous day are very important
 underlines that the univariate AR($p$) and the lasso based models perform well in this time horizon. 

 For the last 7 hours of the day (from 17:00 to 23:00) the most relevant lag 
is always the value of the previous day at the same hour (which corresponds to $P_{d-1,h}$).
 This matches well the observed fact that the 24-dimensional autoregressive models 
 have a good out-of-sample performance for these hours.

Moreover, we see that for the morning and afternoon hours from 6:00 to 16:00 the most important 
parameter is the weekday dummy on Sunday. So the weekday effect seems to be more important than the autoregressive impact.
In general we observe that the weekday dummies have a larger impact during the working hours than in the night and evening hours.
Thus, we can deduce that the weekly seasonal structure of the electricity price is more distinct in the working hours.
This coincides with Figure \ref{fig_wm_epex}, where we see that the difference in the weekly mean is smaller during night and evening than the 
rest of the day. 

Additionally, we observe that 
only the weekday dummies for Sunday, Monday and Saturday occur in Table \ref{tab_importance_EPEX_DE+AT}.
Here it is interesting that the considered expert model from \cite{weron2008forecasting}
uses exactly the same dummies. However, regarding the weekly sample means in Figure \ref{fig_wm_epex} this is not obvious.
From this graph we would expect that the impact of Monday dummies is similar to Friday dummies.
Furthermore the Monday dummies appear only for the hours from 6:00 to 21:00 which is definitively 
not expectable from Figure \ref{fig_wm_epex} which suggests a Monday dummy only from 0:00 to 5:00.  

Another fact that we can derive from Table \ref{tab_importance_EPEX_DE+AT} concerns the weekly lagged impact.
In many models, e.g. the expert model from \cite{weron2008forecasting} a linear dependence on the weekly lagged 
electricity price is considered $P_{d-7,h}$. However, in Table \ref{tab_importance_EPEX_DE+AT}
it never appears in the top 4 for any hour. Only for some hours it is estimated to be the 5th, 6th or 7th most important variable.
This is an indication that given a proper short-term autoregressive and weekday modeling the modeled electricity price 
dependence from the weekly lagged price gets weaker.
The last fact from Table \ref{tab_importance_EPEX_DE+AT} that we want to mention 
concerns the evening hours. We observe that only for the evening hours from 18:00 to 23:00 
a dependence of electricity price of the day before yesterday (lag = 2) appears. 
Again, this dependency is always to the price at the same hour.


\section{Summary and conclusion}

We present a day-ahead electricity spot price model approach for hourly data that is able to
capture the intraday dependency structure well, especially the time-varying cross-hour dependencies.
Even though the considered regression model consists of 24 simple linear regression models
it explains the variety in the price data. This 
relies heavily on the utilized lasso estimation technique. It automatically 
shrinks and sparsifies the model parameters. Thus, the estimated 24 single models can vary over the day and capture 
the dependency in the data.

We show that the dependency structure significantly changes over the days.
For example in the German/Austrian EPEX market, the night hours from 0:00 to 5:00 have a strong dependency to the 
last price at 23:00 of
the previous day. This result holds similarly across all considered European markets. 
In contrast, the price at the evening hours of from 18:00 to 23:00 of the German/Austrian EPEX market depends more on 
the price of the previous day at the same hour. 
We highlight that the majority of available electricity price model approaches are not designed to cover both features, whereas the lasso estimation technique captures both features in an efficient way.

The conducted out-of-sample forecasting study shows 
that the considered lasso based approach performs strong in terms of forecasting accuracy. 
This is remarkable, if we have in mind that the estimation and forecasting procedure is very fast. 
Hence, the model is a suitable candidate for forecast combination approaches as in \cite{bordignon2013combining} or \cite{maciejowska2015probabilistic}.
As the out-of-sample study shows similar performance results for all 10 considered electricity markets, 
it underlines that the flexible model approach can cover the local market specific dependency behavior.

Future research can go in different directions. The likely most important issue is the
covering of the non-linear behavior, especially those effects that induce price spikes.
This should be combined with an approach that takes other sources like renewable energy feed-in or load into account. 
Especially the impact of wind power is very important to cover price spikes, see e.g. \cite{keles2012comparison}.
Furthermore, there should be more research done for probabilistic forecasting.

  \bibliographystyle{apalike}
\bibliography{Bibliothek}

\clearpage 
 
\section*{Appendix}

\section{Figures}

\subsection{Correlation plots}

\begin{figure*}[hbt!]
\centering
\begin{subfigure}[b]{.49\textwidth}
 \includegraphics[width=1\textwidth]{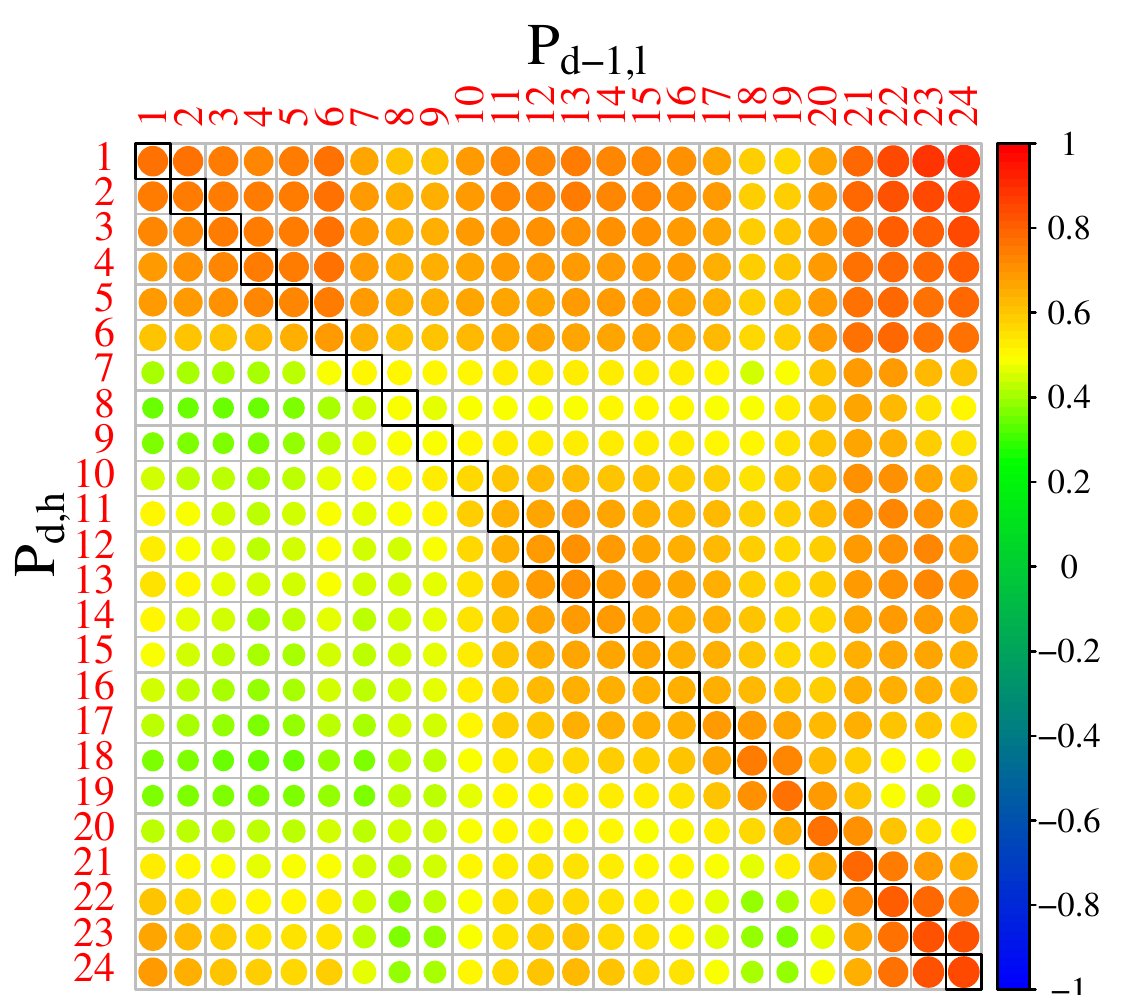} 
  \caption{EXAA - Germany and Austria}
\end{subfigure}
\begin{subfigure}[b]{.49\textwidth}
 \includegraphics[width=1\textwidth]{corrplot_EPEX_DE+AT.pdf} 
  \caption{EPEX - Germany and Austria}
\end{subfigure}
\begin{subfigure}[b]{.49\textwidth}
 \includegraphics[width=1\textwidth]{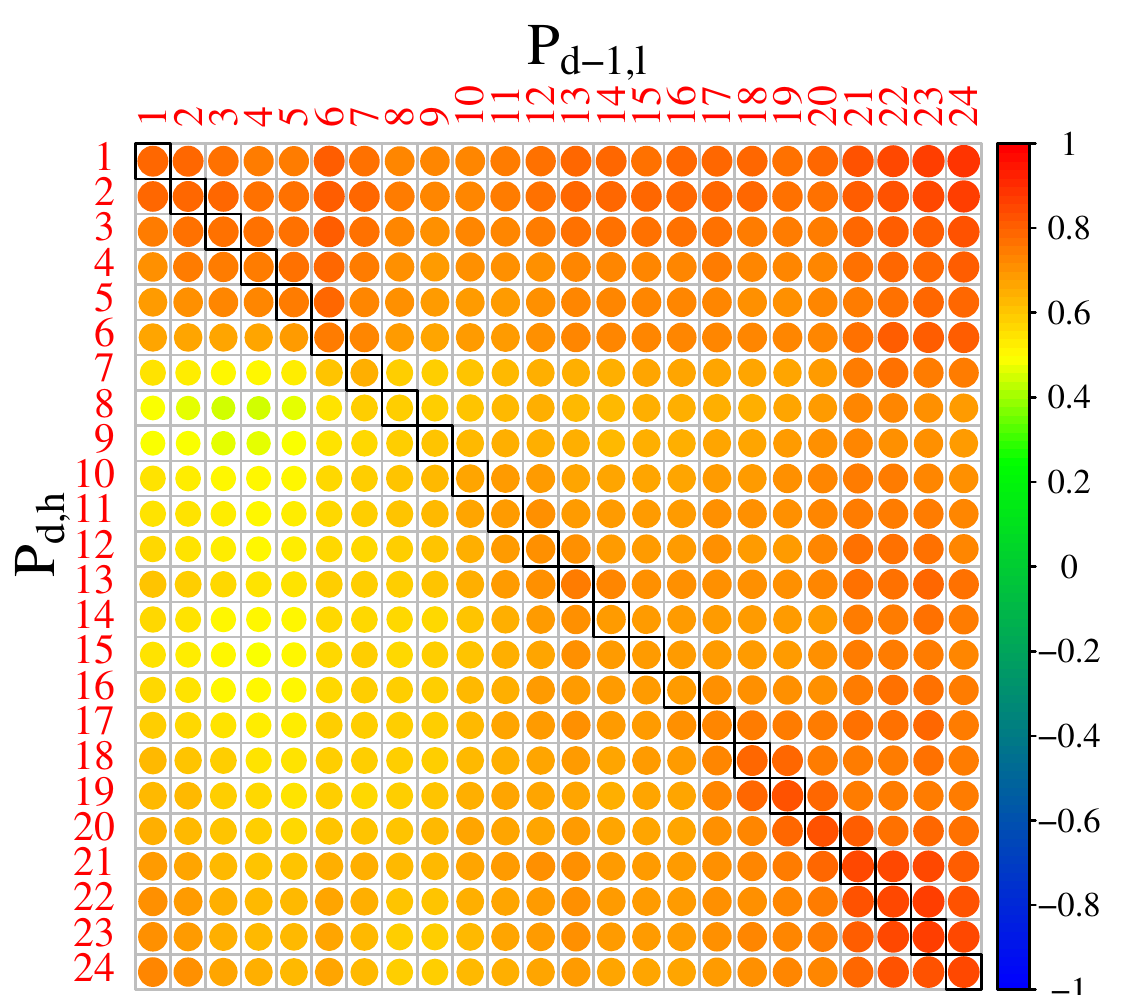} 
  \caption{EPEX - Switzerland}
\end{subfigure}
\begin{subfigure}[b]{.49\textwidth}
 \includegraphics[width=1\textwidth]{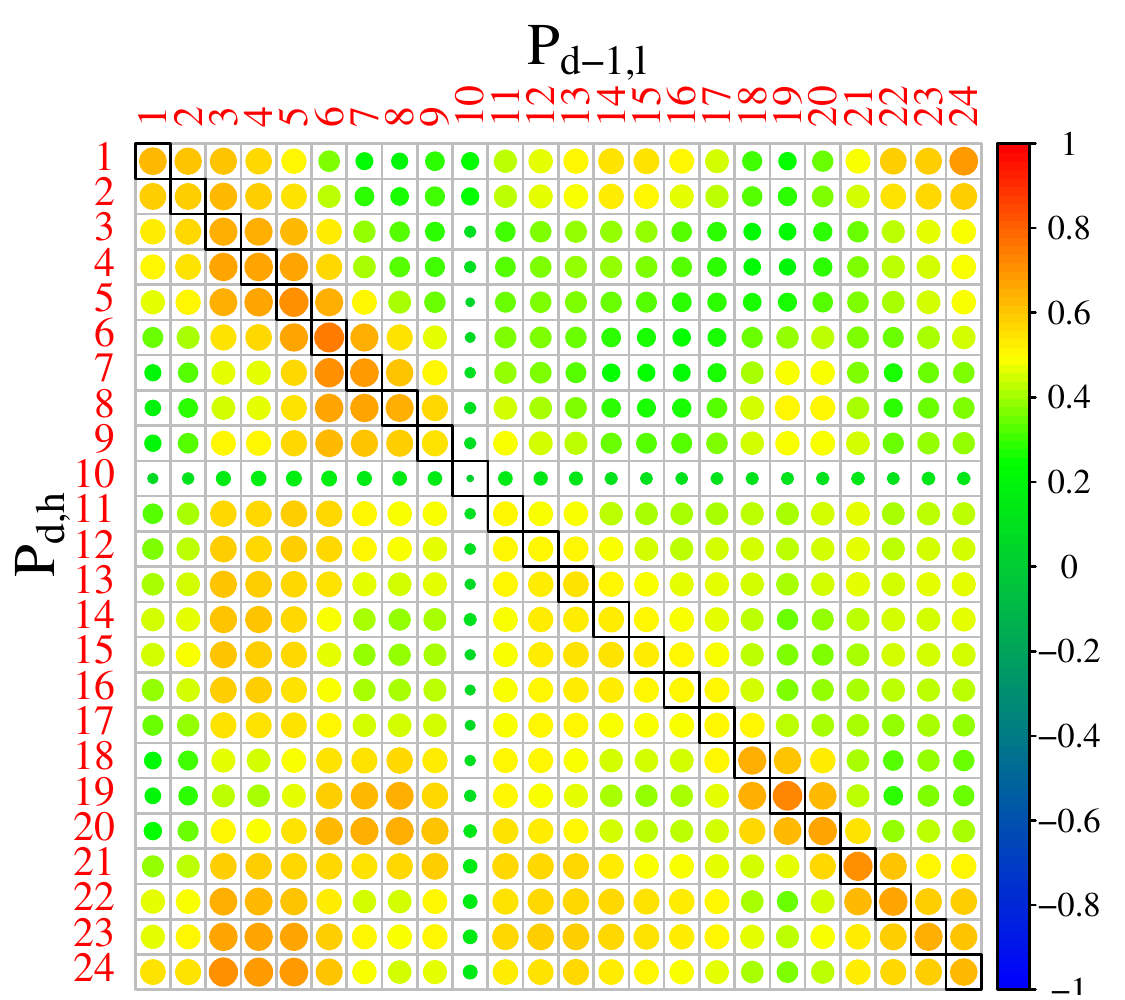} 
  \caption{BELPEX - Belgium}
\end{subfigure}
\caption{Sample correlations of $\mathbb{C}\text{or}( P_{d,h}, P_{d-1,l})$ for $h,l = 0,\ldots, 23$ and selected markets 
from 17.12.2009 to 12.08.2014.}
\end{figure*}

\begin{figure*}[hbt!]
\centering
\begin{subfigure}[b]{.49\textwidth}
 \includegraphics[width=1\textwidth]{corrplot_APX_NL.pdf} 
  \caption{APX - Netherlands}
\end{subfigure}
\begin{subfigure}[b]{.49\textwidth}
 \includegraphics[width=1\textwidth]{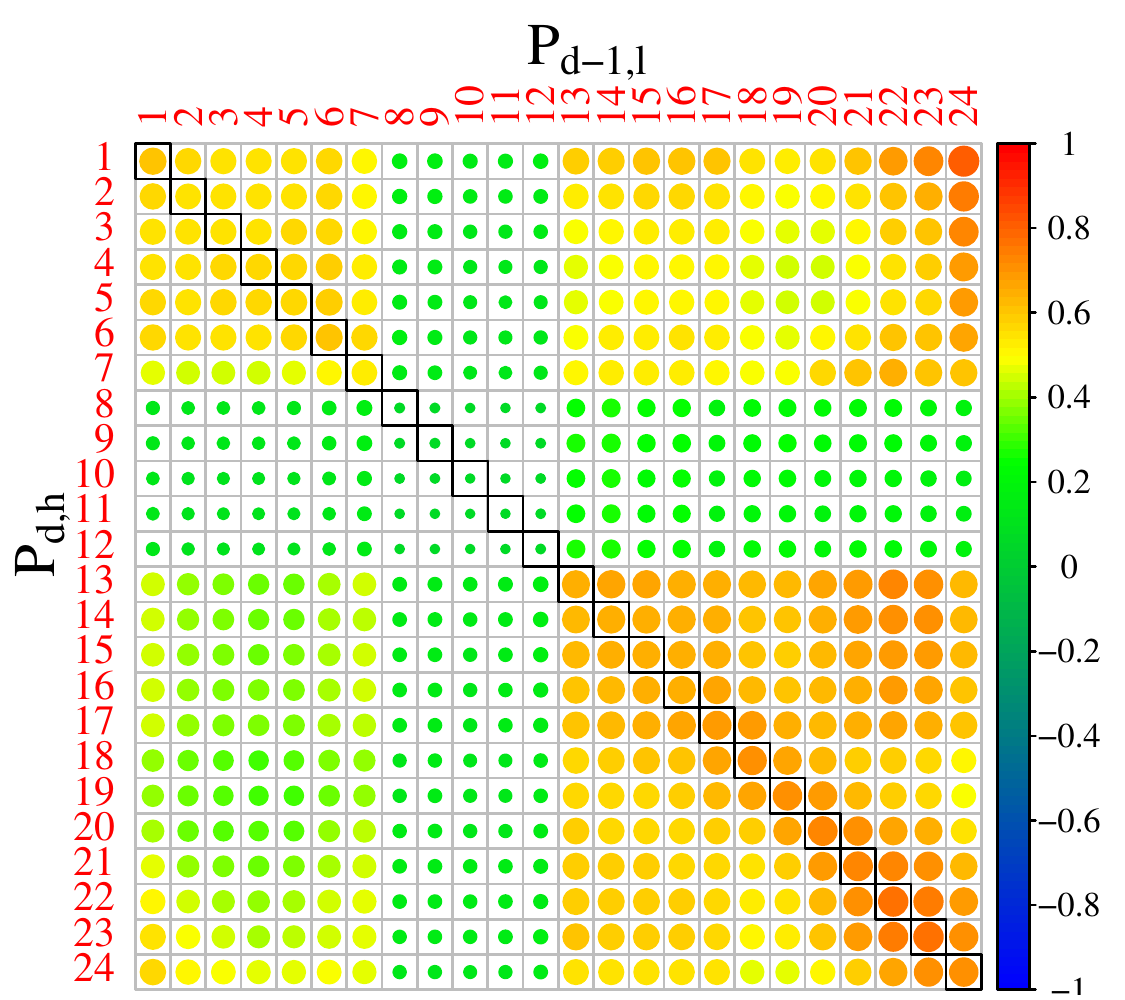} 
  \caption{Nordpool - Denmark West}
\end{subfigure}
\begin{subfigure}[b]{.49\textwidth}
 \includegraphics[width=1\textwidth]{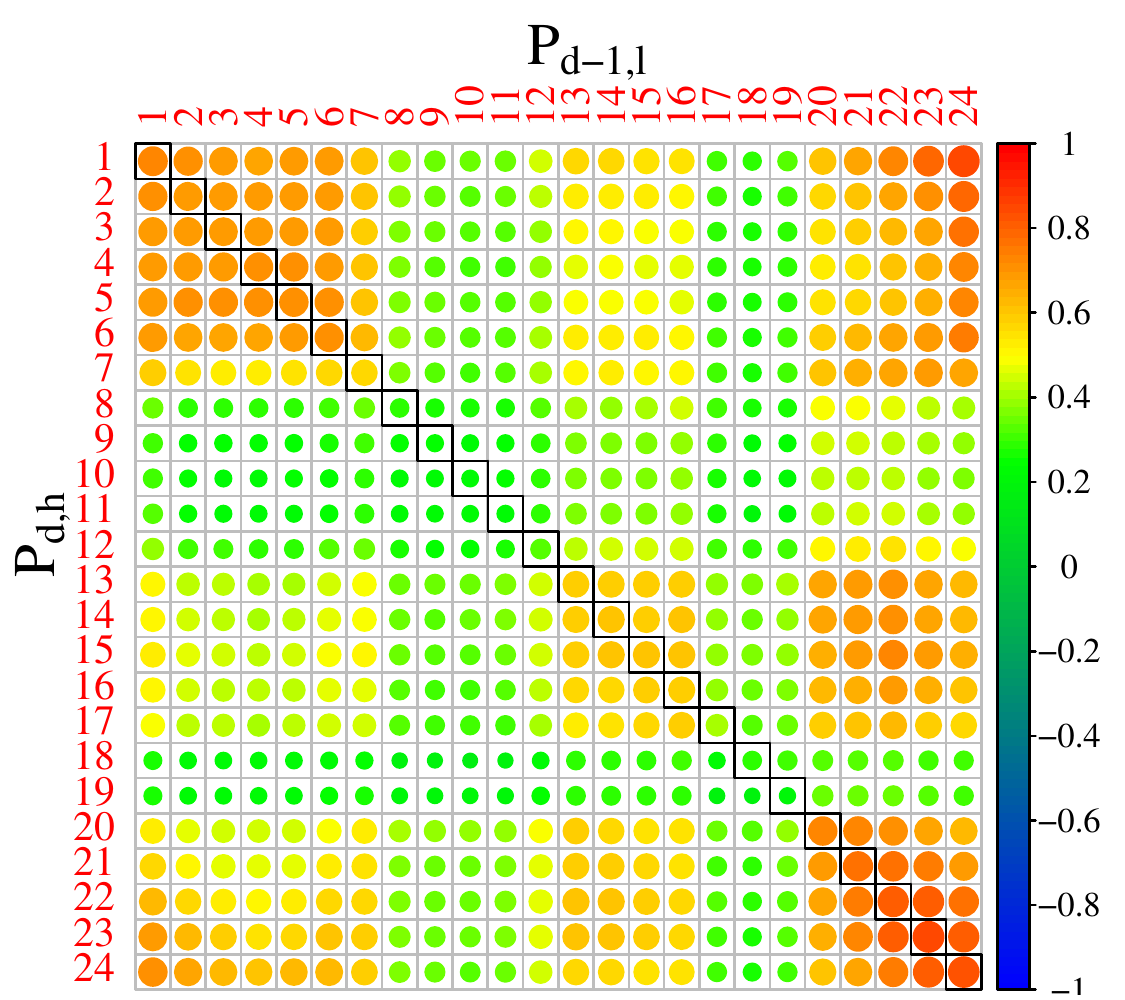} 
  \caption{Nordpool - Denmark East}
\end{subfigure}
\begin{subfigure}[b]{.49\textwidth}
 \includegraphics[width=1\textwidth]{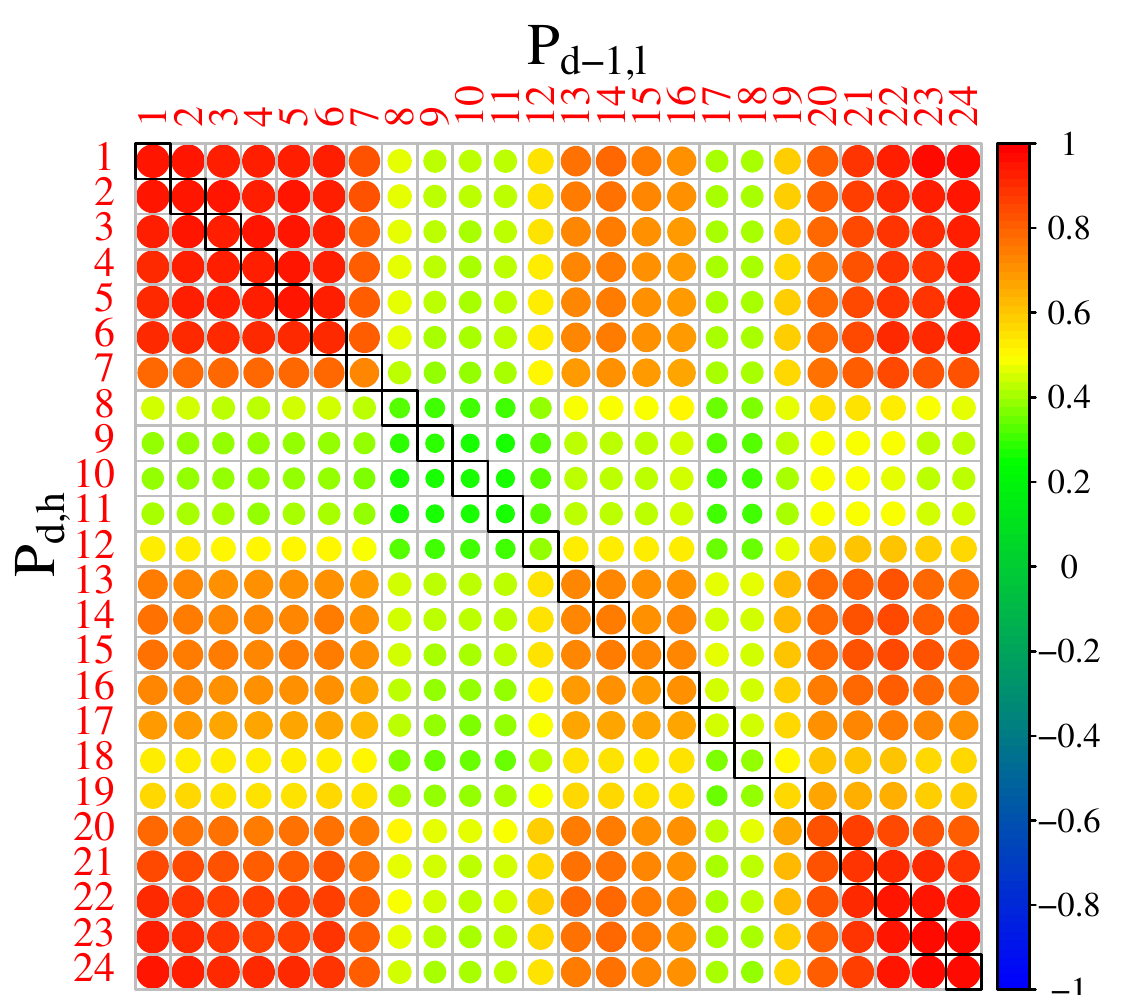} 
  \caption{Nordpool - Sweden(4)}
\end{subfigure}
\begin{subfigure}[b]{.49\textwidth}
 \includegraphics[width=1\textwidth]{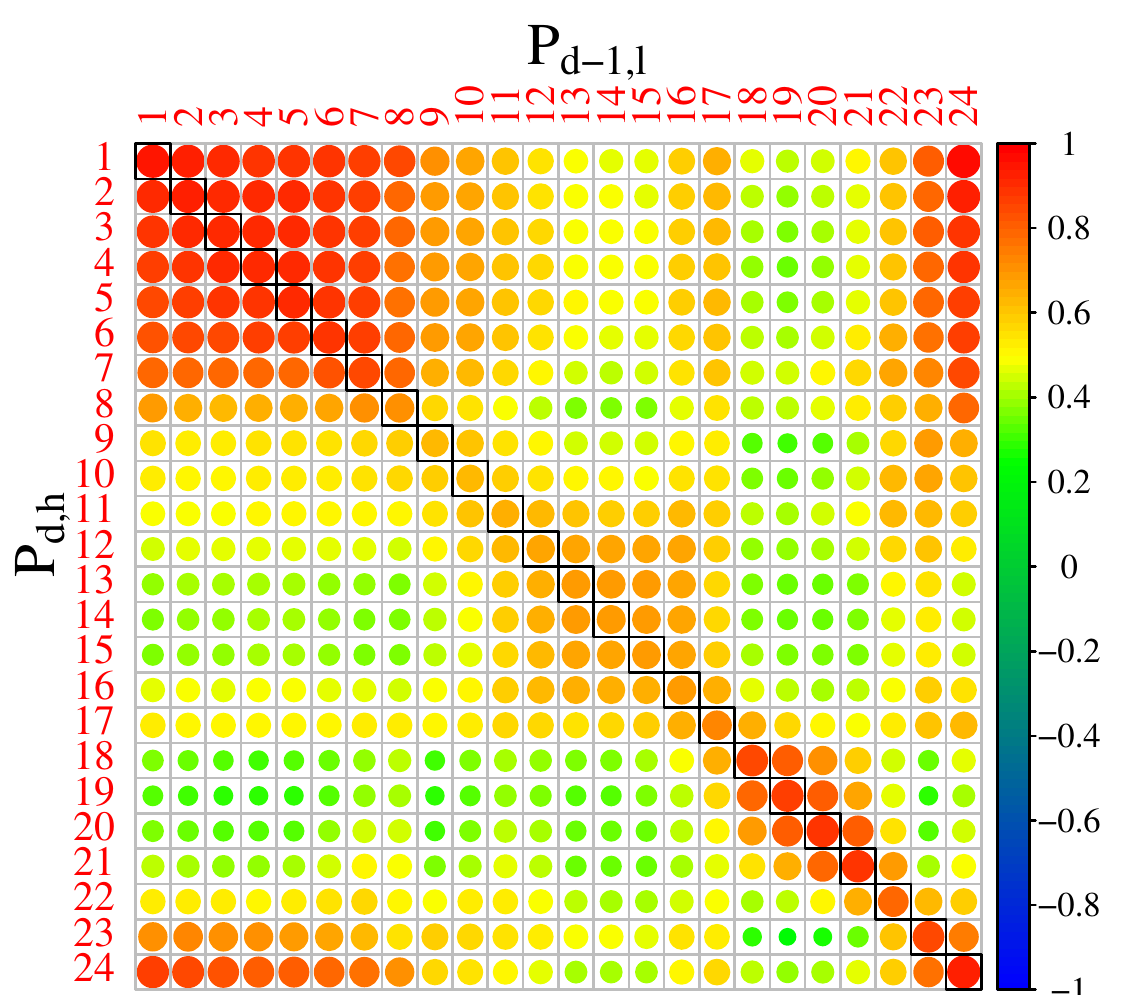} 
  \caption{POLPX - Poland}
\end{subfigure}
\begin{subfigure}[b]{.49\textwidth}
 \includegraphics[width=1\textwidth]{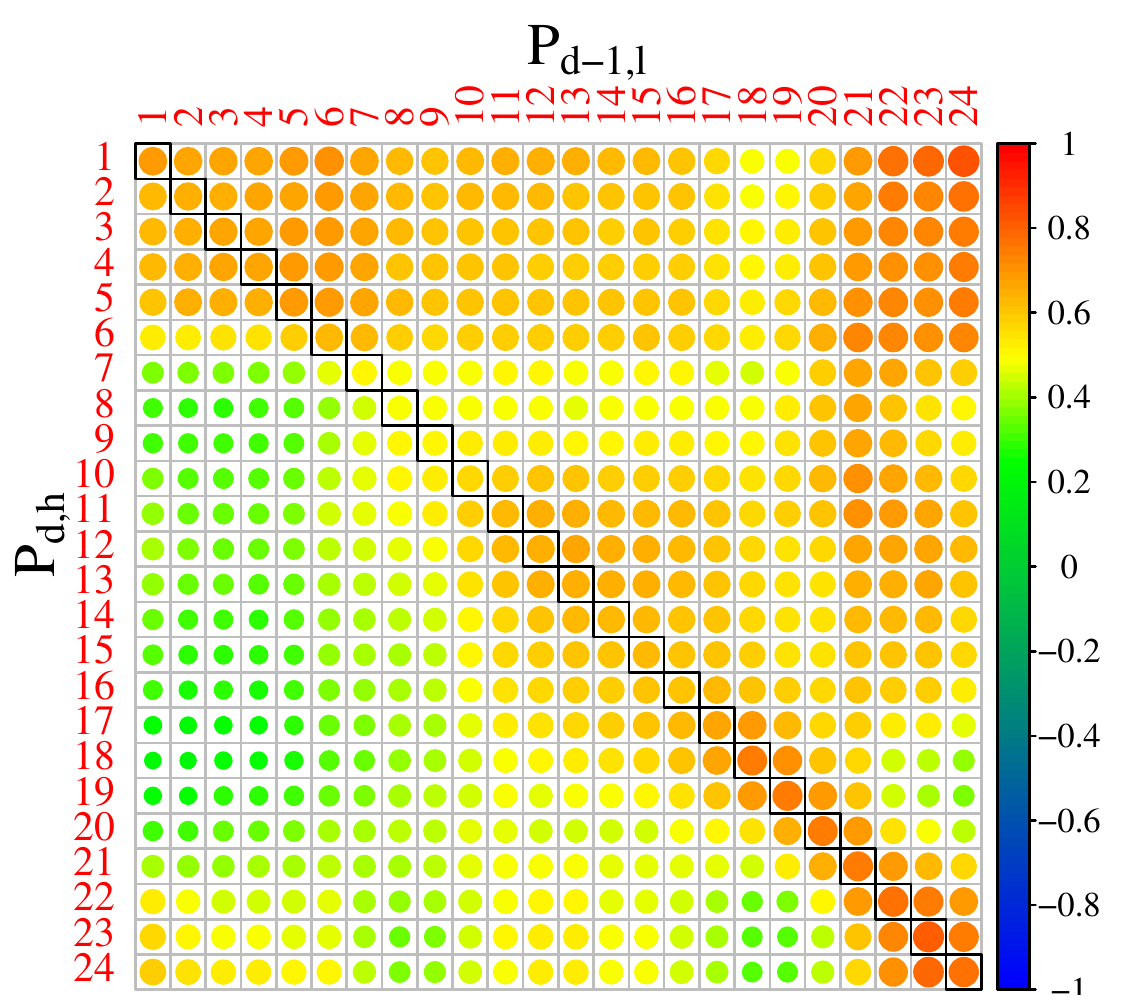} 
  \caption{OTE - Czech}
\end{subfigure}
\caption{Sample correlations of $\mathbb{C}\text{or}( P_{d,h}, P_{d-1,l})$ for $h,l = 0,\ldots, 23$ and selected markets 
from 17.12.2009 to 12.08.2014.}
\end{figure*}

\clearpage

\subsection{Weekly sample mean plots}

\begin{figure*}[hbt!]
\centering
\begin{subfigure}[b]{.49\textwidth}
 \includegraphics[width=1\textwidth]{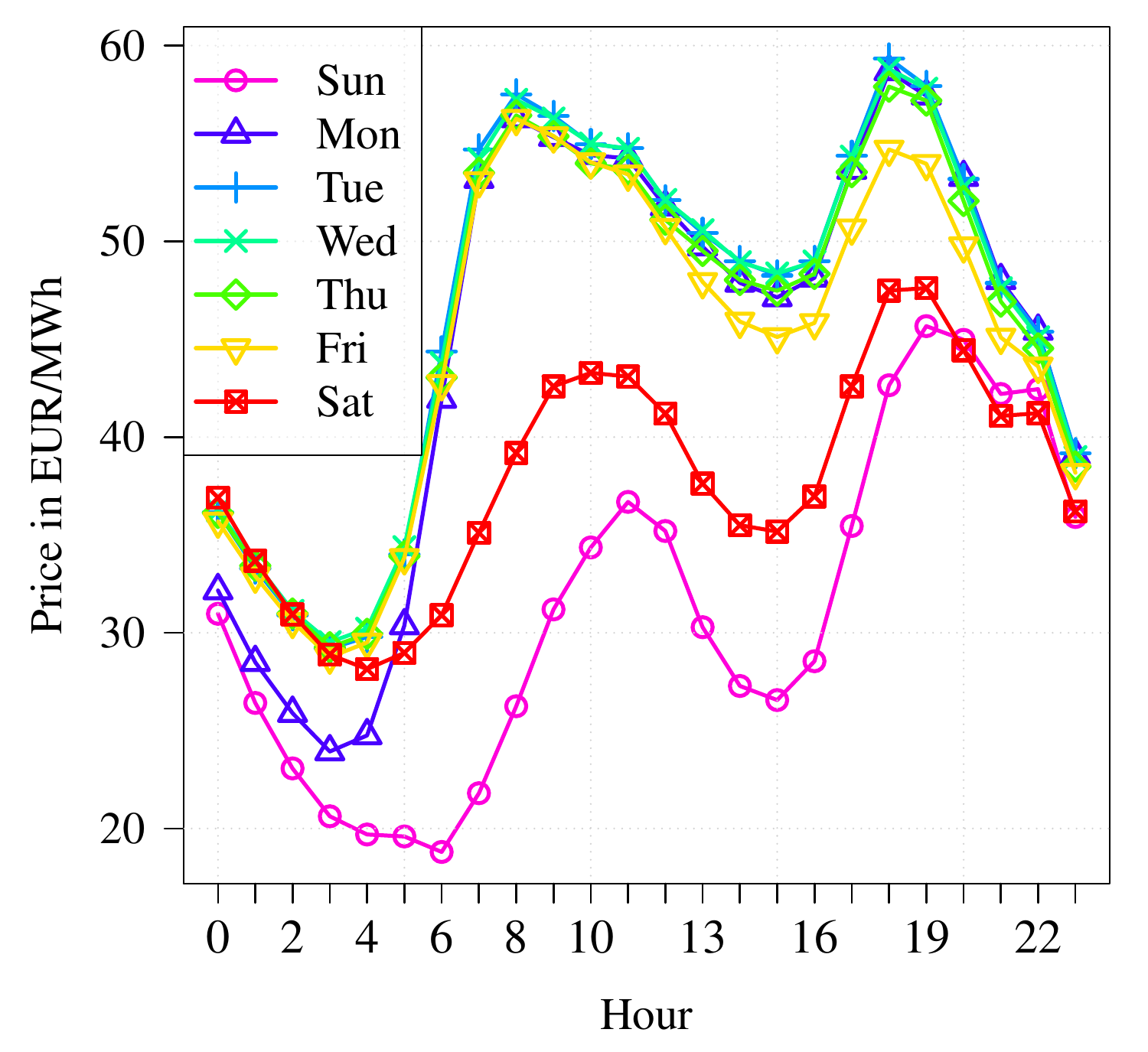} 
  \caption{EXAA - Germany and Austria}
\end{subfigure}
\begin{subfigure}[b]{.49\textwidth}
 \includegraphics[width=1\textwidth]{wm_EPEX_DE+AT.pdf} 
  \caption{EPEX - Germany and Austria}
\end{subfigure}
\begin{subfigure}[b]{.49\textwidth}
 \includegraphics[width=1\textwidth]{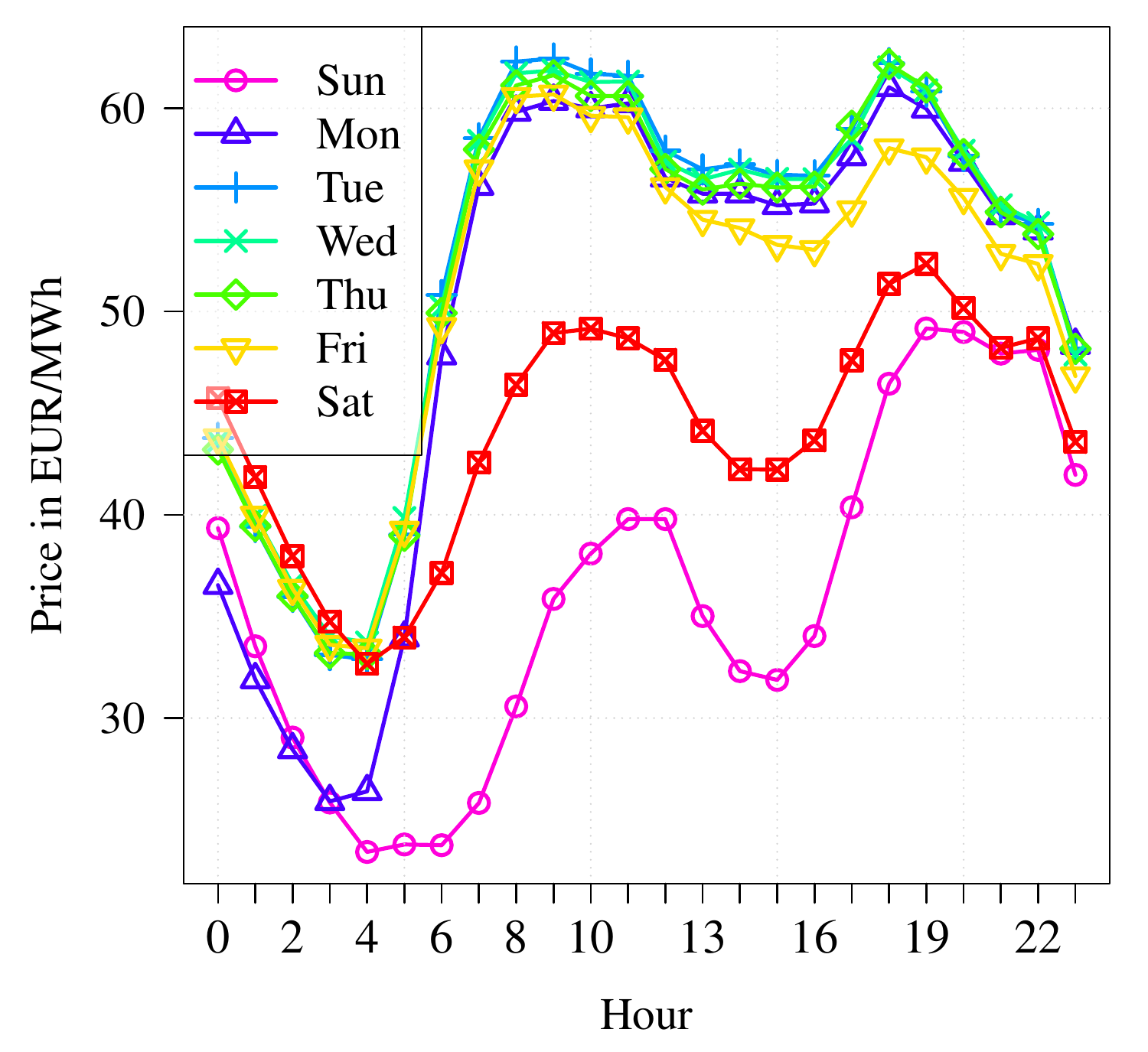} 
  \caption{EPEX - Switzerland}
\end{subfigure}
\begin{subfigure}[b]{.49\textwidth}
 \includegraphics[width=1\textwidth]{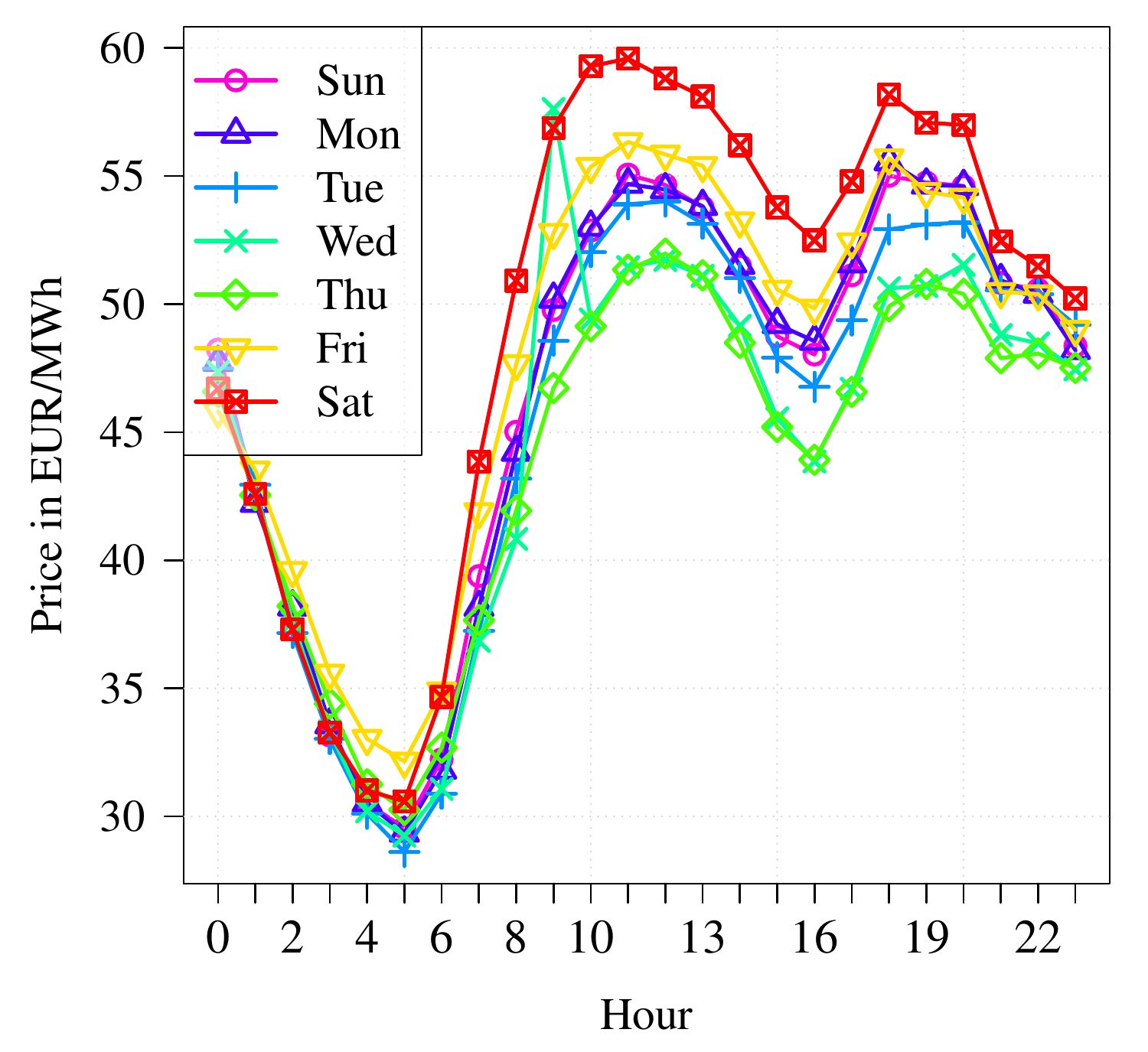} 
  \caption{BELPEX - Belgium}
\end{subfigure}
\caption{Weekly sample mean of $P_{d,h}$ for selected markets. }
\end{figure*}

\begin{figure*}[hbt!]
\centering
\begin{subfigure}[b]{.49\textwidth}
 \includegraphics[width=1\textwidth]{wm_APX_NL.pdf} 
  \caption{APX - Netherlands}
\end{subfigure}
\begin{subfigure}[b]{.49\textwidth}
 \includegraphics[width=1\textwidth]{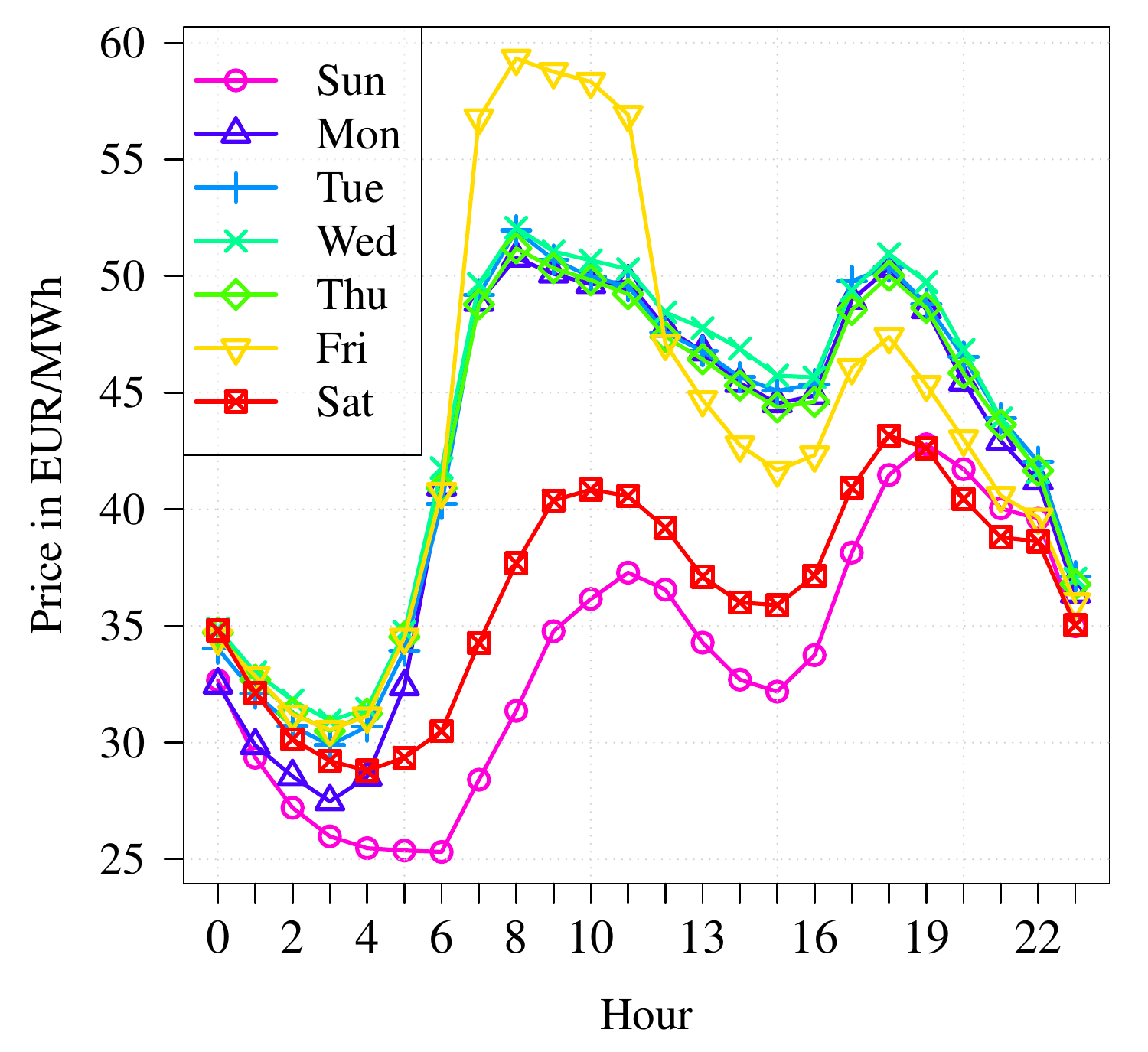} 
  \caption{Nordpool - Denmark West}
\end{subfigure}
\begin{subfigure}[b]{.49\textwidth}
 \includegraphics[width=1\textwidth]{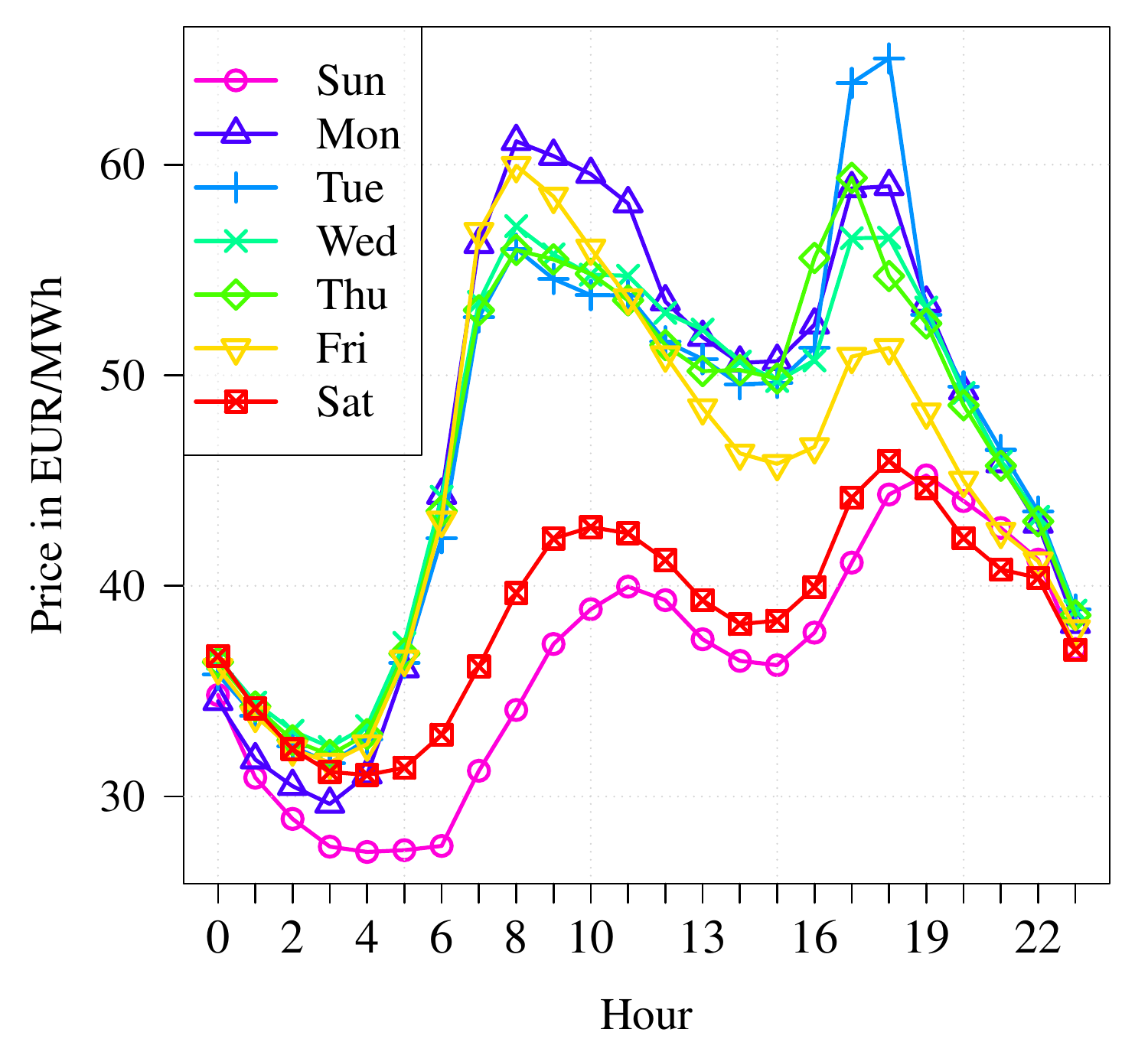} 
  \caption{Nordpool - Denmark East}
\end{subfigure}
\begin{subfigure}[b]{.49\textwidth}
 \includegraphics[width=1\textwidth]{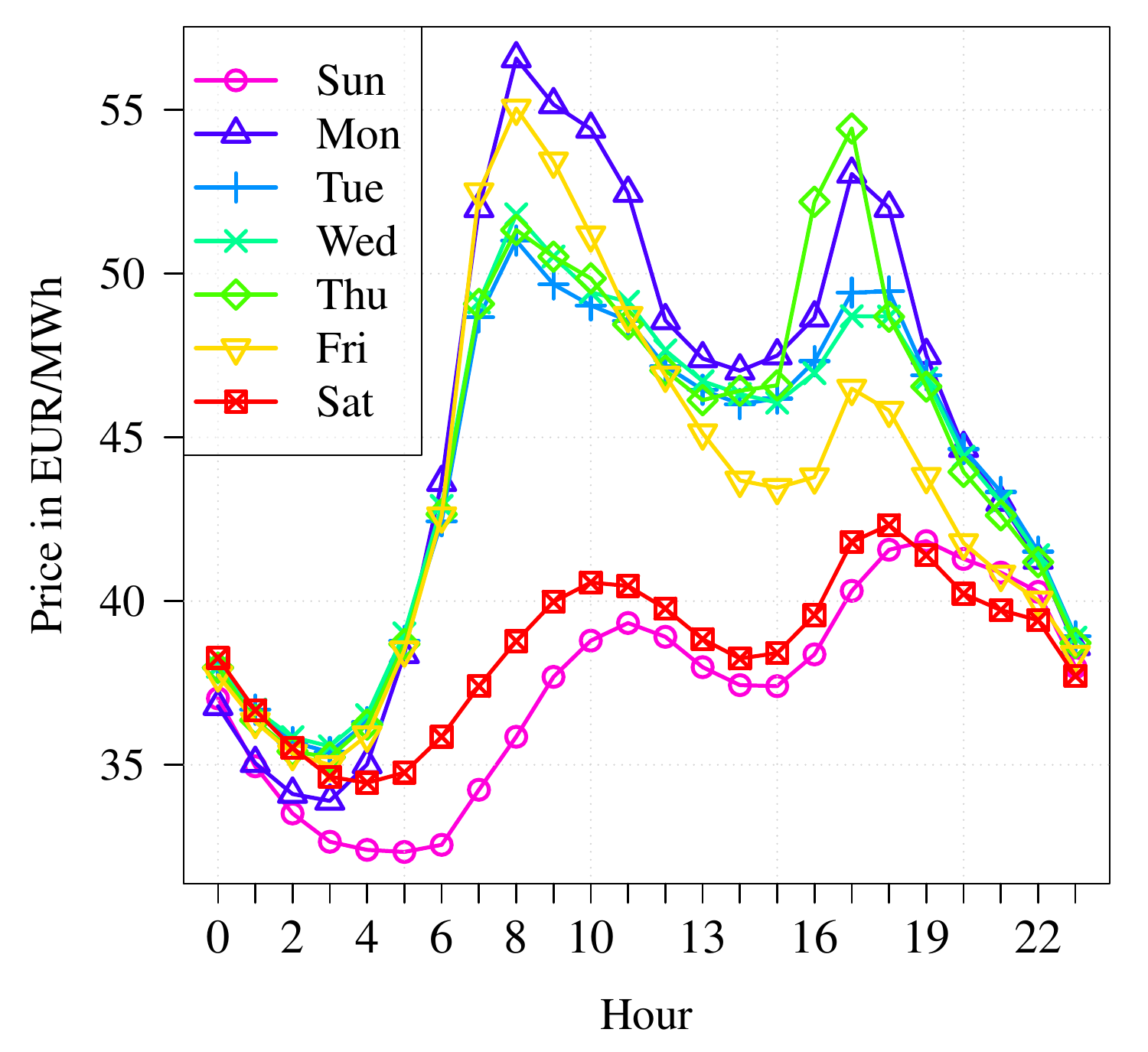} 
  \caption{Nordpool - Sweden(4)}
\end{subfigure}
\begin{subfigure}[b]{.49\textwidth}
 \includegraphics[width=1\textwidth]{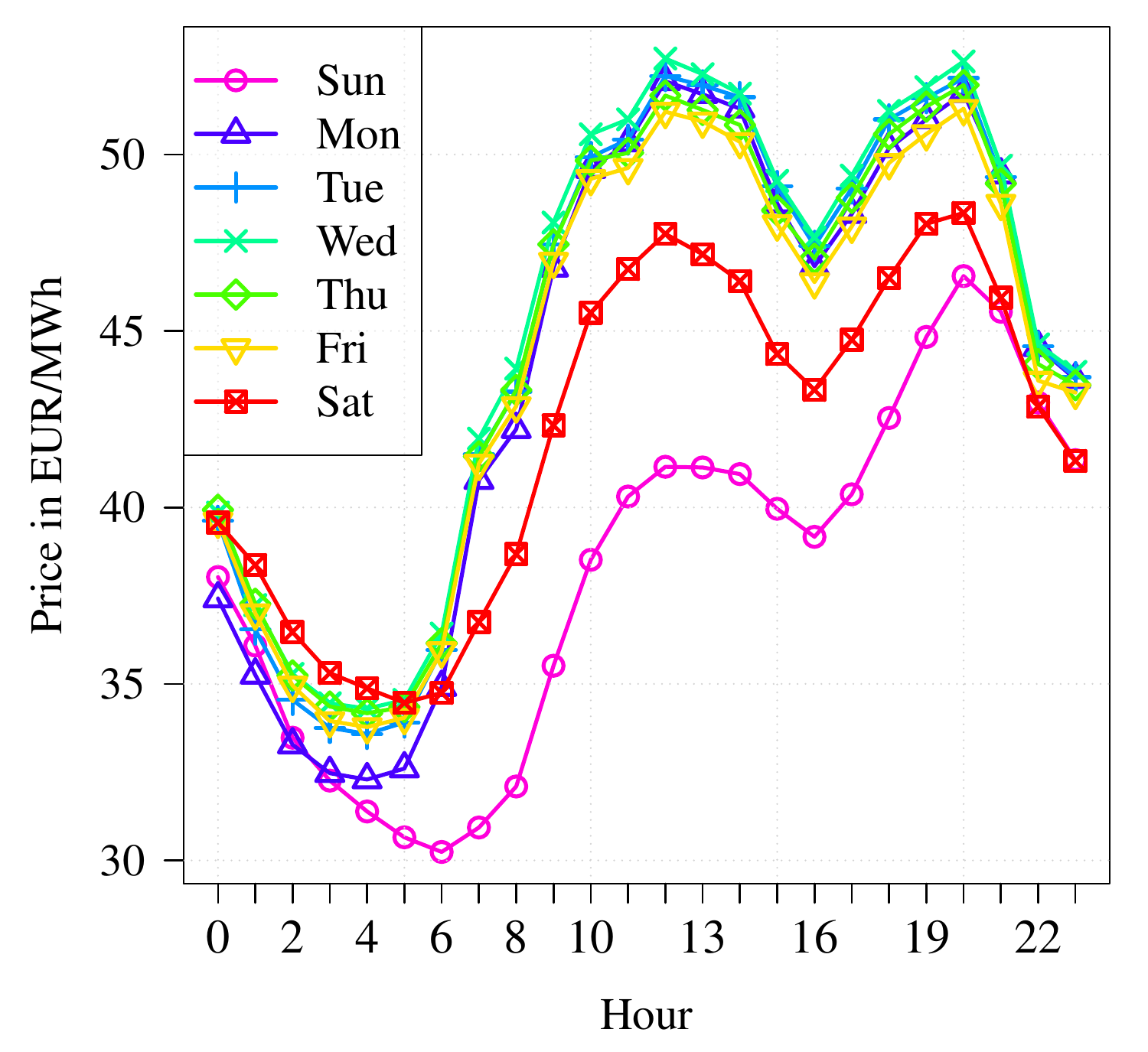} 
  \caption{POLPX - Poland}
\end{subfigure}
\begin{subfigure}[b]{.49\textwidth}
 \includegraphics[width=1\textwidth]{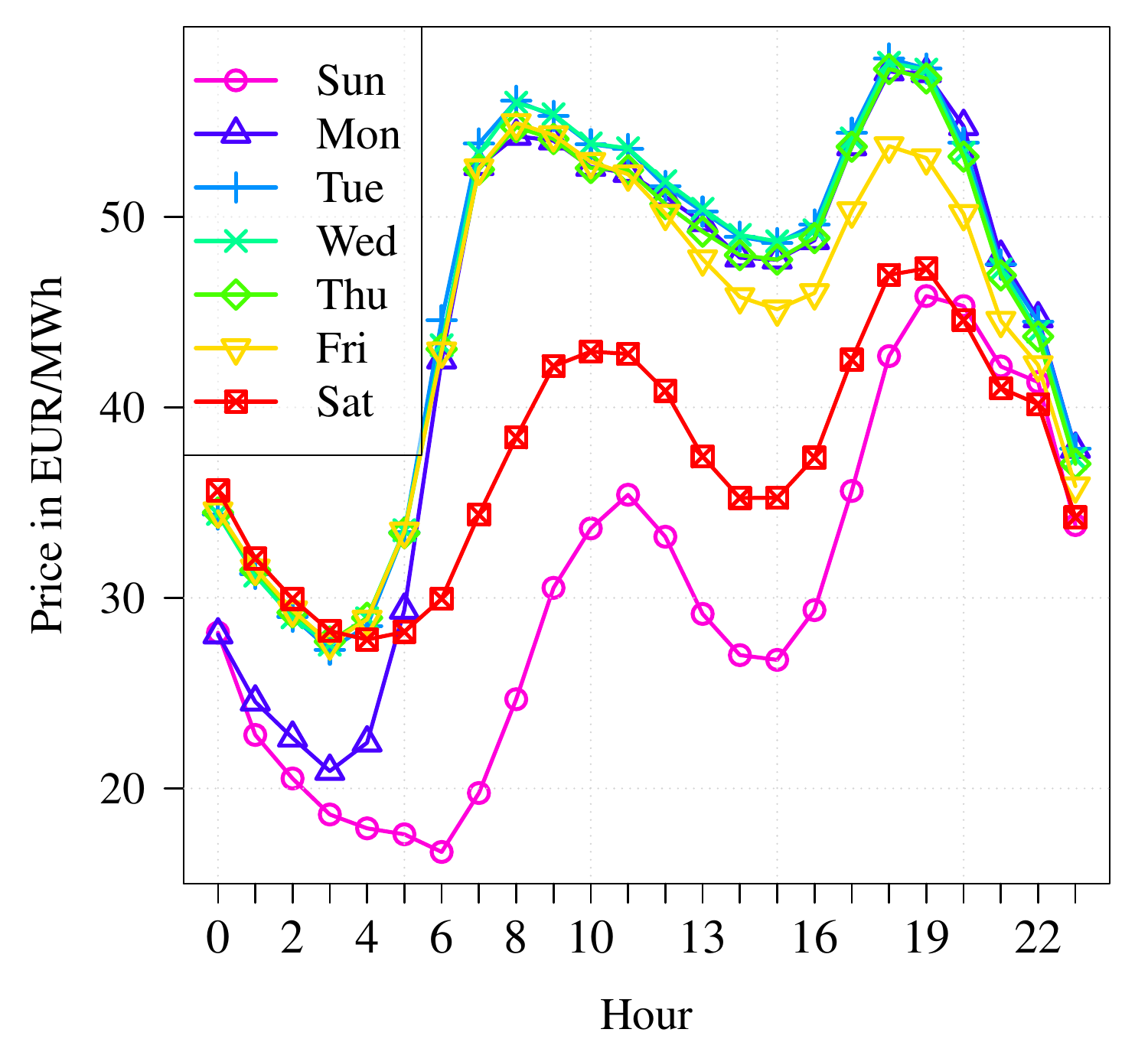} 
  \caption{OTE - Czech}
\end{subfigure}
\caption{Weekly sample mean of $P_{d,h}$ for selected markets. }
\end{figure*}

\clearpage

\subsection{MAE$_h$ plots}

\begin{figure*}[hbt!]
\centering
\begin{subfigure}[b]{.49\textwidth}
 \includegraphics[width=1\textwidth]{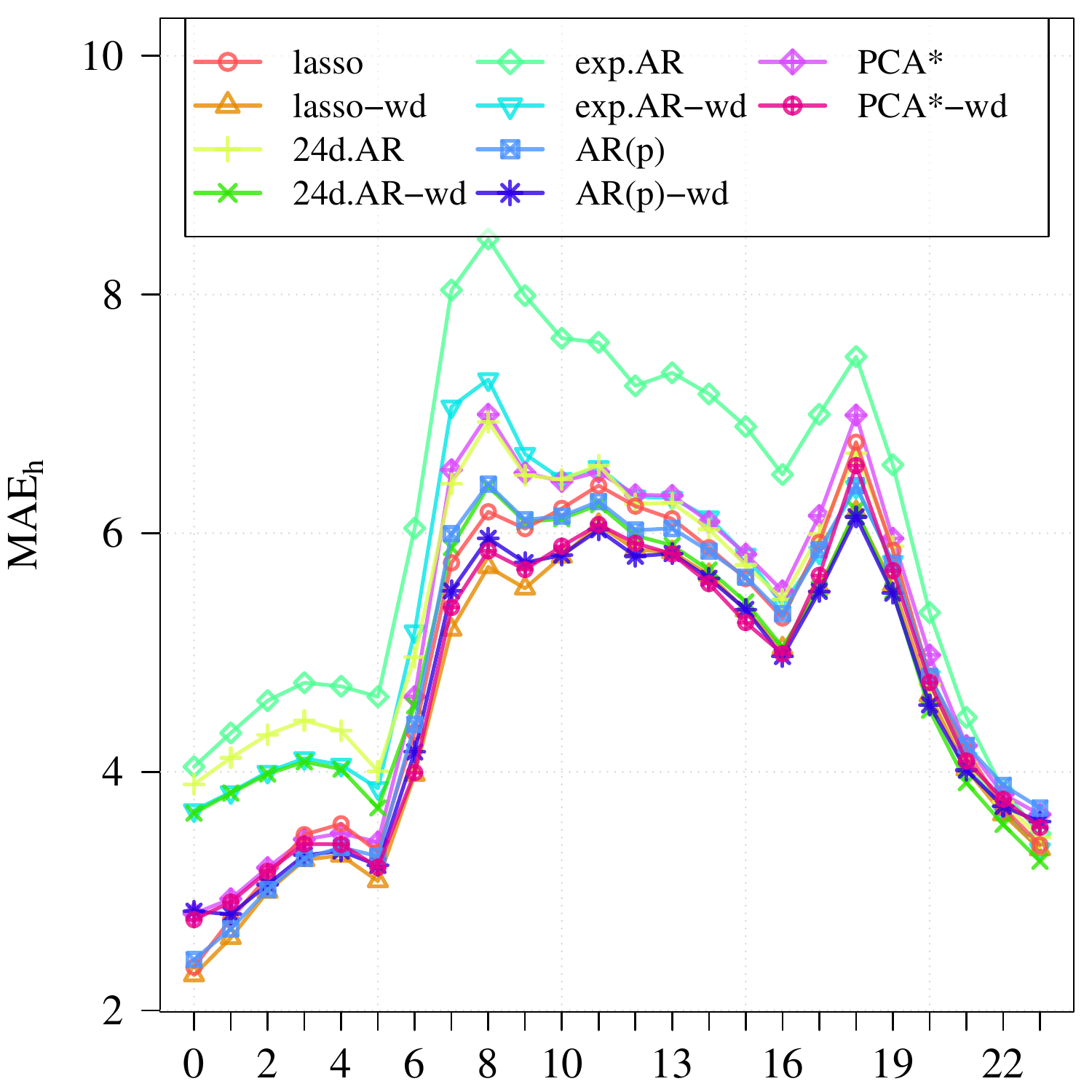} 
  \caption{EXAA - Germany and Austria}
\end{subfigure}
\begin{subfigure}[b]{.49\textwidth}
 \includegraphics[width=1\textwidth]{MAEh_EPEX_DE+AT.pdf} 
  \caption{EPEX - Germany and Austria}
\end{subfigure}
\begin{subfigure}[b]{.49\textwidth}
 \includegraphics[width=1\textwidth]{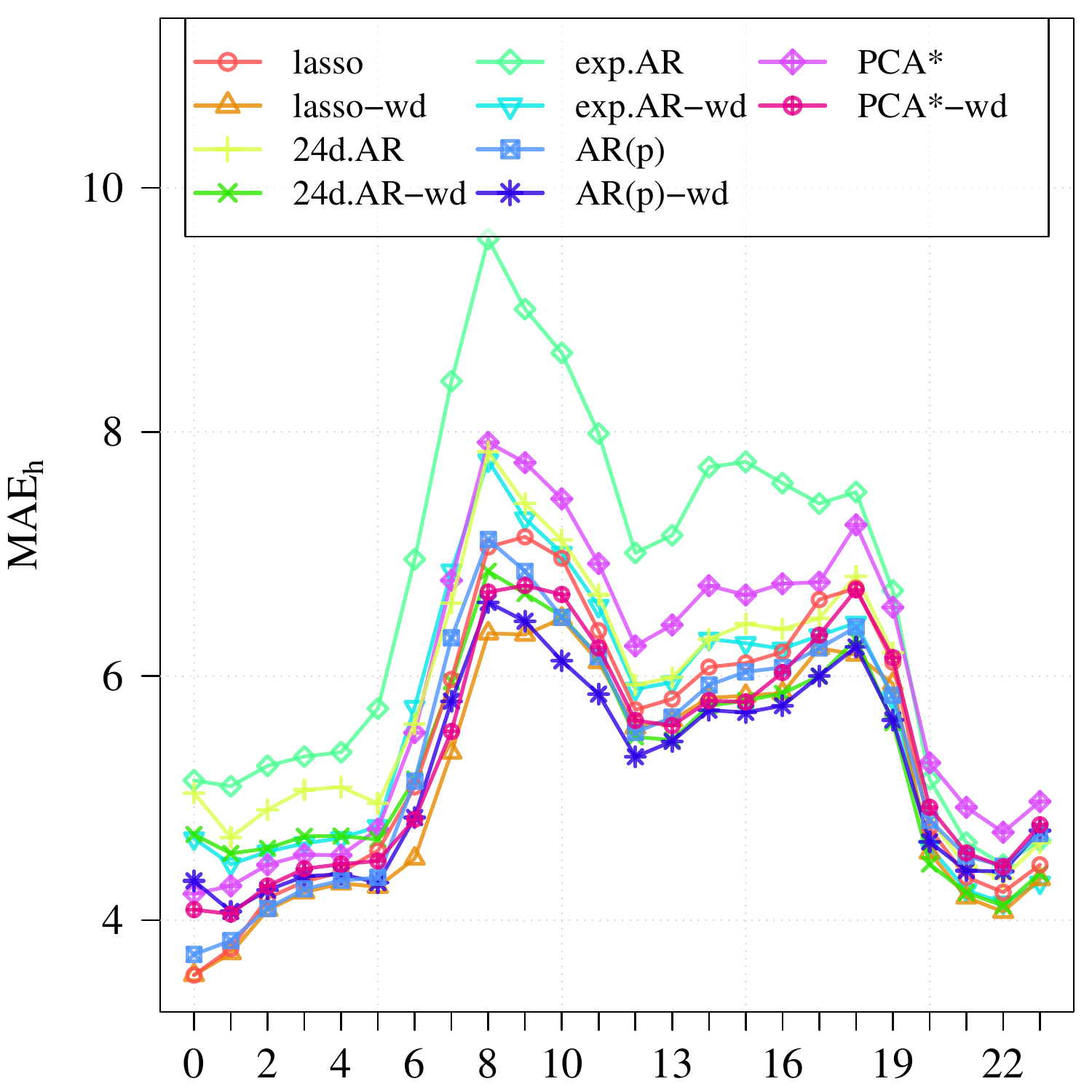} 
  \caption{EPEX - Switzerland}
\end{subfigure}
\begin{subfigure}[b]{.49\textwidth}
 \includegraphics[width=1\textwidth]{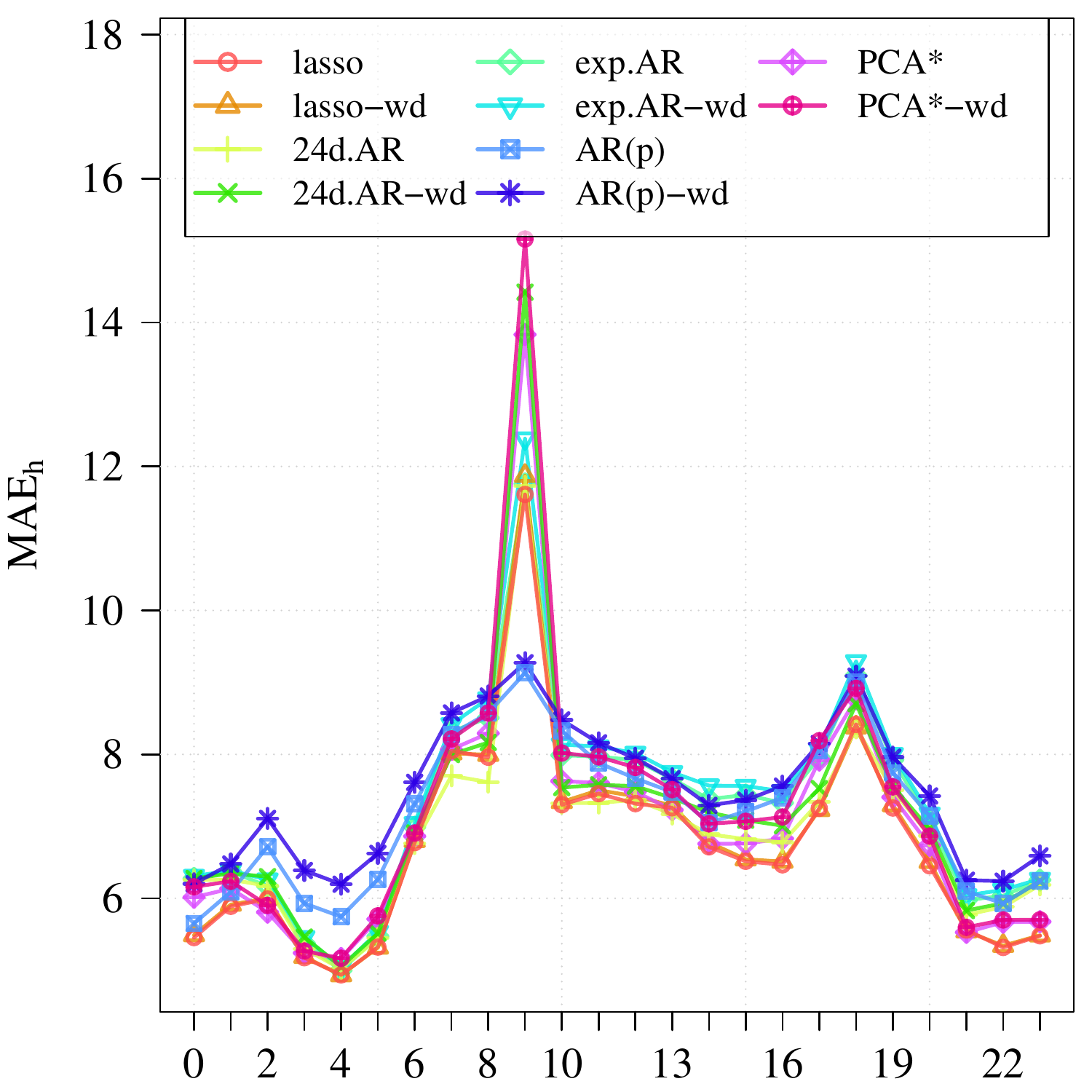} 
  \caption{BELPEX - Belgium}
\end{subfigure}
\caption{MAE$_h$ for selected markets}
\end{figure*}

\begin{figure*}[hbt!]
\centering
\begin{subfigure}[b]{.49\textwidth}
 \includegraphics[width=1\textwidth]{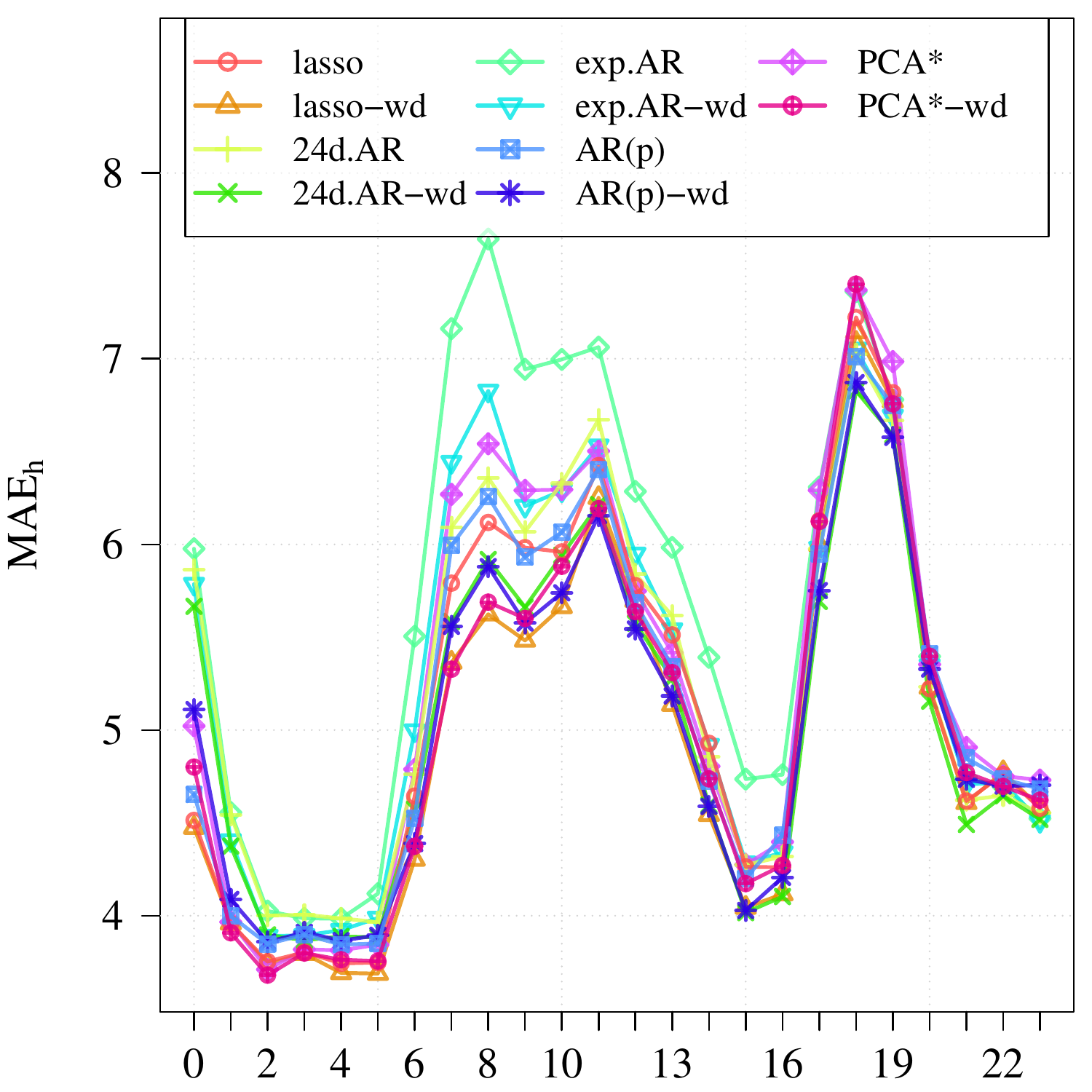} 
  \caption{APX - Netherlands}
\end{subfigure}
\begin{subfigure}[b]{.49\textwidth}
 \includegraphics[width=1\textwidth]{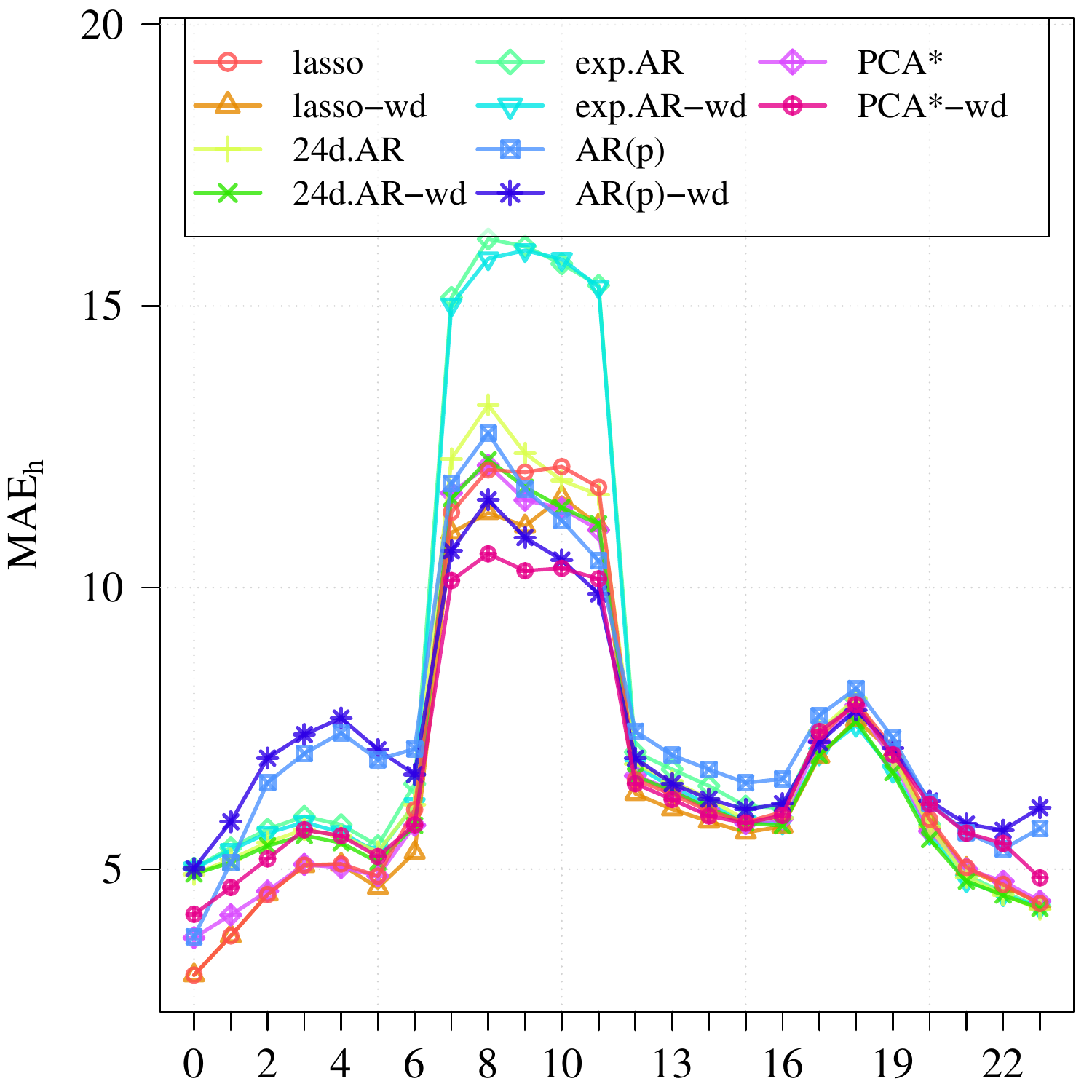} 
  \caption{Nordpool - Denmark West}
\end{subfigure}
\begin{subfigure}[b]{.49\textwidth}
 \includegraphics[width=1\textwidth]{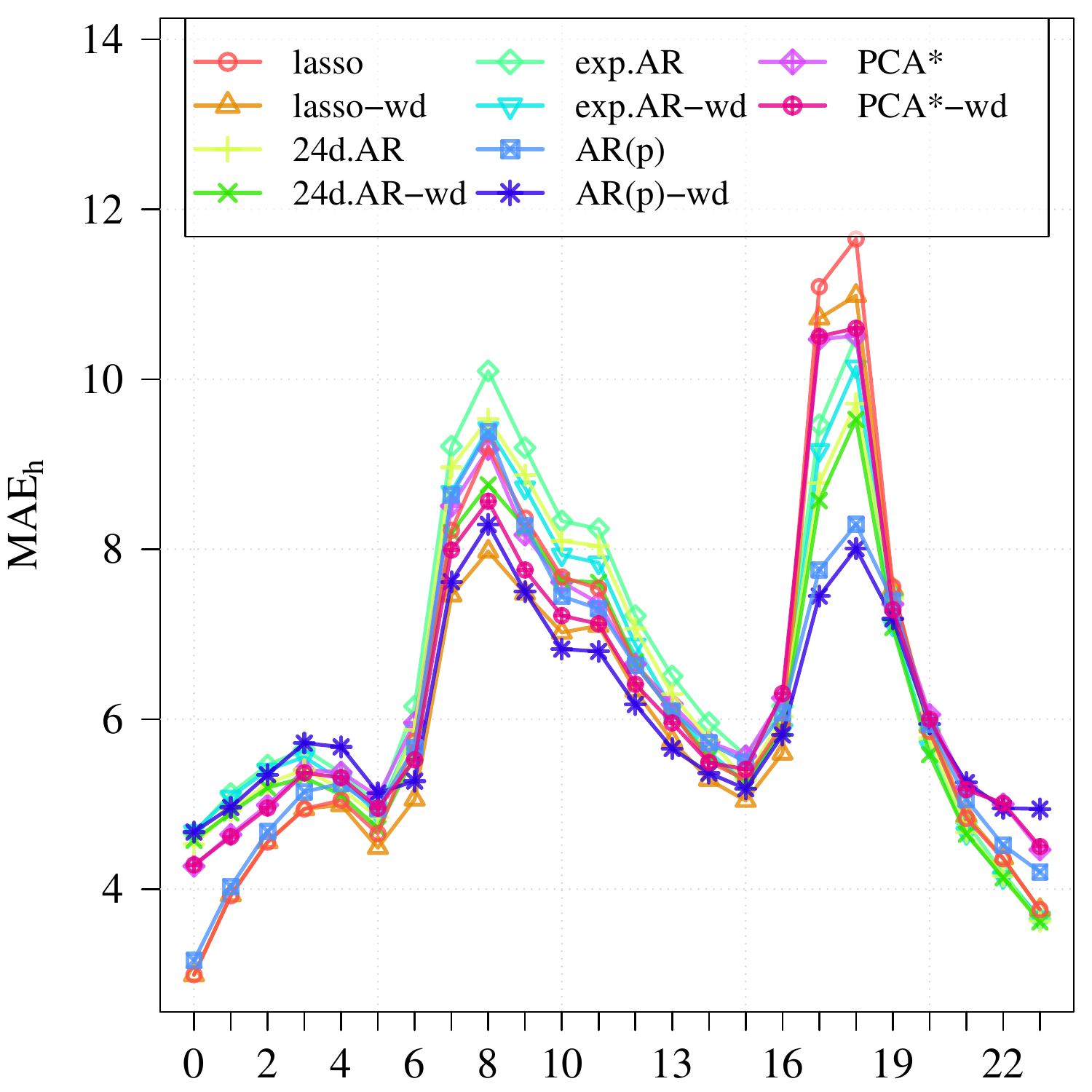} 
  \caption{Nordpool - Denmark East}
\end{subfigure}
\begin{subfigure}[b]{.49\textwidth}
 \includegraphics[width=1\textwidth]{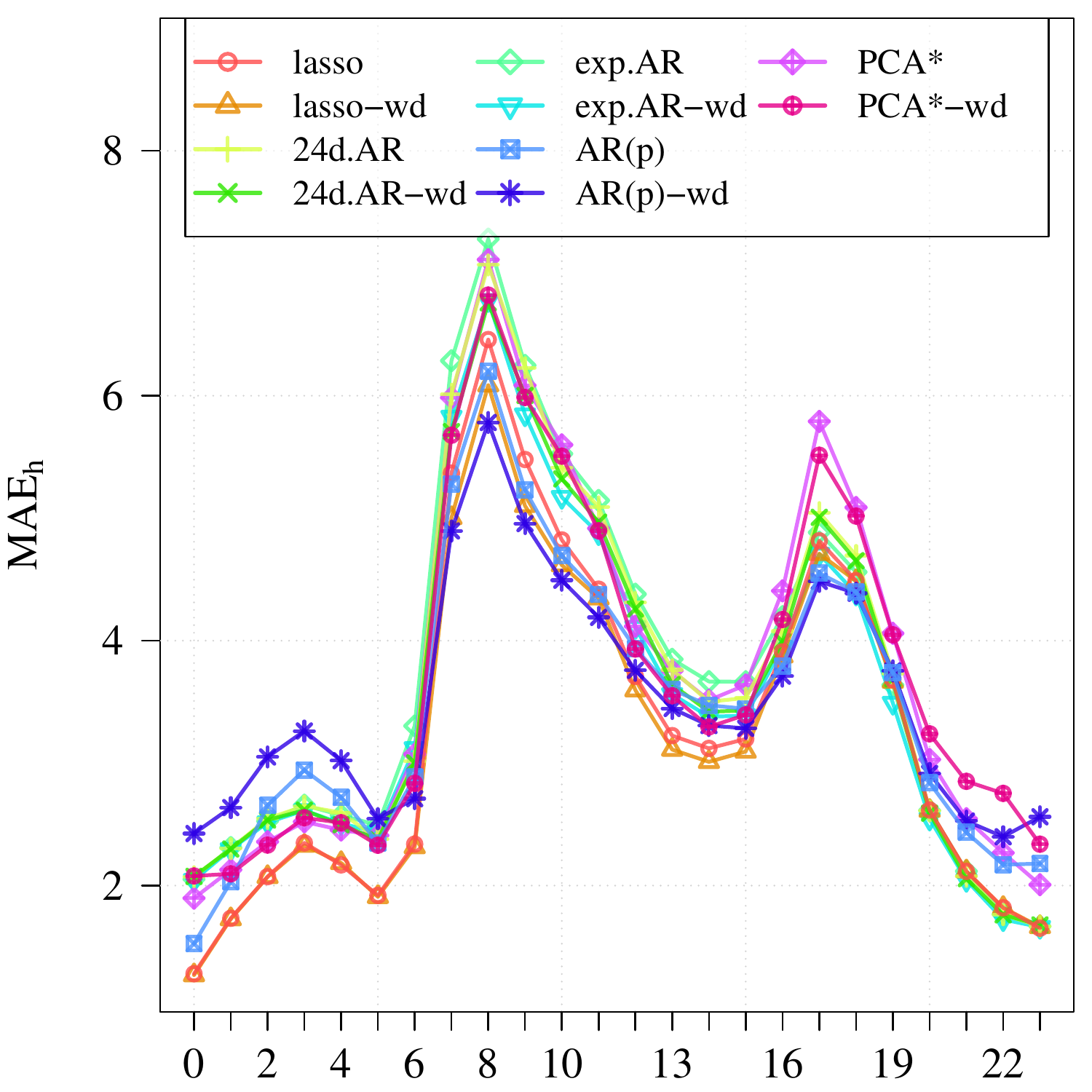} 
  \caption{Nordpool - Sweden(4)}
\end{subfigure}
\begin{subfigure}[b]{.49\textwidth}
 \includegraphics[width=1\textwidth]{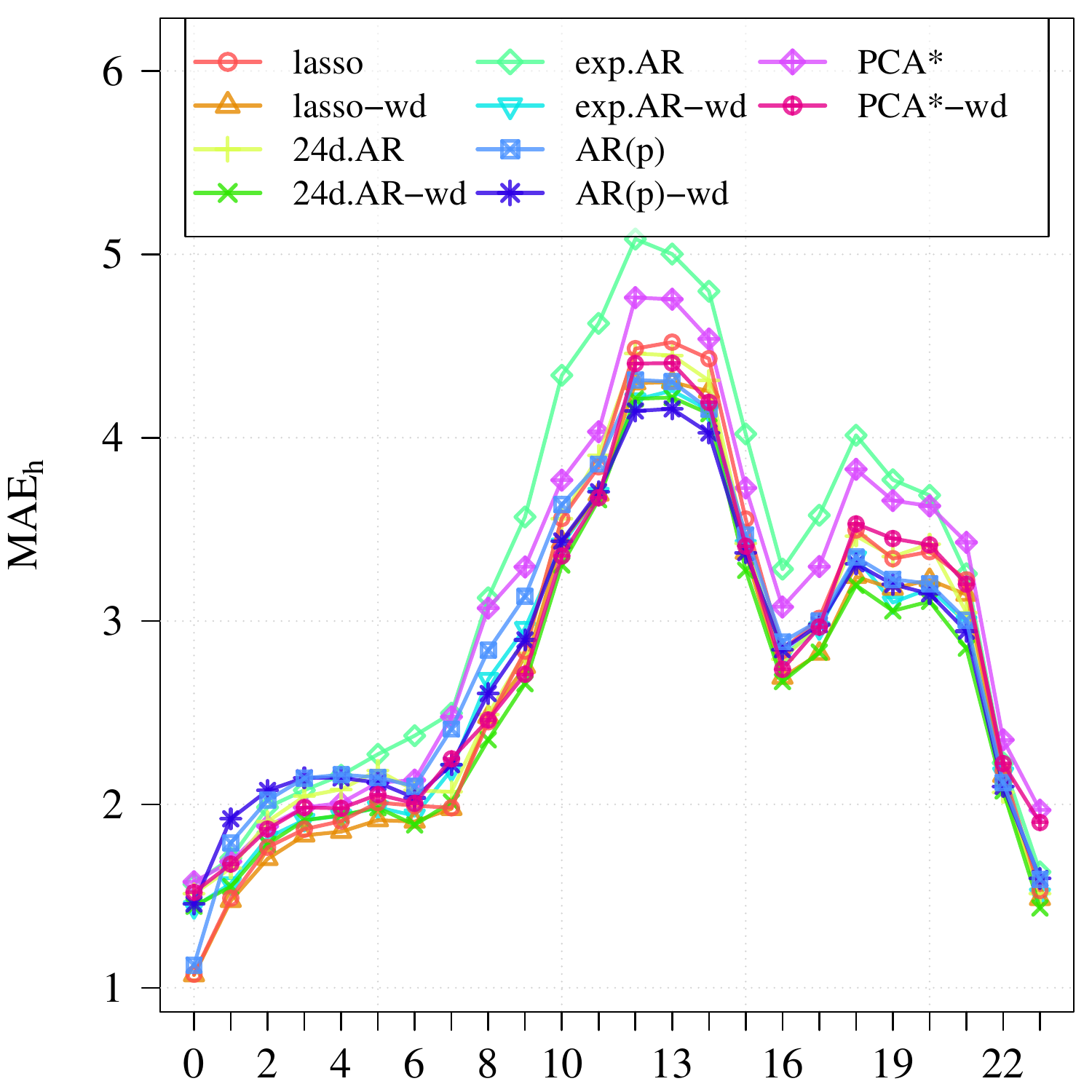} 
  \caption{POLPX - Poland}
\end{subfigure}
\begin{subfigure}[b]{.49\textwidth}
 \includegraphics[width=1\textwidth]{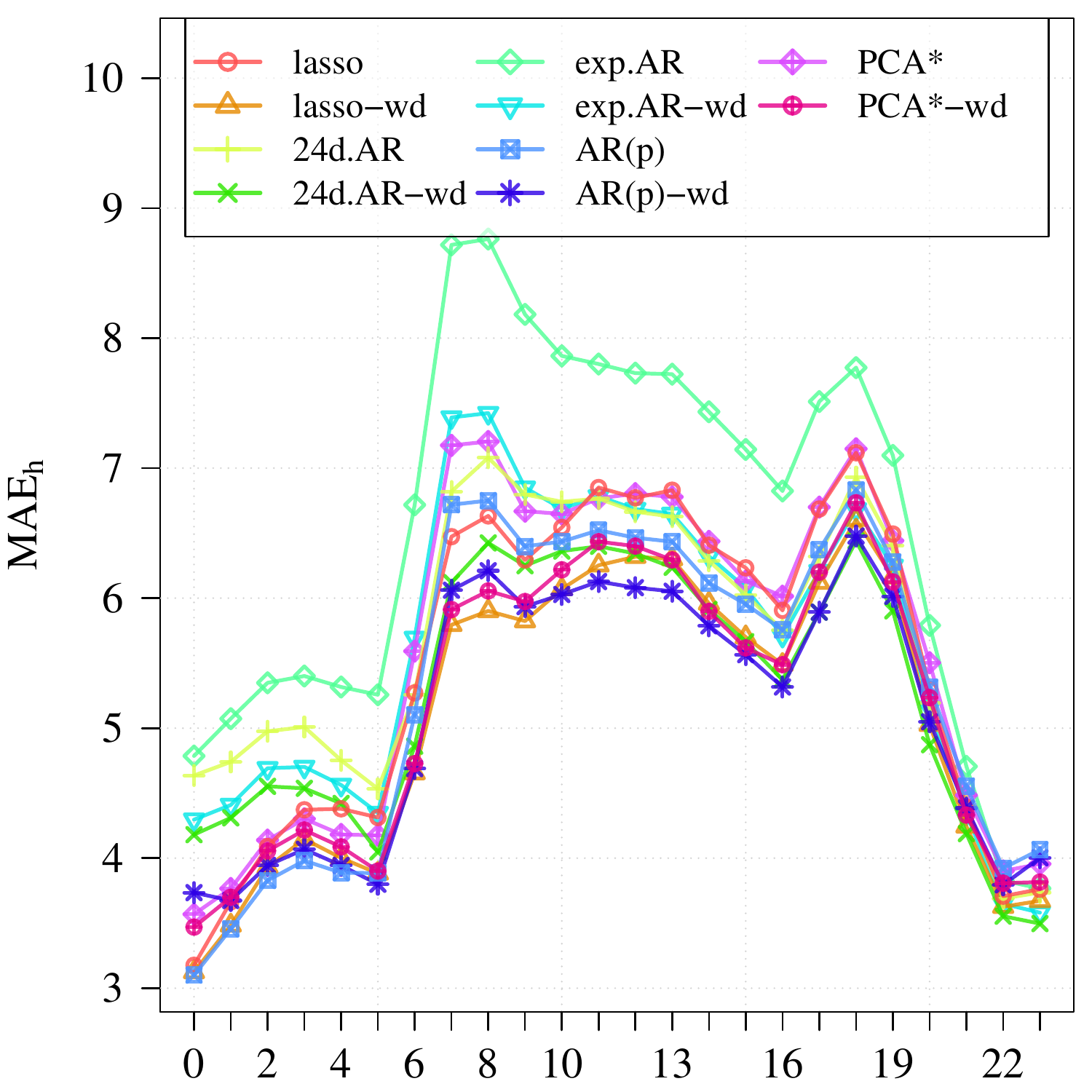} 
  \caption{OTE - Czech}
\end{subfigure}
\caption{MAE$_h$ for selected markets}
\end{figure*}

\clearpage

\subsection{RMSE$_h$ plots}

\begin{figure*}[hbt!]
\centering
\begin{subfigure}[b]{.49\textwidth}
 \includegraphics[width=1\textwidth]{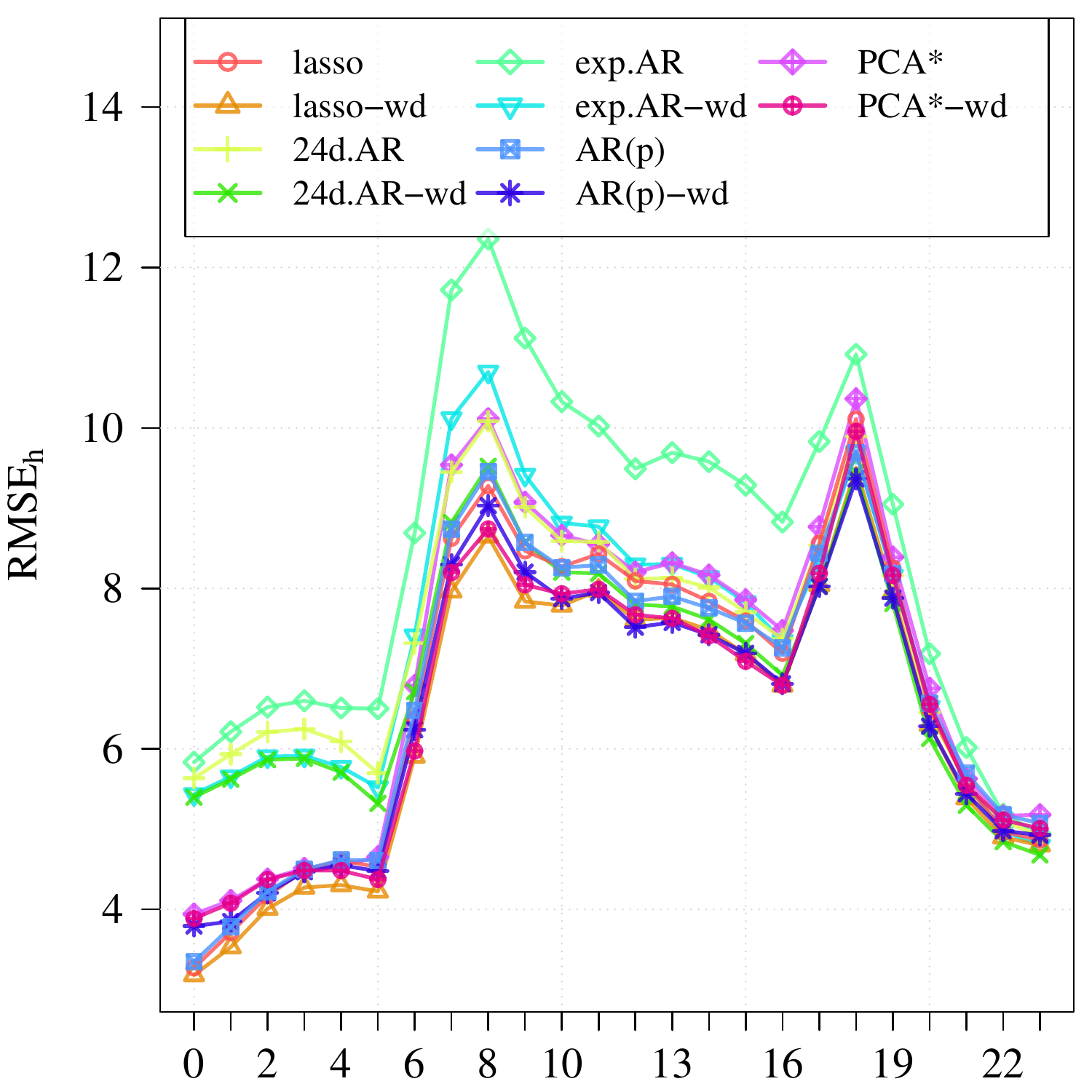} 
  \caption{EXAA - Germany and Austria}
\end{subfigure}
\begin{subfigure}[b]{.49\textwidth}
 \includegraphics[width=1\textwidth]{RMSEh_EPEX_DE+AT.pdf} 
  \caption{EPEX - Germany and Austria}
\end{subfigure}
\begin{subfigure}[b]{.49\textwidth}
 \includegraphics[width=1\textwidth]{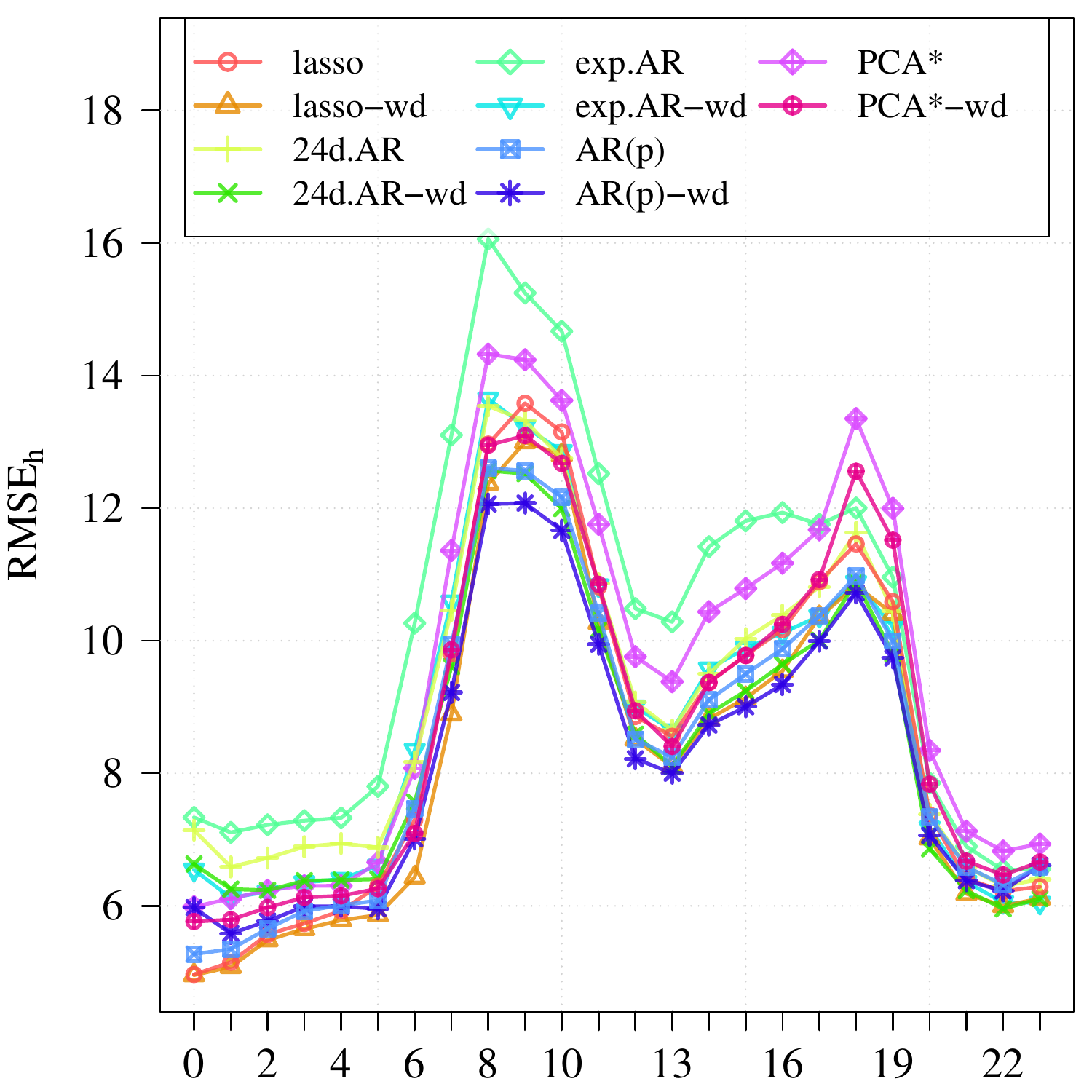} 
  \caption{EPEX - Switzerland}
\end{subfigure}
\begin{subfigure}[b]{.49\textwidth}
 \includegraphics[width=1\textwidth]{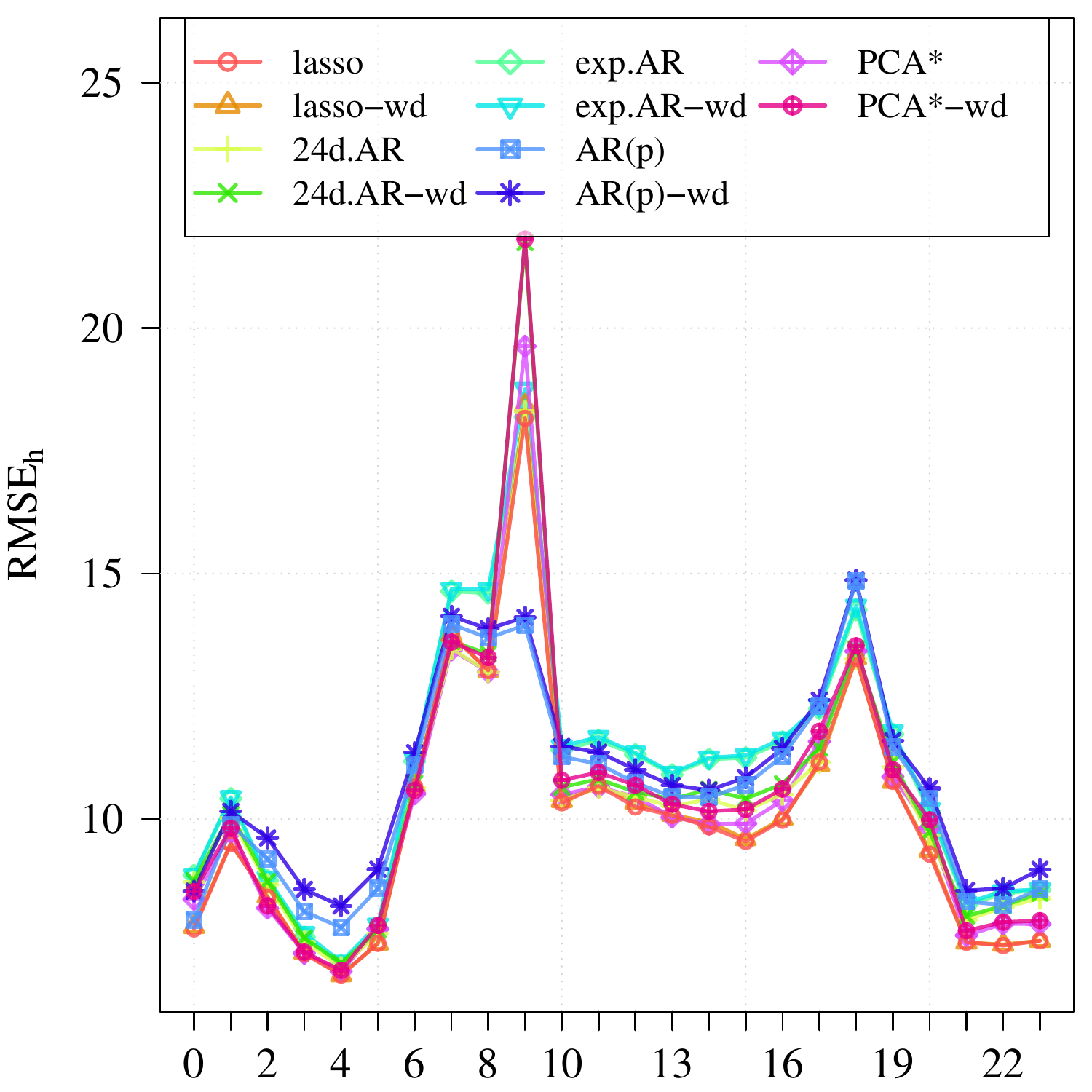} 
  \caption{BELPEX - Belgium}
\end{subfigure}
\caption{RMSE$_h$ for selected markets}
\end{figure*}

\begin{figure*}[hbt!]
\centering
\begin{subfigure}[b]{.49\textwidth}
 \includegraphics[width=1\textwidth]{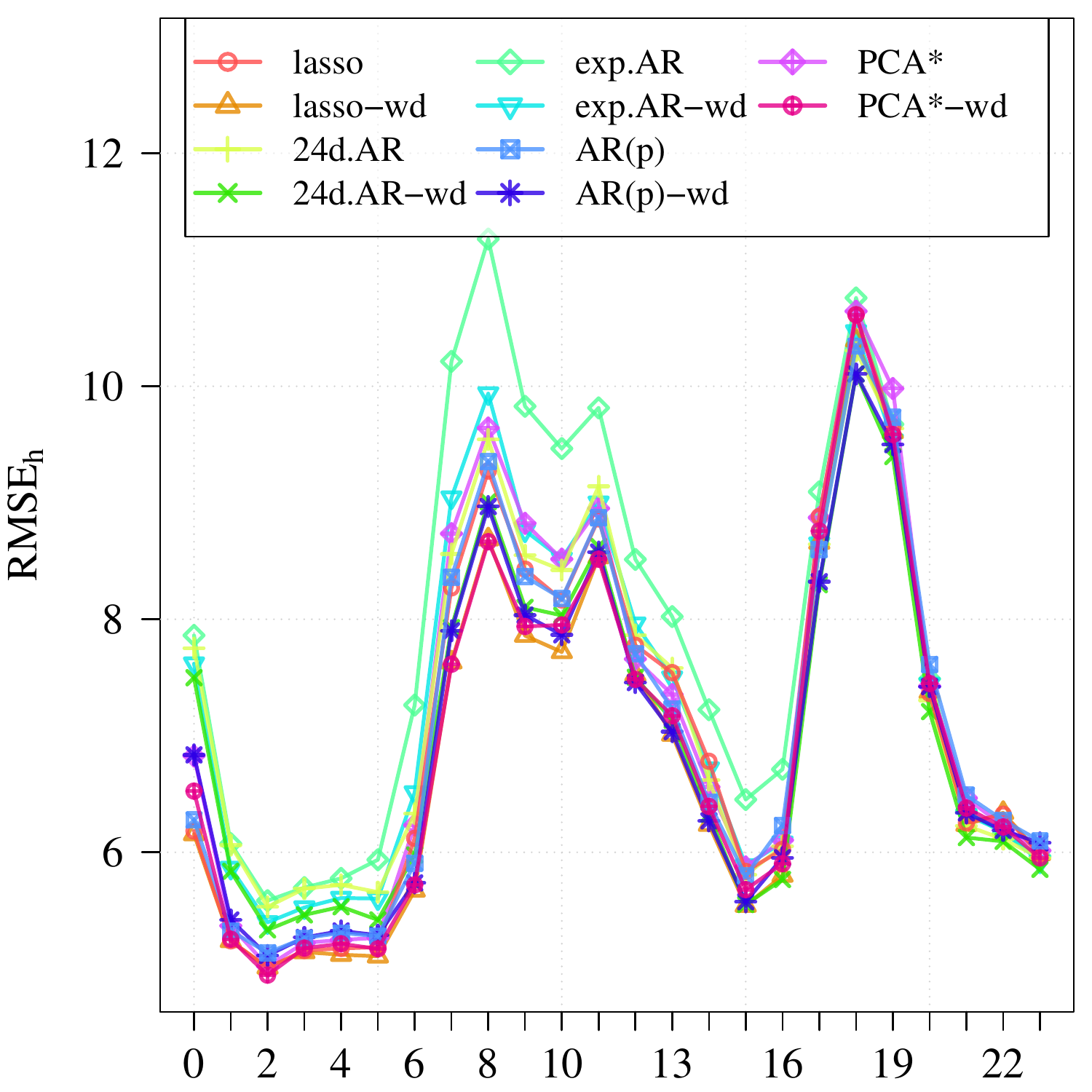} 
  \caption{APX - Netherlands}
\end{subfigure}
\begin{subfigure}[b]{.49\textwidth}
 \includegraphics[width=1\textwidth]{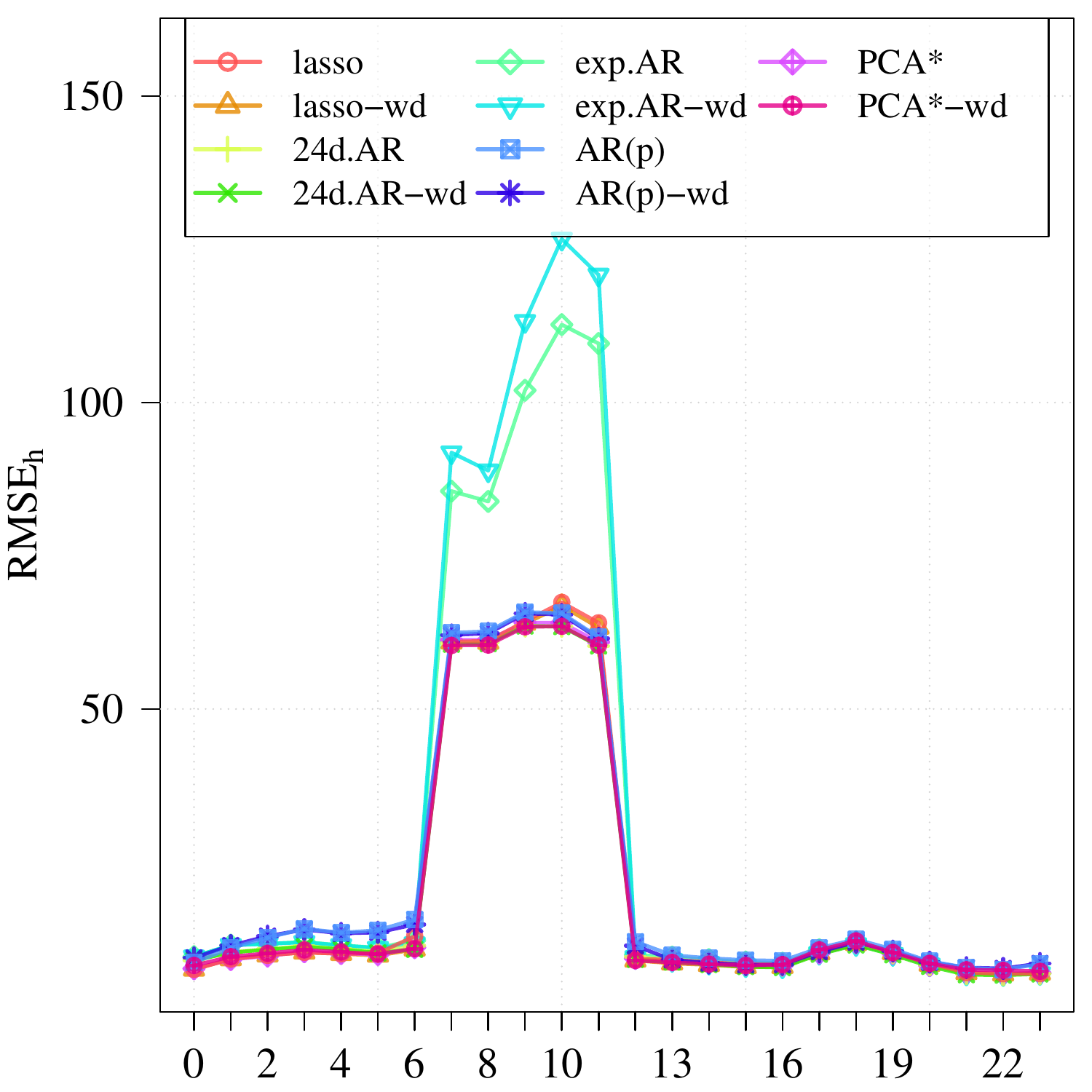} 
  \caption{Nordpool - Denmark West}
\end{subfigure}
\begin{subfigure}[b]{.49\textwidth}
 \includegraphics[width=1\textwidth]{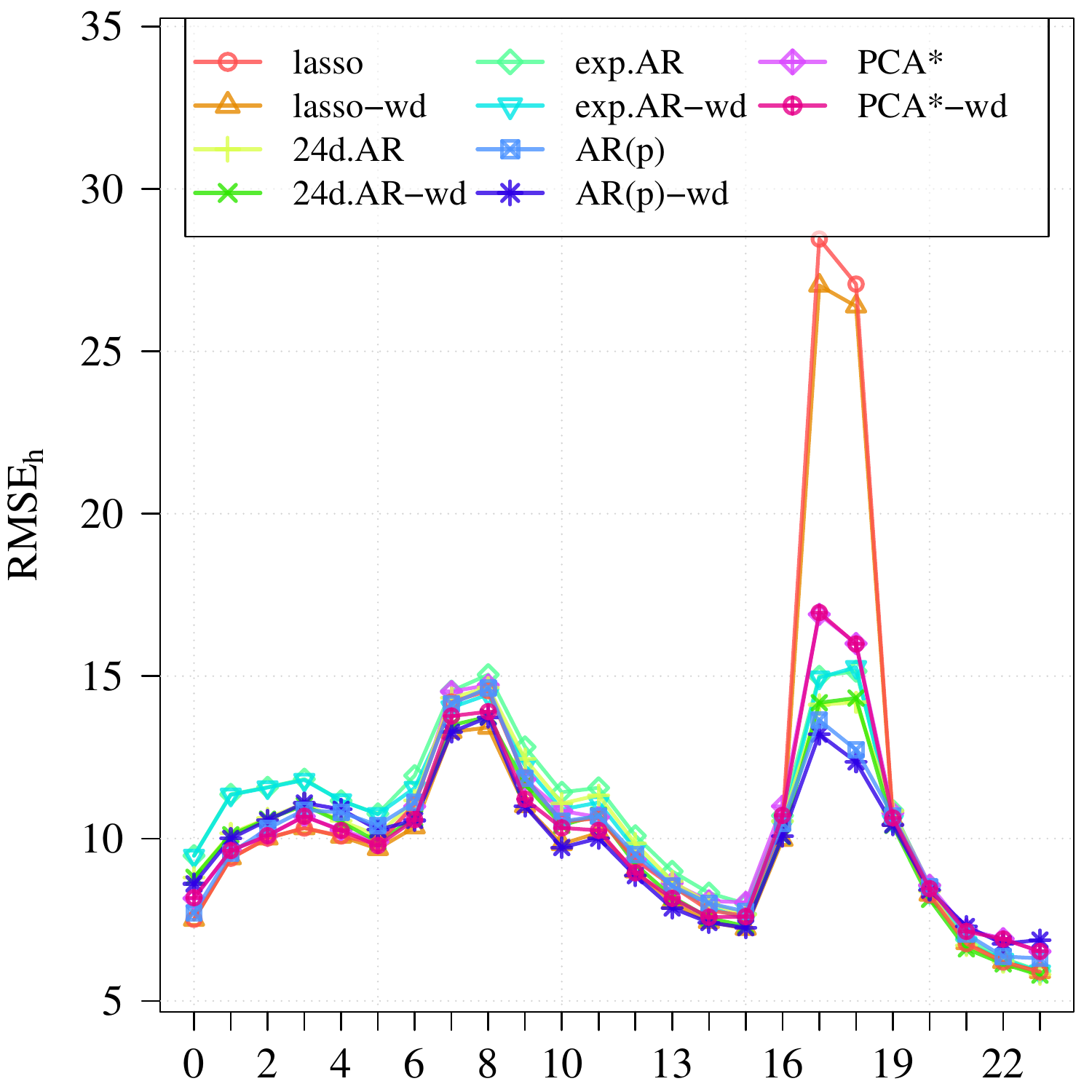} 
  \caption{Nordpool - Denmark East}
\end{subfigure}
\begin{subfigure}[b]{.49\textwidth}
 \includegraphics[width=1\textwidth]{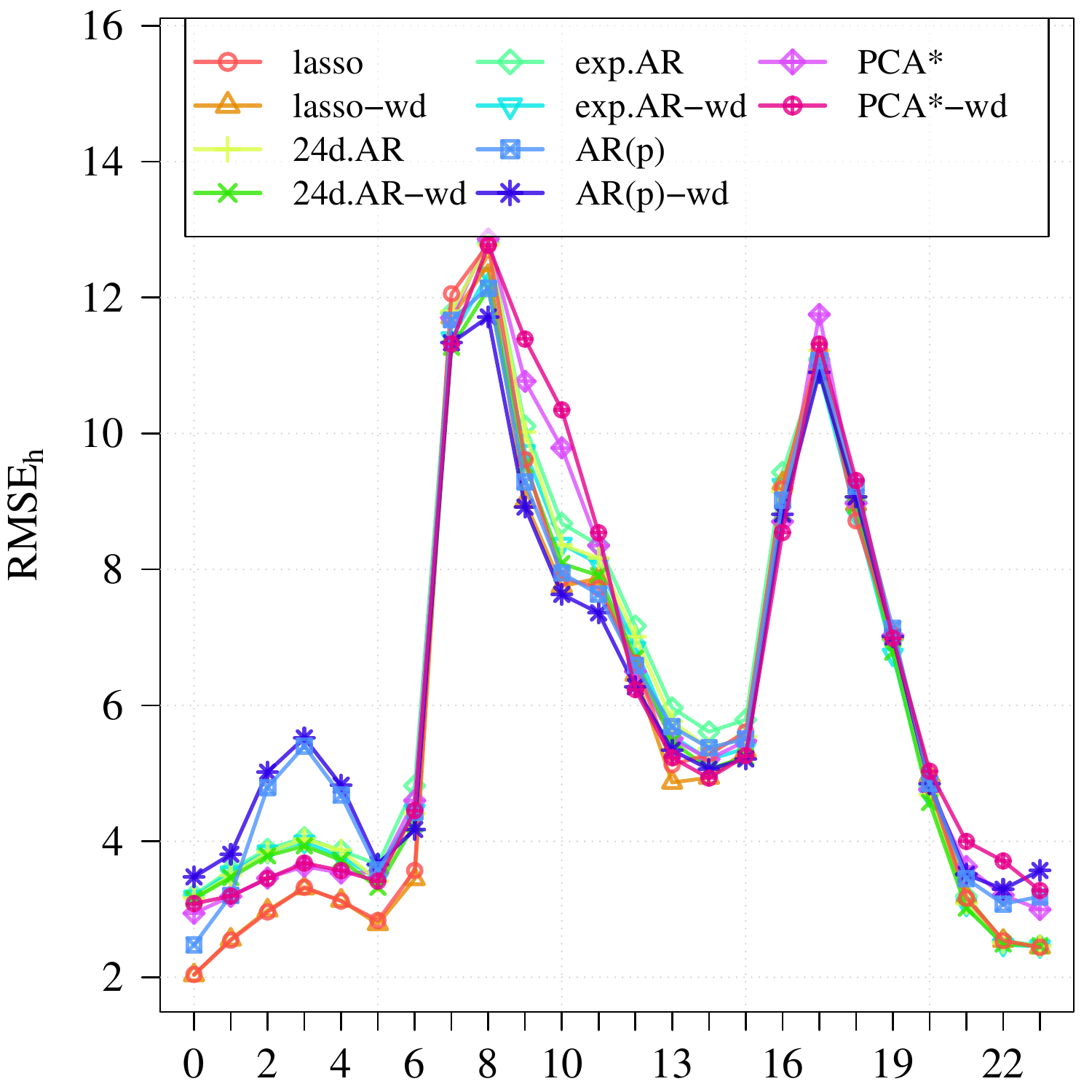} 
  \caption{Nordpool - Sweden(4)}
\end{subfigure}
\begin{subfigure}[b]{.49\textwidth}
 \includegraphics[width=1\textwidth]{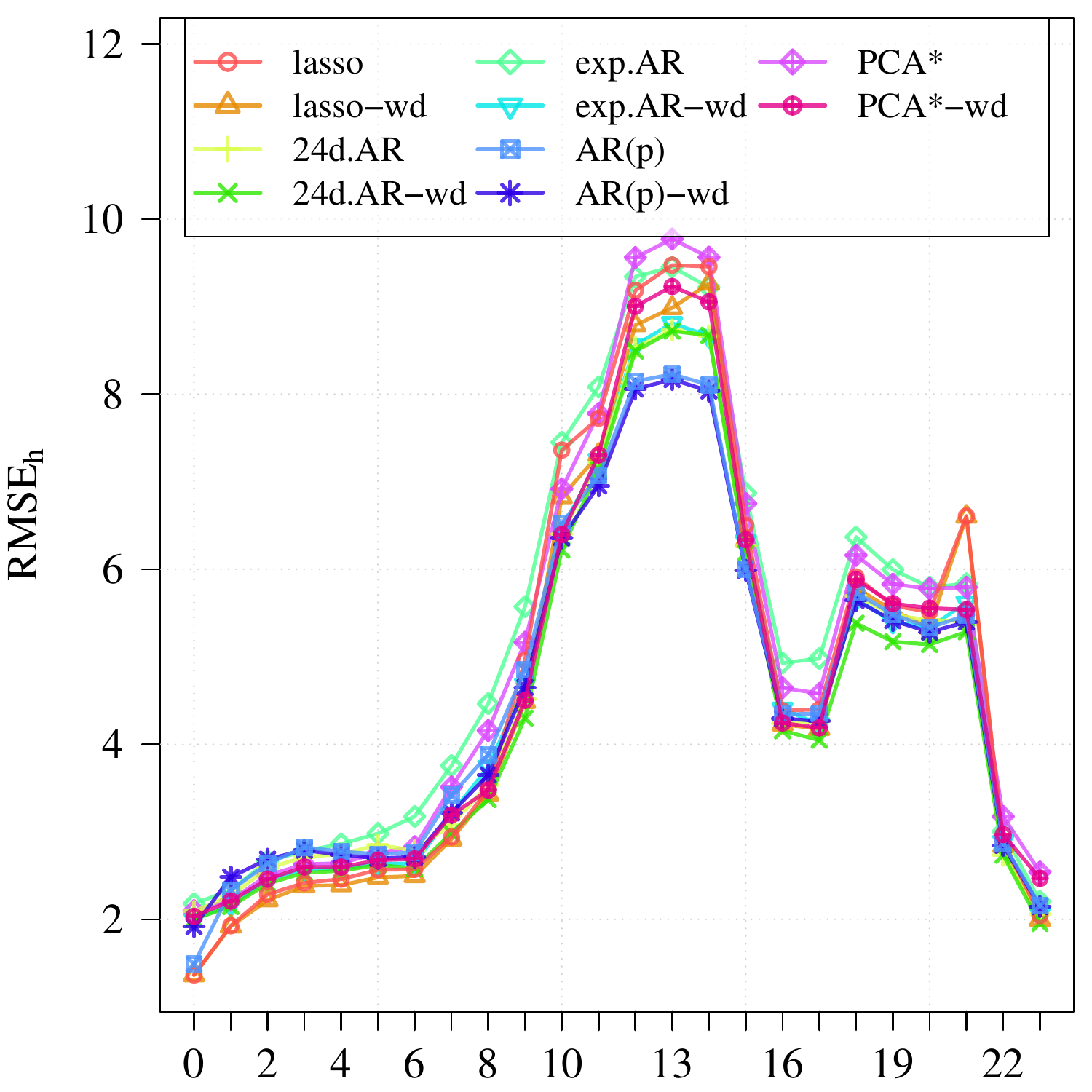} 
  \caption{POLPX - Poland}
\end{subfigure}
\begin{subfigure}[b]{.49\textwidth}
 \includegraphics[width=1\textwidth]{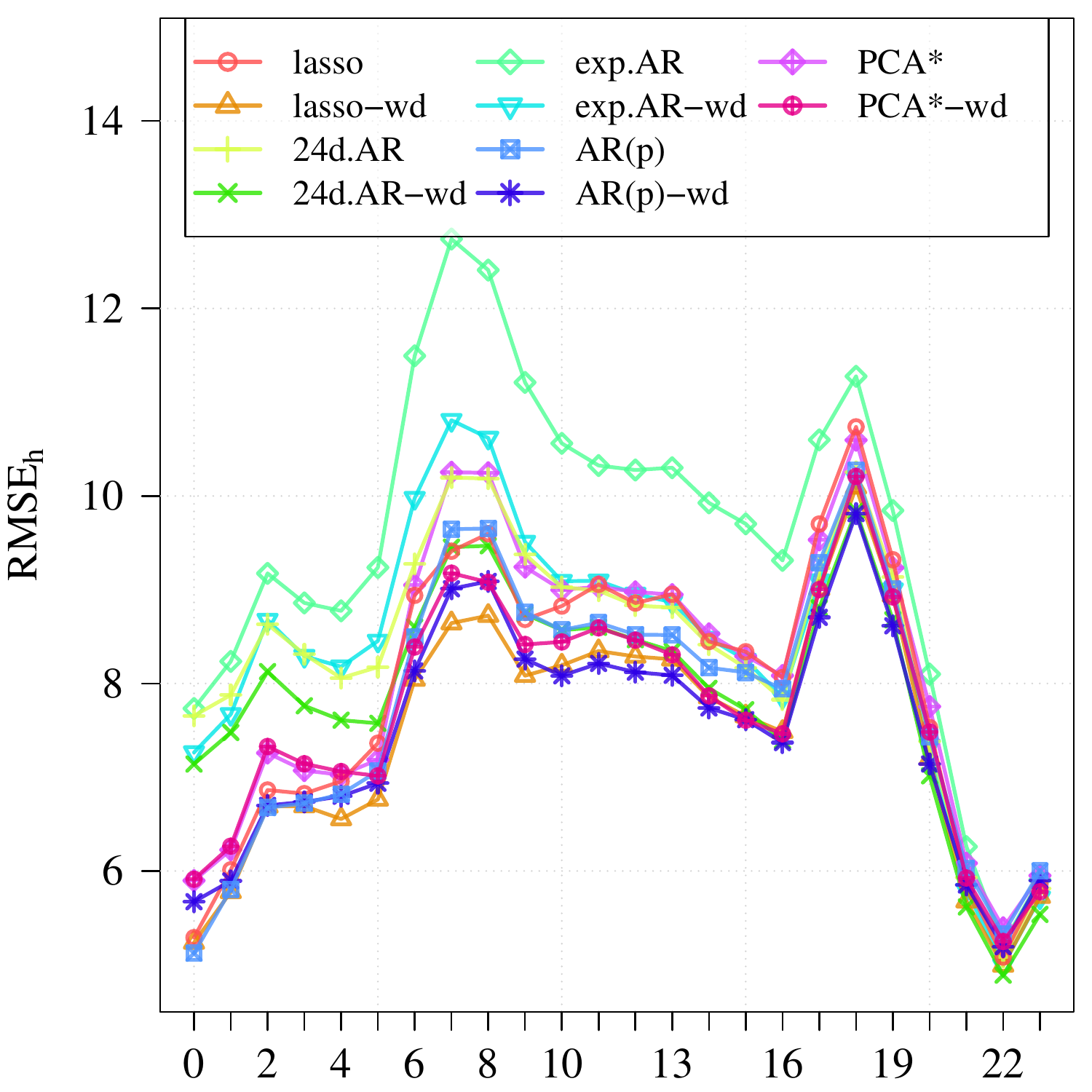} 
  \caption{OTE - Czech}
\end{subfigure}
\caption{RMSE$_h$ for selected markets}
\end{figure*}

\clearpage

\section{Tables of parameter importance}

 \subsection*{EXAA Germany \& Austria}

\begin{table}[ht]
\centering
\begin{tabular}{rrrrrrrr}
  \hline
 & Importance: 1 & Importance: 2 & Importance: 3 & Importance: 4 & Importance: 5 & Importance: 6 & Importance: 7 \\ 
  \hline
0 & 23@1 (65.3) &  3@1 (10.2) & Sun ( 6.6) & 19@1 ( 4.2) & Sat ( 3.7) &  4@1 ( 3.2) & 18@1 ( 3.1) \\ 
  1 & 23@1 (50.0) &  4@1 (10.3) & Sun ( 8.9) &  3@1 ( 7.7) & 19@1 ( 6.3) & 23@2 ( 2.5) & 18@1 ( 2.0) \\ 
  2 & 23@1 (39.1) &  3@1 (13.2) & Sun ( 8.1) & 23@2 ( 7.3) &  4@1 ( 6.9) & 19@1 ( 4.9) & 22@7 ( 3.1) \\ 
  3 & 23@1 (31.9) &  3@1 (10.6) &  4@1 ( 9.7) & 23@2 ( 8.3) & Sun ( 8.3) & 22@7 ( 4.4) & 20@1 ( 3.7) \\ 
  4 & 23@1 (30.6) &  4@1 (19.7) & Sun (11.1) & 23@2 ( 5.8) & 22@7 ( 5.5) &  4@6 ( 5.1) & 20@1 ( 4.6) \\ 
  5 & 23@1 (27.2) & Sun (14.9) &  4@1 ( 7.5) & 20@1 ( 6.7) &  5@7 ( 6.5) & 22@7 ( 6.1) &  5@1 ( 4.6) \\ 
  6 & Sun (25.7) & 20@1 (17.6) & 23@1 (14.2) & Sat (12.6) &  6@7 ( 7.8) & Mon ( 5.8) & 22@7 ( 4.5) \\ 
  7 & Sun (26.0) & 20@1 (17.8) & Sat (14.6) & Mon ( 7.7) & 23@1 ( 7.0) & 18@1 ( 5.0) &  7@7 ( 4.8) \\ 
  8 & Sun (23.6) & 20@1 (15.0) & Sat (13.5) & 23@1 ( 8.8) & Mon ( 8.1) &  7@1 ( 5.4) & 18@1 ( 5.1) \\ 
  9 & Sun (23.8) & 20@1 (15.4) & Sat (13.5) & Mon ( 9.8) & 23@1 ( 8.0) & 11@1 ( 5.1) & 18@1 ( 5.0) \\ 
  10 & Sun (19.5) & 11@1 (13.9) & Sat (12.4) & Mon (10.9) & 20@1 ( 8.6) & 23@1 ( 6.7) & 22@1 ( 6.5) \\ 
  11 & Sun (14.2) & Sat (11.5) & Mon (11.3) & 11@1 (10.6) & 22@1 ( 9.9) & 12@1 ( 6.9) & 23@1 ( 6.0) \\ 
  12 & 12@1 (12.1) & Mon (11.9) & Sun (11.4) & Sat ( 9.3) & 23@1 ( 8.9) & 13@1 ( 8.8) & 22@1 ( 7.6) \\ 
  13 & 13@1 (15.0) & Sun (13.9) & Mon (11.9) & 23@1 (10.1) & Sat (10.0) & 17@1 ( 6.2) & 14@7 ( 5.9) \\ 
  14 & Sun (15.3) & 14@1 (15.0) & Mon (11.2) & 23@1 (10.5) & Sat ( 9.9) & 17@1 ( 7.8) & 14@7 ( 6.2) \\ 
  15 & Sun (17.3) & 15@1 (11.2) & Mon (10.9) & 17@1 (10.6) & Sat ( 9.4) & 23@1 ( 8.7) & 14@7 ( 5.9) \\ 
  16 & Sun (16.9) & 17@1 (15.2) & Mon ( 8.6) & Sat ( 8.3) & 23@1 ( 7.7) & 16@7 ( 6.8) & 17@6 ( 6.2) \\ 
  17 & 17@1 (32.2) & Sun (13.0) & Mon (11.1) & 17@6 (10.3) & 17@7 ( 9.1) & Sat ( 5.9) & 22@7 ( 3.8) \\ 
  18 & 18@1 (42.1) & Mon (14.2) & Sun ( 9.6) & 18@6 ( 9.4) & 18@7 ( 8.8) & Sat ( 4.5) & 17@6 ( 2.8) \\ 
  19 & 19@1 (35.2) & Mon (11.3) & Sun ( 8.8) & 20@1 ( 8.0) & 19@6 ( 7.4) & Sat ( 6.7) & 18@6 ( 4.4) \\ 
  20 & 20@1 (42.6) & 20@6 (11.5) & Mon (10.0) & Sat ( 9.1) & Sun ( 6.5) & 23@2 ( 2.7) & 20@4 ( 2.3) \\ 
  21 & 21@1 (38.5) & 21@6 (12.0) & Mon ( 9.7) & Sat ( 7.7) & 22@1 ( 5.8) & 23@2 ( 4.5) & 21@35 ( 3.3) \\ 
  22 & 22@1 (44.8) & 22@6 ( 9.2) & 23@2 ( 6.6) & Sat ( 6.0) & 23@1 ( 5.7) & Mon ( 5.5) &  0@5 ( 5.4) \\ 
  23 & 23@1 (36.8) & 22@1 ( 9.4) & 23@2 ( 7.5) & 23@6 ( 7.1) & 23@4 ( 5.4) & 23@28 ( 5.2) & Mon ( 4.8) \\ 
   \hline
\end{tabular}
\caption{Most relevant coefficients for each hour.
H@L represents the estimated parameter for $Y_{d-L,H}$ for a model on $Y_{d,h}$. Sun, Mon, $\ldots$, Sat represent the weekday dummies.
The corresponding $\hat{\iota}_{h,i}$ value is given in parenthesis.}
\end{table}

\clearpage
\subsection*{EPEX Germany \& Austria}

\begin{table}[ht]
\centering
\begin{tabular}{rrrrrrrr}
  \hline
 & Importance: 1 & Importance: 2 & Importance: 3 & Importance: 4 & Importance: 5 & Importance: 6 & Importance: 7 \\ 
  \hline
0 & 23@1 (70.6) & 19@1 ( 5.9) &  1@1 ( 5.7) & 18@1 ( 4.1) & 21@1 ( 3.6) &  4@1 ( 3.1) & Sat ( 2.2) \\ 
  1 & 23@1 (67.3) & 19@1 ( 8.6) &  3@1 ( 7.1) &  4@1 ( 5.9) & Sun ( 4.7) &  5@1 ( 2.6) &  6@1 ( 2.0) \\ 
  2 & 23@1 (56.3) &  4@1 ( 9.4) & 19@1 ( 8.9) &  3@1 ( 7.9) & Sun ( 6.6) & 22@7 ( 5.1) &  2@25 ( 1.6) \\ 
  3 & 23@1 (45.0) &  4@1 (14.3) &  3@1 (12.2) & 19@1 ( 8.5) & Sun ( 7.3) & 22@7 ( 4.3) &  3@6 ( 2.5) \\ 
  4 & 23@1 (44.8) &  4@1 (13.8) & 19@1 (10.3) & Sun (10.2) &  3@1 ( 7.4) & 22@7 ( 2.5) &  4@25 ( 2.4) \\ 
  5 & 23@1 (34.1) & Sun (15.5) &  5@1 ( 9.7) &  4@1 ( 8.2) & Sat ( 7.4) & 19@1 ( 7.2) & 22@7 ( 3.8) \\ 
  6 & Sun (20.4) & 23@1 (20.3) & Sat (12.8) &  6@1 ( 9.4) & Mon ( 8.7) & 20@1 ( 6.0) & 23@2 ( 4.1) \\ 
  7 & Sun (23.9) & Sat (15.9) & 23@1 (12.1) & Mon ( 7.8) & 20@1 ( 6.3) &  6@1 ( 5.3) & 19@1 ( 4.5) \\ 
  8 & Sun (28.3) & Sat (15.9) & 20@1 ( 8.5) & 23@1 ( 7.9) & 18@1 ( 7.0) & Mon ( 5.4) & 21@1 ( 5.2) \\ 
  9 & Sun (25.9) & 20@1 (13.3) & Sat (13.3) & 23@1 ( 7.0) & 22@1 ( 6.3) & Mon ( 6.1) & 18@1 ( 5.5) \\ 
  10 & Sun (21.8) & Sat (11.9) & 22@1 ( 9.6) & 20@1 ( 9.0) & Mon ( 6.6) & 13@1 ( 6.1) & 23@1 ( 5.6) \\ 
  11 & Sun (17.2) & 22@1 (12.3) & Sat (11.7) & 13@1 ( 9.2) & Mon ( 8.0) & 23@1 ( 5.8) & 11@7 ( 4.1) \\ 
  12 & Sun (15.6) & 13@1 (12.0) & Sat (10.7) & 22@1 (10.2) & Mon ( 8.1) & 23@1 ( 8.0) & 17@1 ( 4.9) \\ 
  13 & Sun (18.6) & Sat (11.2) & 13@1 (11.1) & 22@1 (10.0) & Mon ( 7.9) & 23@1 ( 7.3) & 17@1 ( 6.8) \\ 
  14 & Sun (18.7) & Sat (10.1) & Mon ( 8.2) & 23@1 ( 8.2) & 16@1 ( 8.1) & 22@1 ( 7.6) & 13@1 ( 4.2) \\ 
  15 & Sun (19.8) & 16@1 (12.5) & Sat ( 8.9) & Mon ( 8.0) & 23@1 ( 7.1) & 17@6 ( 6.0) & 17@1 ( 4.9) \\ 
  16 & Sun (19.3) & 17@1 (12.5) & Sat ( 8.8) & 16@1 ( 8.5) & Mon ( 8.4) & 23@1 ( 7.1) & 17@6 ( 6.7) \\ 
  17 & 17@1 (26.4) & Sun (16.1) & Mon (10.1) & 17@6 (10.0) & Sat ( 6.7) & 17@7 ( 6.3) & 18@1 ( 6.0) \\ 
  18 & 18@1 (33.8) & Sun (14.1) & Mon ( 9.9) & 18@6 ( 9.4) & Sat ( 6.0) & 18@2 ( 5.9) & 18@5 ( 4.6) \\ 
  19 & 19@1 (34.0) & Mon ( 9.4) & 19@6 ( 8.5) & Sun ( 8.2) & 19@2 ( 7.6) &  8@6 ( 5.0) & 19@7 ( 4.7) \\ 
  20 & 20@1 (23.9) & 20@2 (10.4) & Sat ( 9.5) & Mon ( 7.8) & 21@1 ( 7.5) & 20@6 ( 7.5) & Sun ( 7.0) \\ 
  21 & 21@1 (22.8) & Sat (10.9) & 21@2 ( 9.9) & Mon ( 6.2) & 21@28 ( 4.7) & Sun ( 4.6) & 23@1 ( 4.6) \\ 
  22 & 22@1 (30.6) & 22@2 ( 9.5) & Sat ( 7.7) & 23@3 ( 6.5) & 22@28 ( 6.1) & 22@5 ( 5.1) & 22@35 ( 5.0) \\ 
  23 & 23@1 (24.8) & 22@1 (17.5) & 23@3 (10.7) & 22@2 ( 9.6) & Sat ( 5.7) &  0@5 ( 5.0) & 12@7 ( 4.4) \\ 
   \hline
\end{tabular}
\caption{Most relevant coefficients for each hour.
H@L represents the estimated parameter for $Y_{d-L,H}$ for a model on $Y_{d,h}$. Sun, Mon, $\ldots$, Sat represent the weekday dummies.
The corresponding $\hat{\iota}_{h,i}$ value is given in parenthesis.}
\end{table}

\clearpage
\subsection*{EPEX Switzerland}

\begin{table}[ht]
\centering
\begin{tabular}{rrrrrrrr}
  \hline
 & Importance: 1 & Importance: 2 & Importance: 3 & Importance: 4 & Importance: 5 & Importance: 6 & Importance: 7 \\ 
  \hline
0 & 23@1 (50.8) & 22@1 (11.4) &  1@1 ( 6.1) & Sat ( 5.6) &  6@1 ( 5.2) &  5@1 ( 4.2) & 18@1 ( 3.7) \\ 
  1 & 23@1 (47.6) &  5@1 (13.6) &  2@1 (12.0) & 18@1 ( 6.5) & Sun ( 5.6) & Sat ( 4.4) &  2@3 ( 3.9) \\ 
  2 & 23@1 (32.6) &  2@1 (12.7) &  4@1 (10.6) & Sun ( 8.4) & 18@1 ( 8.1) &  5@1 ( 5.4) &  3@7 ( 3.5) \\ 
  3 & 23@1 (30.7) &  4@1 (23.9) & Sun ( 9.6) & 18@1 ( 6.0) &  4@7 ( 4.3) &  4@4 ( 4.2) &  5@1 ( 3.7) \\ 
  4 & 23@1 (22.6) &  4@1 (20.8) & Sun (12.0) &  5@1 ( 6.7) & 22@1 ( 6.6) &  4@4 ( 5.1) & 19@5 ( 4.8) \\ 
  5 & 23@1 (25.9) & Sun (19.2) &  5@1 (13.4) & Sat ( 6.9) &  5@7 ( 5.6) & 18@1 ( 4.9) &  4@1 ( 4.0) \\ 
  6 & Sun (27.4) & 23@1 (20.8) & Sat (12.0) & 20@1 ( 9.5) & 22@1 ( 4.8) &  6@6 ( 4.8) &  5@1 ( 4.7) \\ 
  7 & Sun (28.1) & Sat (12.4) & 22@1 (12.0) & 20@1 ( 9.9) & 23@1 ( 8.5) & 18@1 ( 5.4) & Mon ( 4.5) \\ 
  8 & Sun (22.8) & 22@1 (11.3) & 19@1 (10.3) & Sat ( 8.8) & Mon ( 6.7) & 10@1 ( 6.5) &  8@6 ( 5.6) \\ 
  9 & Sun (16.9) & 10@1 (10.7) & 22@1 ( 9.8) & 19@1 ( 8.2) & Mon ( 7.7) & Sat ( 7.0) &  9@2 ( 6.5) \\ 
  10 & Sun (15.8) & 22@1 (14.2) & Mon (10.1) & 11@1 ( 9.7) & 19@1 ( 9.5) & Sat ( 7.8) & 10@1 ( 6.2) \\ 
  11 & Sun ( 8.3) & 22@1 ( 6.9) & 10@2 ( 6.3) & 14@1 ( 5.0) & Sat ( 4.7) & Mon ( 4.4) & 13@7 ( 4.3) \\ 
  12 & Sun ( 8.2) & 10@2 ( 7.6) & 22@1 ( 7.2) & 13@7 ( 7.1) & 23@1 ( 5.1) & 13@1 ( 4.8) & 12@1 ( 4.5) \\ 
  13 & 12@1 (16.8) & Sun (15.8) & 22@1 (11.5) & 13@7 ( 9.0) & 23@1 ( 9.0) & Sat ( 8.8) & Mon ( 8.4) \\ 
  14 & Sun (17.0) & 22@1 (14.5) & Sat (10.2) & Mon ( 9.9) & 13@7 ( 8.6) & 12@1 ( 8.3) & 14@1 ( 7.3) \\ 
  15 & Sun (12.7) & 22@1 (10.6) & 10@2 ( 6.9) & Mon ( 6.7) & Sat ( 6.2) & 23@1 ( 5.3) & 13@7 ( 5.3) \\ 
  16 & Sun (14.5) & 22@1 (13.3) & Mon ( 9.1) & 17@1 ( 8.9) & Sat ( 6.8) & 14@1 ( 6.1) & 18@1 ( 5.1) \\ 
  17 & 17@1 (15.0) & 18@1 (10.3) & Sun ( 8.4) & 22@1 ( 6.5) & 10@2 ( 5.6) & 17@6 ( 4.7) & Mon ( 4.7) \\ 
  18 & 18@1 (40.6) & 18@6 ( 9.9) & Mon ( 8.9) & Sun ( 8.8) & Sat ( 5.8) & 23@1 ( 4.8) & 22@1 ( 4.4) \\ 
  19 & 19@1 (28.9) & 10@2 (11.2) & 14@2 (11.0) & Sun ( 5.2) & 22@1 ( 4.9) & 23@1 ( 4.7) & 18@6 ( 4.2) \\ 
  20 & 20@1 (26.7) & 23@1 ( 7.7) & Mon ( 6.8) & 22@1 ( 6.5) &  9@2 ( 5.9) & 15@2 ( 5.5) & Sun ( 5.4) \\ 
  21 & 22@1 (18.1) & 23@1 (17.7) & 21@1 (15.9) & Mon ( 9.8) & 20@1 ( 9.7) & 13@7 ( 4.6) &  6@6 ( 4.1) \\ 
  22 & 22@1 (29.6) & 23@1 (22.1) & Mon ( 9.0) & 13@6 ( 6.1) & 13@7 ( 4.7) & Sat ( 4.0) & 20@1 ( 3.8) \\ 
  23 & 23@1 (39.6) & 22@1 (20.8) & Mon ( 9.8) & 13@7 ( 6.3) &  6@6 ( 4.6) & 18@1 ( 4.6) &  0@2 ( 4.3) \\ 
   \hline
\end{tabular}
\caption{Most relevant coefficients for each hour.
H@L represents the estimated parameter for $Y_{d-L,H}$ for a model on $Y_{d,h}$. Sun, Mon, $\ldots$, Sat represent the weekday dummies.
The corresponding $\hat{\iota}_{h,i}$ value is given in parenthesis.}
\end{table}

\clearpage
\subsection*{BELPEX Belgium}

\begin{table}[ht]
\centering
\begin{tabular}{rrrrrrrr}
  \hline
 & Importance: 1 & Importance: 2 & Importance: 3 & Importance: 4 & Importance: 5 & Importance: 6 & Importance: 7 \\ 
  \hline
0 & 23@1 (29.3) &  0@1 (10.4) &  2@1 ( 8.8) &  7@7 ( 6.9) &  7@8 ( 5.1) &  1@1 ( 4.3) &  0@6 ( 4.3) \\ 
  1 &  2@1 (19.9) &  0@1 (12.2) & 23@1 ( 8.8) &  9@1 ( 6.4) &  1@1 ( 5.6) & 21@1 ( 5.4) & 13@1 ( 5.3) \\ 
  2 &  4@1 (23.2) &  2@1 (22.9) &  0@1 ( 7.6) &  1@1 ( 7.1) & 12@6 ( 6.2) & 10@6 ( 5.9) & 17@8 ( 3.2) \\ 
  3 &  4@1 (22.8) &  2@1 (11.9) &  3@1 ( 8.6) & 10@6 ( 5.4) &  0@1 ( 5.4) &  8@7 ( 4.2) &  1@1 ( 3.6) \\ 
  4 &  4@1 (33.7) &  5@1 (10.7) & 10@6 (10.4) &  3@1 ( 8.5) &  0@1 ( 6.1) &  4@7 ( 5.0) &  1@1 ( 4.2) \\ 
  5 &  5@1 (38.7) &  4@1 (12.2) &  6@6 (11.5) &  5@7 ( 8.6) & 15@8 ( 5.1) & 10@6 ( 5.1) &  6@2 ( 4.1) \\ 
  6 &  5@1 (40.0) &  6@7 (13.1) &  6@1 (11.2) &  7@6 ( 7.0) &  6@14 ( 5.9) &  6@6 ( 5.7) &  7@7 ( 5.5) \\ 
  7 &  5@1 (29.9) &  7@7 (16.1) &  7@1 (12.9) &  7@14 ( 9.9) &  7@6 ( 7.1) &  7@21 ( 4.6) &  8@7 ( 3.8) \\ 
  8 &  7@1 (12.0) & 10@7 (11.5) &  5@1 (10.9) &  8@7 (10.6) &  8@14 ( 7.7) &  3@1 ( 6.5) &  2@1 ( 6.2) \\ 
  9 & 10@7 (59.7) &  4@1 (21.9) &  5@1 (18.3) &  0@1 ( 0.0) &  0@2 ( 0.0) &  0@3 ( 0.0) &  0@4 ( 0.0) \\ 
  10 & 10@7 (17.1) &  2@1 (13.5) & 10@14 ( 7.4) & 10@35 ( 6.2) &  4@1 ( 6.0) &  6@1 ( 5.5) & 10@1 ( 5.1) \\ 
  11 &  2@1 (11.6) & 10@7 ( 9.1) & 11@14 ( 7.8) & 12@7 ( 6.6) & 11@35 ( 6.0) & 11@21 ( 4.4) &  4@1 ( 4.1) \\ 
  12 &  2@1 (15.0) & 12@1 ( 8.9) & 12@14 ( 8.4) & 13@7 ( 7.4) & 12@35 ( 6.3) & 10@7 ( 4.8) & 12@7 ( 3.7) \\ 
  13 &  2@1 (20.3) & 13@7 (13.4) & 13@14 ( 8.6) &  3@1 ( 7.5) & 13@1 ( 7.0) & 16@7 ( 6.8) & 13@35 ( 6.7) \\ 
  14 &  3@1 (11.4) &  2@1 (10.6) & 16@7 ( 9.6) & 13@1 ( 8.1) & 14@14 ( 7.8) & 14@21 ( 6.4) & 14@35 ( 5.3) \\ 
  15 &  2@1 (11.3) & 15@14 (10.1) &  3@1 (10.0) & 16@7 ( 6.7) & 15@21 ( 5.9) & 16@1 ( 5.6) & 15@35 ( 5.2) \\ 
  16 &  2@1 (13.4) & 16@14 (10.6) & 17@1 ( 9.0) &  3@1 ( 8.1) & 16@7 ( 7.5) & 17@7 ( 4.9) & 16@28 ( 4.7) \\ 
  17 & 17@1 (14.2) & 17@7 (12.4) &  2@1 (10.4) & 17@14 ( 7.8) & 17@6 ( 5.7) &  4@1 ( 4.3) &  7@1 ( 3.8) \\ 
  18 & 18@1 (26.2) &  2@1 (11.0) & 18@7 ( 7.1) & 18@14 ( 6.5) & 17@7 ( 6.0) &  7@1 ( 5.6) & 21@1 ( 4.4) \\ 
  19 & 19@1 (14.1) &  2@1 (13.4) & 19@7 ( 8.2) & 12@7 ( 7.1) &  7@1 ( 6.7) & 19@14 ( 4.9) & 19@2 ( 4.5) \\ 
  20 & 20@1 (19.0) &  2@1 (10.4) & 12@7 ( 5.6) & 20@2 ( 5.3) & 20@6 ( 5.3) &  3@1 ( 4.7) &  7@1 ( 4.3) \\ 
  21 &  2@1 (15.9) & 21@7 (15.3) & 21@1 (13.0) & 20@1 ( 7.4) & 21@6 ( 6.2) &  4@1 ( 5.7) & 21@14 ( 4.8) \\ 
  22 &  2@1 (18.6) & 22@6 (11.4) & 22@1 ( 9.8) &  4@1 ( 8.9) & 21@7 ( 7.9) & 22@2 ( 4.8) & 10@1 ( 4.6) \\ 
  23 &  2@1 (26.9) &  4@1 (17.3) & 23@1 ( 8.5) & 23@2 ( 7.0) & 21@7 ( 5.5) & 12@1 ( 5.4) & 12@7 ( 5.2) \\ 
   \hline
\end{tabular}
\caption{Most relevant coefficients for each hour.
H@L represents the estimated parameter for $Y_{d-L,H}$ for a model on $Y_{d,h}$. Sun, Mon, $\ldots$, Sat represent the weekday dummies.
The corresponding $\hat{\iota}_{h,i}$ value is given in parenthesis.}
\end{table}

\clearpage
\subsection*{APX Netherlands}

\begin{table}[ht]
\centering
\begin{tabular}{rrrrrrrr}
  \hline
 & Importance: 1 & Importance: 2 & Importance: 3 & Importance: 4 & Importance: 5 & Importance: 6 & Importance: 7 \\ 
  \hline
0 & 23@1 (52.2) &  2@1 (14.7) & Sat ( 5.8) &  9@1 ( 5.3) & 22@1 ( 5.0) & 21@1 ( 2.7) &  1@8 ( 2.1) \\ 
  1 & 23@1 (22.9) &  2@1 (12.0) &  5@1 ( 7.2) & 16@1 ( 5.3) &  1@6 ( 5.3) & 19@1 ( 5.0) & 22@1 ( 4.7) \\ 
  2 & 23@1 (22.5) &  2@1 (20.8) &  5@1 (11.1) & 19@1 ( 7.8) &  3@1 ( 6.3) & 16@1 ( 5.6) & 22@1 ( 5.3) \\ 
  3 &  3@1 (20.3) & 23@1 (16.5) &  5@1 ( 9.1) &  4@6 ( 5.6) & Sun ( 4.6) &  4@1 ( 4.5) & 21@1 ( 4.2) \\ 
  4 &  4@1 (12.3) &  3@1 (12.3) & 23@1 (12.2) &  5@1 (11.5) & Sun ( 9.3) &  5@7 ( 5.6) & 21@1 ( 5.6) \\ 
  5 &  5@1 (23.5) & Sun (14.9) & 23@1 (10.5) &  6@7 ( 9.9) & 21@1 ( 8.1) & 19@1 ( 6.9) &  5@6 ( 6.7) \\ 
  6 & Sun (23.2) &  6@7 (12.5) & Sat (12.4) & 23@1 ( 7.0) & 19@1 ( 6.9) &  5@1 ( 6.6) &  6@1 ( 5.9) \\ 
  7 & Sun (27.5) & Sat (17.3) & 20@1 ( 6.9) & 19@1 ( 6.6) &  7@1 ( 6.1) & Mon ( 5.5) &  7@7 ( 3.8) \\ 
  8 & Sun (27.2) & Sat (17.0) &  7@1 ( 6.9) &  7@7 ( 6.0) & Mon ( 5.7) & 19@1 ( 5.7) & 20@1 ( 5.7) \\ 
  9 & Sun (26.3) & Sat (11.6) & 23@1 ( 5.2) & 20@1 ( 4.3) & 18@1 ( 4.1) &  7@7 ( 4.0) & 21@1 ( 3.7) \\ 
  10 & Sun (25.9) &  9@7 (10.8) & 12@1 (10.5) & 18@1 ( 8.0) & Sat ( 6.9) & 22@1 ( 5.4) & 10@6 ( 4.8) \\ 
  11 & Sun (23.4) & 12@1 (11.8) & Sat ( 9.6) & 22@1 ( 7.0) & Mon ( 5.4) & 16@1 ( 5.2) & 10@7 ( 4.1) \\ 
  12 & Sun (14.3) & 12@1 (11.5) & 12@6 ( 7.6) & Sat ( 6.1) & Mon ( 5.8) & 22@1 ( 4.2) & 12@14 ( 4.2) \\ 
  13 & Sun (16.2) & 12@1 (11.2) & Mon (10.6) & Sat (10.0) & 16@1 ( 9.8) & 12@6 ( 6.1) & 22@1 ( 5.6) \\ 
  14 & Sun (17.8) & Sat (12.3) & 16@1 (10.1) & 12@1 ( 8.9) & Mon ( 8.5) & 15@7 ( 7.3) & 22@1 ( 6.2) \\ 
  15 & Sun (18.7) & 16@1 (14.6) & Sat (11.3) & 15@7 ( 9.0) & Mon ( 6.5) & 22@1 ( 6.5) & 15@6 ( 5.7) \\ 
  16 & Sun (16.7) & 16@1 (13.7) & Sat (10.3) & Mon ( 7.1) & 16@7 ( 6.4) & 17@2 ( 6.1) & 17@1 ( 5.5) \\ 
  17 & 17@1 (27.3) & Sun (11.5) & 17@7 (10.1) & Sat ( 9.2) & 17@6 ( 7.5) & 17@2 ( 5.6) & Mon ( 5.4) \\ 
  18 & 18@1 (34.5) & 18@2 ( 8.2) & Mon ( 7.9) & 18@7 ( 6.1) & 18@3 ( 6.0) & 17@6 ( 5.3) & 18@6 ( 4.4) \\ 
  19 & 19@1 (44.1) & 19@2 (15.4) & 19@6 ( 8.6) & 19@3 ( 6.6) & Mon ( 5.2) & 19@14 ( 5.1) &  7@6 ( 3.9) \\ 
  20 & 20@1 (33.8) & 20@2 (16.3) & 20@6 ( 8.2) & 20@3 ( 7.8) & 20@13 ( 5.2) & 15@6 ( 5.1) &  6@6 ( 4.3) \\ 
  21 & 21@1 (25.4) & 21@2 (15.0) & 21@4 ( 6.0) & 21@13 ( 5.9) & Sat ( 5.3) & 21@6 ( 5.1) &  0@1 ( 4.7) \\ 
  22 & 22@1 (21.3) & 22@2 (13.2) & 23@1 (12.7) & 22@13 ( 5.4) & 22@3 ( 4.7) & Sat ( 4.4) & 23@8 ( 4.2) \\ 
  23 & 23@1 (27.2) & 22@1 (15.0) & 23@2 ( 8.8) & 22@2 ( 6.1) & 23@3 ( 5.4) & 23@8 ( 5.4) & 23@28 ( 5.0) \\ 
   \hline
\end{tabular}
\caption{Most relevant coefficients for each hour.
H@L represents the estimated parameter for $Y_{d-L,H}$ for a model on $Y_{d,h}$. Sun, Mon, $\ldots$, Sat represent the weekday dummies.
The corresponding $\hat{\iota}_{h,i}$ value is given in parenthesis.}
\end{table}

\clearpage
\subsection*{Nordpool Denmark West}

\begin{table}[ht]
\centering
\begin{tabular}{rrrrrrrr}
  \hline
 & Importance: 1 & Importance: 2 & Importance: 3 & Importance: 4 & Importance: 5 & Importance: 6 & Importance: 7 \\ 
  \hline
0 & 23@1 (70.2) &  4@1 ( 9.4) & 17@1 ( 5.4) & 23@2 ( 4.4) &  4@2 ( 4.1) & 15@1 ( 2.6) &  3@4 ( 2.4) \\ 
  1 & 23@1 (61.8) &  4@1 (15.3) & 17@1 ( 6.6) &  4@2 ( 4.3) &  3@4 ( 4.2) & 16@6 ( 2.7) &  3@1 ( 1.8) \\ 
  2 & 23@1 (62.0) &  4@1 (19.1) &  3@1 ( 5.3) & 17@1 ( 4.4) &  3@4 ( 2.7) & Sun ( 2.4) &  4@6 ( 2.0) \\ 
  3 & 23@1 (48.8) &  4@1 (21.0) &  4@6 (11.1) & 17@1 ( 3.9) &  3@1 ( 3.7) &  5@1 ( 3.7) & 16@6 ( 3.2) \\ 
  4 & 23@1 (48.3) &  4@1 (23.8) & Sun ( 6.7) &  5@1 ( 6.2) &  4@6 ( 4.7) &  3@1 ( 4.0) & 17@1 ( 2.5) \\ 
  5 & 23@1 (37.2) &  5@1 (16.1) & Sun (12.8) &  4@1 ( 9.9) & Sat ( 6.9) &  6@7 ( 5.2) & 18@1 ( 3.7) \\ 
  6 & Sun (16.0) & 23@1 (14.6) & 21@1 (13.9) &  6@1 (12.0) & Sat (11.1) & Mon ( 7.2) & 19@1 ( 6.4) \\ 
  7 & 13@1 (40.0) & 16@1 (11.2) & 18@1 (10.5) & Sat (10.4) & Sun ( 8.9) & Mon ( 8.4) & Fri ( 6.3) \\ 
  8 & 13@1 (45.7) & Sat (13.0) & Sun (12.5) & 18@1 (11.1) & Mon (10.2) & Fri ( 7.4) &  0@1 ( 0.0) \\ 
  9 & 13@1 (41.2) & 16@1 (25.2) & 17@1 ( 8.2) & Mon ( 6.7) & Fri ( 5.1) & 18@1 ( 4.7) & Sat ( 4.0) \\ 
  10 & 16@1 (33.1) & 13@1 (28.7) & 17@1 (13.0) & 15@1 (10.4) & Mon ( 5.3) & Fri ( 3.8) & Sat ( 2.1) \\ 
  11 & 16@1 (32.6) & 13@1 (29.8) & 17@1 (13.1) & 15@1 ( 9.6) & Mon ( 5.7) & Fri ( 3.5) & Sat ( 2.4) \\ 
  12 & 22@1 (13.0) & 21@1 (10.0) & Sun ( 9.7) & 13@1 ( 8.8) & Sat ( 8.1) & 18@1 ( 7.8) & Mon ( 6.0) \\ 
  13 & Sun (10.9) & 22@1 (10.4) & 21@1 (10.2) & Sat ( 8.8) & 17@1 ( 7.0) & 13@7 ( 6.3) & Mon ( 6.0) \\ 
  14 & Sun (12.3) & 21@1 (10.4) & Sat ( 9.1) & 17@1 ( 9.0) & 23@1 ( 7.7) & 14@28 ( 7.0) & 22@1 ( 6.5) \\ 
  15 & Sun (13.1) & 21@1 (10.1) & 17@1 ( 9.4) & Sat ( 8.9) & 23@1 ( 8.6) & 15@28 ( 6.7) & 16@7 ( 6.6) \\ 
  16 & 17@1 (16.9) & 16@7 (11.3) & 16@1 (10.8) & 23@1 (10.4) & Sun ( 9.6) & 17@6 ( 6.8) & Mon ( 6.7) \\ 
  17 & 17@1 (22.4) & 18@1 (12.1) & 16@7 ( 9.1) & 17@6 ( 8.8) & Mon ( 8.7) & Sun ( 8.2) & Sat ( 5.9) \\ 
  18 & 18@1 (40.1) & Mon ( 8.0) & 18@2 ( 7.5) & 19@1 ( 7.3) & 17@6 ( 6.0) & 18@6 ( 4.9) & Sun ( 4.9) \\ 
  19 & 19@1 (47.7) & 12@7 (10.0) & 19@2 ( 8.0) & 20@5 ( 5.6) & 18@6 ( 5.4) & 14@6 ( 4.0) & 20@1 ( 3.5) \\ 
  20 & 21@1 (22.0) & 19@1 (10.8) & 20@1 ( 7.7) & 20@6 ( 7.6) & 19@2 ( 6.9) & 12@7 ( 6.8) & 20@5 ( 5.4) \\ 
  21 & 21@1 (34.8) & 12@7 ( 7.0) & 23@1 ( 6.7) & 22@1 ( 6.6) & 21@5 ( 6.0) &  6@6 ( 5.7) & 21@8 ( 4.7) \\ 
  22 & 22@1 (31.6) & 22@5 (11.5) & 23@1 ( 8.2) & 22@2 ( 8.1) & 21@1 ( 5.8) & 12@7 ( 5.3) & 22@4 ( 4.5) \\ 
  23 & 23@1 (43.3) & 22@2 ( 8.8) & 23@3 ( 4.4) & 21@1 ( 4.4) & 23@6 ( 3.5) &  5@4 ( 3.0) & 22@5 ( 3.0) \\ 
   \hline
\end{tabular}
\caption{Most relevant coefficients for each hour.
H@L represents the estimated parameter for $Y_{d-L,H}$ for a model on $Y_{d,h}$. Sun, Mon, $\ldots$, Sat represent the weekday dummies.
The corresponding $\hat{\iota}_{h,i}$ value is given in parenthesis.}
\end{table}

\clearpage
\subsection*{Nordpool Denmark East}

\begin{table}[ht]
\centering
\begin{tabular}{rrrrrrrr}
  \hline
 & Importance: 1 & Importance: 2 & Importance: 3 & Importance: 4 & Importance: 5 & Importance: 6 & Importance: 7 \\ 
  \hline
0 & 23@1 (68.12) &  4@1 ( 8.53) &  4@2 ( 8.08) & 19@1 ( 4.19) &  0@1 ( 3.25) & 23@2 ( 2.11) &  4@5 ( 1.73) \\ 
  1 & 23@1 (64.05) &  4@1 (14.74) &  4@2 ( 7.74) &  3@1 ( 7.09) &  4@5 ( 3.61) & 19@1 ( 2.44) &  4@6 ( 0.33) \\ 
  2 & 23@1 (55.15) &  4@1 (17.31) &  3@1 (12.03) &  4@2 ( 5.01) &  4@5 ( 3.51) &  4@6 ( 3.10) & Sun ( 1.65) \\ 
  3 & 23@1 (43.57) &  4@1 (22.45) &  3@1 (12.87) &  4@6 (11.09) &  4@2 ( 2.93) & Sun ( 2.82) &  4@5 ( 2.73) \\ 
  4 & 23@1 (42.37) &  4@1 (28.57) &  3@1 ( 8.57) &  4@6 ( 6.15) & Sun ( 3.94) &  4@2 ( 3.01) &  4@5 ( 2.87) \\ 
  5 & 23@1 (33.17) &  4@1 (15.31) & Sun (11.31) &  5@1 (10.39) &  5@6 ( 5.68) & Sat ( 5.64) & 19@1 ( 5.58) \\ 
  6 & 23@1 (19.80) & Sun (13.82) &  6@1 (10.20) & 19@1 ( 9.52) & Sat ( 9.12) & 22@1 ( 6.43) & 21@1 ( 6.40) \\ 
  7 & 19@1 (23.52) & 23@1 (16.48) & Sun (12.43) & 21@1 (11.61) & Sat ( 9.63) & 16@1 ( 8.89) & Mon ( 7.23) \\ 
  8 & 19@1 (23.30) & 21@1 (17.47) & 23@1 (14.63) & Sun (10.59) & Mon ( 7.90) & Sat ( 7.64) & 16@1 ( 7.44) \\ 
  9 & 19@1 (24.02) & 21@1 (23.68) & 23@1 (13.26) & Mon (10.61) & Sun ( 8.49) &  8@7 ( 5.58) & Sat ( 5.51) \\ 
  10 & 21@1 (33.92) & 19@1 (18.90) & 23@1 (11.54) & Mon (11.25) &  8@7 ( 6.11) & 16@1 ( 5.55) & 22@1 ( 5.41) \\ 
  11 & 21@1 (22.86) & 16@1 (12.18) & 19@1 (10.54) & 23@1 ( 9.81) & Mon ( 8.30) & Sun ( 5.61) &  8@1 ( 4.87) \\ 
  12 & 21@1 (20.38) & 16@1 (13.66) & 19@1 ( 8.47) & Sun ( 6.78) & Mon ( 6.72) & Sat ( 6.44) & 23@1 ( 6.30) \\ 
  13 & 21@1 (23.12) & 14@1 ( 9.47) & Mon ( 7.04) & Sat ( 6.94) & 19@1 ( 6.93) & Sun ( 6.88) & 16@1 ( 6.79) \\ 
  14 & 21@1 (21.38) & 16@1 (14.27) & 23@1 ( 9.69) & Sun ( 7.40) & Sat ( 6.98) & 17@6 ( 6.70) &  8@1 ( 5.72) \\ 
  15 & 21@1 (17.41) & 16@1 (15.43) & 23@1 ( 8.86) &  8@1 ( 8.04) & Sun ( 6.61) & Mon ( 5.94) & Sat ( 5.70) \\ 
  16 & 16@1 (27.37) & 23@1 ( 8.70) & 21@1 ( 7.92) & Mon ( 6.52) & 17@6 ( 6.11) & 17@7 ( 5.18) & 16@6 ( 4.64) \\ 
  17 & 16@6 (34.01) & 17@1 (24.75) & 16@1 (24.74) & 16@7 (13.44) & 23@1 ( 2.41) & Sun ( 0.64) &  0@1 ( 0.00) \\ 
  18 & 16@6 (40.52) & 19@1 (30.44) & 16@1 (19.99) & 16@7 ( 9.05) &  0@1 ( 0.00) &  0@2 ( 0.00) &  0@3 ( 0.00) \\ 
  19 & 19@1 (39.15) & 22@1 ( 9.71) & 23@1 ( 6.63) & 19@2 ( 6.60) & Mon ( 6.52) & 17@6 ( 5.66) & 16@7 ( 3.73) \\ 
  20 & 20@1 (24.92) & 22@1 (14.63) & 21@5 ( 6.47) & 19@1 ( 6.00) & 23@1 ( 5.52) & Sat ( 5.08) & 17@6 ( 4.95) \\ 
  21 & 21@1 (22.91) & 22@1 (15.72) & 23@1 (11.11) & 21@5 ( 9.02) & 17@7 ( 5.04) & Sat ( 4.53) & Fri ( 3.43) \\ 
  22 & 22@1 (37.78) & 23@1 (17.21) & 22@2 (11.15) & 22@5 (10.50) & 14@6 ( 3.15) & 22@6 ( 2.97) & 22@7 ( 2.77) \\ 
  23 & 23@1 (52.37) & 22@2 ( 6.60) &  0@1 ( 5.79) & 22@5 ( 5.78) & 22@1 ( 5.64) &  0@5 ( 3.39) &  3@3 ( 3.17) \\ 
   \hline
\end{tabular}
\caption{Most relevant coefficients for each hour.
H@L represents the estimated parameter for $Y_{d-L,H}$ for a model on $Y_{d,h}$. Sun, Mon, $\ldots$, Sat represent the weekday dummies.
The corresponding $\hat{\iota}_{h,i}$ value is given in parenthesis.}
\end{table}

\clearpage
\subsection*{Nordpool Sweden(4)}

\begin{table}[ht]
\centering
\begin{tabular}{rrrrrrrr}
  \hline
 & Importance: 1 & Importance: 2 & Importance: 3 & Importance: 4 & Importance: 5 & Importance: 6 & Importance: 7 \\ 
  \hline
0 & 23@1 (74.57) &  1@1 (10.65) &  4@1 ( 9.78) & 18@3 ( 2.33) & 17@2 ( 0.96) &  7@4 ( 0.63) & Sat ( 0.50) \\ 
  1 & 23@1 (46.44) &  4@1 (13.77) &  1@1 ( 9.20) & 22@6 ( 5.28) & 22@2 ( 4.92) &  3@3 ( 3.10) & 21@2 ( 2.90) \\ 
  2 & 23@1 (41.80) &  4@1 (16.20) & 22@2 ( 6.91) & 22@6 ( 6.81) &  3@1 ( 6.64) &  3@3 ( 4.90) & 22@3 ( 3.90) \\ 
  3 & 23@1 (35.98) &  4@1 (22.09) & 22@6 ( 8.95) & 22@2 ( 6.30) & 22@3 ( 4.94) &  3@3 ( 4.83) &  4@5 ( 3.39) \\ 
  4 & 23@1 (27.78) &  4@1 (16.84) & 22@2 ( 5.81) & 23@6 ( 5.18) &  4@5 ( 4.80) & 22@6 ( 3.42) &  5@6 ( 2.89) \\ 
  5 & 23@1 (26.23) &  4@1 (17.77) & 21@1 (10.58) & 22@6 ( 8.55) & 22@7 ( 5.45) &  5@7 ( 4.66) &  5@6 ( 4.40) \\ 
  6 & 21@1 (25.42) & 23@1 (14.07) & 22@6 ( 8.11) &  8@7 ( 7.58) & 11@7 ( 6.84) & 11@1 ( 6.48) & 19@1 ( 5.70) \\ 
  7 & 21@1 (17.21) &  8@7 (16.37) & 11@7 ( 9.53) & 11@1 ( 9.26) & 10@7 ( 5.11) & 16@1 ( 5.00) & 18@1 ( 4.56) \\ 
  8 & 21@1 (24.41) &  8@7 (15.96) & 11@7 (12.62) & 11@1 ( 7.48) & 19@1 ( 4.13) & 21@3 ( 3.76) & 16@1 ( 3.73) \\ 
  9 & 21@1 (30.62) &  8@7 (16.67) & 11@7 (12.07) & 19@1 ( 4.19) & 11@1 ( 4.03) & 21@3 ( 4.00) & 20@7 ( 2.73) \\ 
  10 & 21@1 (26.27) &  8@7 (17.43) & 11@7 (10.60) & 18@1 ( 6.24) & 11@1 ( 5.22) & 10@7 ( 3.11) &  7@1 ( 3.06) \\ 
  11 & 21@1 (30.07) &  8@7 (13.90) & 11@7 ( 9.53) & 11@1 ( 5.41) &  6@7 ( 4.19) & 18@1 ( 3.62) & 16@1 ( 3.40) \\ 
  12 & 21@1 (19.11) & 22@1 (11.14) & 16@1 ( 7.47) &  8@7 ( 5.79) & 11@1 ( 5.62) & 19@1 ( 4.79) &  7@2 ( 4.32) \\ 
  13 & 21@1 (17.35) &  8@7 ( 7.06) & 22@1 ( 5.96) &  7@2 ( 5.10) &  8@1 ( 4.23) & 23@1 ( 3.94) & 18@1 ( 3.45) \\ 
  14 & 21@1 (18.97) &  8@7 ( 5.70) &  7@2 ( 4.75) & 18@1 ( 4.40) & 16@1 ( 4.23) &  8@1 ( 4.14) & 11@1 ( 3.88) \\ 
  15 & 21@1 (20.49) & 16@1 ( 8.22) &  8@1 ( 7.50) &  8@7 ( 4.90) & 11@7 ( 4.86) & 23@1 ( 4.70) & 22@1 ( 4.44) \\ 
  16 & 21@1 (13.72) & 16@1 (12.11) & 11@1 ( 5.77) &  7@2 ( 5.64) & 11@7 ( 4.82) & 22@6 ( 4.33) & 22@1 ( 4.25) \\ 
  17 & 21@1 (18.05) &  8@7 (12.49) & 17@1 (10.94) &  9@1 (10.56) & 11@7 (10.54) & 11@1 ( 4.33) & 21@3 ( 3.53) \\ 
  18 & 21@1 (19.43) &  8@7 (17.33) & 11@7 (10.77) & 19@1 ( 8.85) & 18@1 ( 6.29) & 11@1 ( 5.24) & 21@3 ( 4.49) \\ 
  19 & 19@1 (11.97) & 21@1 (10.90) &  8@7 ( 9.15) & 11@7 ( 8.18) & 11@1 ( 7.31) & 20@1 ( 5.64) &  7@5 ( 5.00) \\ 
  20 & 21@1 (23.22) &  8@7 ( 9.37) & 11@7 ( 8.54) & 20@1 ( 7.37) & 22@1 ( 5.51) & 19@1 ( 5.50) &  7@5 ( 4.95) \\ 
  21 & 21@1 (54.22) & 22@1 (26.19) & 23@1 ( 7.75) & 14@6 ( 3.83) & Mon ( 2.46) & 16@6 ( 1.92) & Fri ( 1.51) \\ 
  22 & 22@1 (66.34) & 23@1 (10.96) &  0@1 ( 4.78) & 22@5 ( 3.76) & 21@1 ( 2.87) &  9@4 ( 2.06) & 21@6 ( 1.94) \\ 
  23 & 23@1 (54.20) & 22@1 (26.52) & 23@2 ( 6.36) &  0@1 ( 3.83) &  1@3 ( 2.82) &  4@5 ( 2.70) & 23@5 ( 1.72) \\ 
   \hline
\end{tabular}
\caption{Most relevant coefficients for each hour.
H@L represents the estimated parameter for $Y_{d-L,H}$ for a model on $Y_{d,h}$. Sun, Mon, $\ldots$, Sat represent the weekday dummies.
The corresponding $\hat{\iota}_{h,i}$ value is given in parenthesis.}
\end{table}

\clearpage
\subsection*{POLPX Poland}

\begin{table}[ht]
\centering
\begin{tabular}{rrrrrrrr}
  \hline
 & Importance: 1 & Importance: 2 & Importance: 3 & Importance: 4 & Importance: 5 & Importance: 6 & Importance: 7 \\ 
  \hline
0 & 23@1 (23.8) &  0@1 (13.9) &  7@1 ( 8.5) &  1@1 ( 7.0) & 23@2 ( 6.8) &  0@7 ( 6.0) & 22@1 ( 3.5) \\ 
  1 & 23@1 (28.1) &  1@1 (23.1) &  1@7 ( 9.6) &  3@1 ( 8.9) & 22@1 ( 5.8) & Sat ( 5.6) &  6@1 ( 4.5) \\ 
  2 & 23@1 (23.6) &  3@1 (15.2) &  2@1 (10.4) & 22@1 ( 8.9) &  3@7 ( 8.1) &  4@1 ( 7.7) &  5@1 ( 6.7) \\ 
  3 &  4@1 (23.3) & 23@1 (17.5) &  3@1 (15.3) &  3@7 ( 9.5) & 22@1 ( 7.7) & Sun ( 7.2) & Sat ( 4.2) \\ 
  4 &  4@1 (29.4) & 23@1 (13.2) &  5@1 (11.6) & Sun (10.4) &  3@7 ( 8.6) & 22@1 ( 8.2) &  4@5 ( 4.6) \\ 
  5 &  5@1 (32.6) & Sun (10.9) & 22@1 (10.6) & 23@1 (10.1) &  5@7 ( 8.7) &  6@5 ( 5.1) & 22@2 ( 3.4) \\ 
  6 &  6@1 (23.2) & Sun (11.4) & 23@1 ( 9.0) &  6@7 ( 8.2) & 22@1 ( 7.7) & Mon ( 4.8) &  6@6 ( 3.8) \\ 
  7 & Sun (17.8) & 23@1 (14.5) &  7@1 (14.2) &  7@7 (12.0) & Mon ( 7.4) & Sat ( 6.7) &  7@6 ( 5.3) \\ 
  8 &  8@1 (17.9) & Sun (12.9) & 22@1 (11.3) &  8@7 (10.5) & Mon ( 5.7) & 22@8 ( 4.1) & Sat ( 4.0) \\ 
  9 &  9@1 (20.1) & Sun (16.6) & 22@1 ( 7.5) & Mon ( 7.2) & Sat ( 5.3) &  9@7 ( 4.8) &  9@14 ( 4.5) \\ 
  10 & 10@1 (20.4) & Sun (17.1) & Mon (10.1) & 21@1 ( 6.8) & 22@1 ( 6.1) & 12@1 ( 5.2) &  8@6 ( 3.9) \\ 
  11 & 12@1 (19.7) & 14@7 ( 9.7) & Sun ( 7.4) & 14@8 ( 5.6) & 22@1 ( 4.7) & 12@3 ( 4.0) & Mon ( 3.9) \\ 
  12 & 12@1 (24.1) & 14@7 ( 9.1) & 12@21 ( 5.9) & 12@14 ( 5.2) & 13@8 ( 4.7) & Mon ( 4.1) & Sun ( 4.0) \\ 
  13 & 12@1 (19.1) & 14@7 ( 8.3) & 13@14 ( 5.6) & 13@21 ( 5.3) & 15@2 ( 4.0) & Mon ( 3.9) & 13@8 ( 3.8) \\ 
  14 & 14@1 (12.3) & 14@7 ( 9.8) & 12@1 ( 9.7) & 11@8 ( 6.5) & 14@21 ( 6.5) & 14@14 ( 5.1) & 15@1 ( 4.7) \\ 
  15 & 15@1 (26.2) & 14@7 (11.5) & Sun ( 6.7) & 14@8 ( 6.3) & 15@14 ( 5.5) & Mon ( 5.4) & 22@1 ( 5.3) \\ 
  16 & 16@1 (14.5) & 16@7 ( 8.9) & 17@1 ( 7.8) & 22@1 ( 7.2) & Sun ( 6.5) & 16@21 ( 4.9) & 15@1 ( 4.7) \\ 
  17 & 17@1 (41.4) & 17@6 (10.2) & Sun ( 9.0) & 18@1 ( 7.9) & 16@7 ( 7.7) & Mon ( 7.5) & 16@8 ( 5.6) \\ 
  18 & 18@1 (51.5) & 18@6 ( 8.6) & Sun ( 7.0) & Mon ( 6.5) & 16@7 ( 4.8) & 16@8 ( 4.0) & 18@8 ( 3.8) \\ 
  19 & 19@1 (46.2) & 18@8 ( 7.1) & 18@6 ( 5.1) & Sun ( 4.2) & Mon ( 4.1) & 19@7 ( 3.6) & 20@1 ( 3.4) \\ 
  20 & 20@1 (56.0) & Mon ( 5.5) & 20@6 ( 4.6) & 22@8 ( 4.5) & 22@7 ( 3.9) & Sun ( 2.9) & 20@20 ( 2.6) \\ 
  21 & 21@1 (43.1) & 22@7 ( 8.6) & 20@1 ( 5.6) & 11@8 ( 5.4) & 21@6 ( 5.2) & 21@7 ( 5.2) & Sat ( 4.8) \\ 
  22 & 22@1 (34.6) & 22@7 (20.7) & 22@8 ( 6.7) & 22@6 ( 5.5) & 22@35 ( 4.5) & 22@21 ( 4.1) & 23@1 ( 3.8) \\ 
  23 & 23@1 (51.4) & 23@7 (10.8) & 23@6 ( 9.1) & Sat ( 5.6) & Mon ( 5.2) & 23@21 ( 4.2) &  1@3 ( 2.4) \\ 
   \hline
\end{tabular}
\caption{Most relevant coefficients for each hour.
H@L represents the estimated parameter for $Y_{d-L,H}$ for a model on $Y_{d,h}$. Sun, Mon, $\ldots$, Sat represent the weekday dummies.
The corresponding $\hat{\iota}_{h,i}$ value is given in parenthesis.}
\end{table}

\clearpage
\subsection*{OTE Czech}

\begin{table}[ht]
\centering
\begin{tabular}{rrrrrrrr}
  \hline
 & Importance: 1 & Importance: 2 & Importance: 3 & Importance: 4 & Importance: 5 & Importance: 6 & Importance: 7 \\ 
  \hline
0 & 23@1 (55.3) & 21@1 (10.1) &  4@1 ( 9.2) &  7@1 ( 8.2) & Sat ( 5.1) & Sun ( 3.9) &  3@1 ( 3.1) \\ 
  1 & 23@1 (43.4) &  4@1 (14.1) & Sun (11.7) & 21@1 ( 9.7) &  7@1 ( 9.4) &  3@1 ( 4.5) & 19@1 ( 2.4) \\ 
  2 & 23@1 (37.3) &  4@1 (18.5) & Sun (10.4) & 20@1 ( 7.3) & 18@1 ( 4.7) &  3@6 ( 4.5) & 19@1 ( 4.4) \\ 
  3 & 23@1 (33.3) &  4@1 (22.4) & Sun (11.9) & 20@1 ( 9.2) & 19@1 ( 4.5) &  3@1 ( 4.1) & 18@1 ( 4.1) \\ 
  4 & 23@1 (28.4) &  4@1 (21.6) & Sun (15.3) & 20@1 ( 8.3) &  3@6 ( 5.1) & 18@1 ( 4.5) & 19@1 ( 4.2) \\ 
  5 & 23@1 (25.5) & Sun (20.2) & 20@1 (11.1) &  4@1 ( 9.4) &  5@6 ( 7.9) & 19@1 ( 4.2) &  6@7 ( 3.1) \\ 
  6 & Sun (28.6) & 23@1 (16.8) & 20@1 (15.4) & Sat (12.5) & Mon ( 7.6) &  6@1 ( 5.4) & 19@1 ( 4.5) \\ 
  7 & Sun (25.7) & 20@1 (17.3) & Sat (14.1) & 23@1 ( 8.1) & Mon ( 7.0) &  7@1 ( 4.6) & 18@1 ( 4.4) \\ 
  8 & Sun (25.4) & 20@1 (16.4) & Sat (13.7) & Mon ( 7.7) &  8@1 ( 6.7) & 23@1 ( 5.9) &  8@7 ( 3.9) \\ 
  9 & Sun (22.4) & 20@1 (18.4) & Sat (11.0) & 11@1 ( 9.8) & Mon ( 9.5) & 17@1 ( 6.9) &  8@7 ( 6.4) \\ 
  10 & Sun (17.5) & 20@1 (12.0) & 11@1 (10.5) & Mon (10.4) & 22@1 ( 9.4) & Sat ( 9.1) & 11@7 ( 5.3) \\ 
  11 & 22@1 (14.1) & Sun (12.6) & Mon (11.3) & 11@1 (11.2) & Sat ( 8.3) & 20@1 ( 6.3) & 12@1 ( 5.7) \\ 
  12 & Sun (13.0) & 22@1 (11.1) & Mon (10.3) & 14@1 ( 9.7) & Sat ( 8.0) & 12@1 ( 6.6) & 14@7 ( 4.5) \\ 
  13 & 14@1 (14.3) & Sun (14.1) & 22@1 (13.3) & Mon (12.1) & Sat ( 8.6) & 14@7 ( 7.1) & 17@1 ( 5.4) \\ 
  14 & Sun (16.3) & 14@1 (12.2) & Mon (10.6) & 22@1 ( 9.7) & Sat ( 8.2) & 14@7 ( 7.2) & 16@1 ( 6.9) \\ 
  15 & Sun (17.7) & 16@1 (14.5) & Mon (11.1) & 22@1 ( 9.9) & Sat ( 7.6) & 14@7 ( 6.8) & 17@1 ( 6.7) \\ 
  16 & Sun (16.6) & 16@1 (14.9) & 17@1 (12.3) & Mon (10.3) & 22@1 ( 9.4) & 16@7 ( 7.1) & Sat ( 6.7) \\ 
  17 & 17@1 (35.0) & Sun (12.7) & Mon (12.0) & 17@7 ( 9.5) & 17@6 ( 5.7) & Sat ( 5.3) & 22@1 ( 3.8) \\ 
  18 & 18@1 (30.6) & Mon (10.9) & 18@6 (10.5) & Sun ( 9.4) & 18@7 ( 7.7) & 17@1 ( 7.1) & Sat ( 4.9) \\ 
  19 & 19@1 (39.4) & Mon (12.0) & 19@6 ( 7.5) & 19@7 ( 6.9) & 20@1 ( 6.7) & Sun ( 6.1) & Sat ( 5.8) \\ 
  20 & 20@1 (45.8) & Mon (13.2) & 20@6 (10.9) & Sat ( 6.5) & 20@7 ( 6.3) & 22@1 ( 4.2) & 20@21 ( 3.4) \\ 
  21 & 21@1 (30.6) & 22@1 (14.0) & 21@6 (10.1) & Mon ( 9.5) & Sat ( 6.2) & 21@7 ( 5.6) & 20@1 ( 4.9) \\ 
  22 & 22@1 (45.7) & 22@6 ( 8.6) & Mon ( 5.6) & 22@4 ( 5.5) & Sat ( 4.2) & 22@21 ( 3.9) & 22@7 ( 3.8) \\ 
  23 & 22@1 (30.6) & 23@1 (26.1) & 23@6 ( 7.9) & Mon ( 6.9) & 22@2 ( 4.1) & 23@4 ( 3.7) & 23@35 ( 2.7) \\ 
   \hline
\end{tabular}
\caption{Most relevant coefficients for each hour.
H@L represents the estimated parameter for $Y_{d-L,H}$ for a model on $Y_{d,h}$. Sun, Mon, $\ldots$, Sat represent the weekday dummies.
The corresponding $\hat{\iota}_{h,i}$ value is given in parenthesis.}
\end{table}

\end{document}